\documentclass[runningheads]{llncs}

\usepackage{eccv}

\usepackage{eccvabbrv}

\usepackage{graphicx}
\usepackage{booktabs}

\usepackage[accsupp]{axessibility}  

\usepackage{hyperref}

\usepackage{orcidlink}

\usepackage{booktabs}
\usepackage{color, colortbl}
\usepackage{arydshln}
\definecolor{LightGreen}{RGB}{204, 255, 204}
\definecolor{LightRed}{RGB}{255, 204, 204}
\usepackage{float}
\usepackage{capt-of}
\usepackage{amsmath}
\usepackage{comment}
\DeclareMathOperator*{\argmax}{arg\,max}

\begin{document}

\title{EGIC: Enhanced Low-Bit-Rate Generative Image Compression Guided by Semantic Segmentation} 

\titlerunning{EGIC}

\author{Nikolai Körber\inst{1, 2}\orcidlink{0000-0002-1034-7898} \and
Eduard Kromer\inst{2}\orcidlink{0000-0003-4540-8061} \and
Andreas Siebert\inst{2}\orcidlink{0009-0002-5144-1482} \and \\
Sascha Hauke\inst{2}\orcidlink{0000-0001-7822-0191} \and
Daniel Mueller-Gritschneder\inst{3}\orcidlink{0000-0003-0903-631X} \and
Björn Schuller\inst{1}\orcidlink{0000-0002-6478-8699}}

\authorrunning{N.~Körber et al.}


\institute{Technical University of Munich, Munich, Germany \\
\email{\{nikolai.koerber, schuller\}@tum.de}\\ \and
University of Applied Sciences Landshut, Landshut, Germany \\
\email{\{eduard.kromer, andreas.siebert, sascha.hauke\}@haw-landshut.de}\\ \and
TU Wien, Vienna, Austria \\
\email{daniel.mueller-gritschneder@tuwien.ac.at}
}

\maketitle

\begin{abstract}

  We introduce EGIC, an enhanced generative image compression method that allows traversing the distortion-perception curve efficiently from a single model. EGIC is based on two novel building blocks: i) OASIS-C, a conditional pre-trained semantic segmentation-guided discriminator, which provides both spatially and semantically-aware gradient feedback to the generator, conditioned on the latent image distribution, and ii) Output Residual Prediction (ORP), a retrofit solution for multi-realism image compression that allows control over the synthesis process by adjusting the impact of the residual between an MSE-optimized and GAN-optimized decoder output on the GAN-based reconstruction. Together, EGIC forms a powerful codec, outperforming state-of-the-art diffusion and GAN-based methods (\eg, HiFiC, MS-ILLM, and DIRAC-$100$), while performing almost on par with VTM-20.0 on the distortion end. EGIC is simple to implement, very lightweight, and provides excellent interpolation characteristics, which makes it a promising candidate for practical applications targeting the low bit range.

\begin{figure}[tb]
    \centering
    \includegraphics[height=4.8cm]{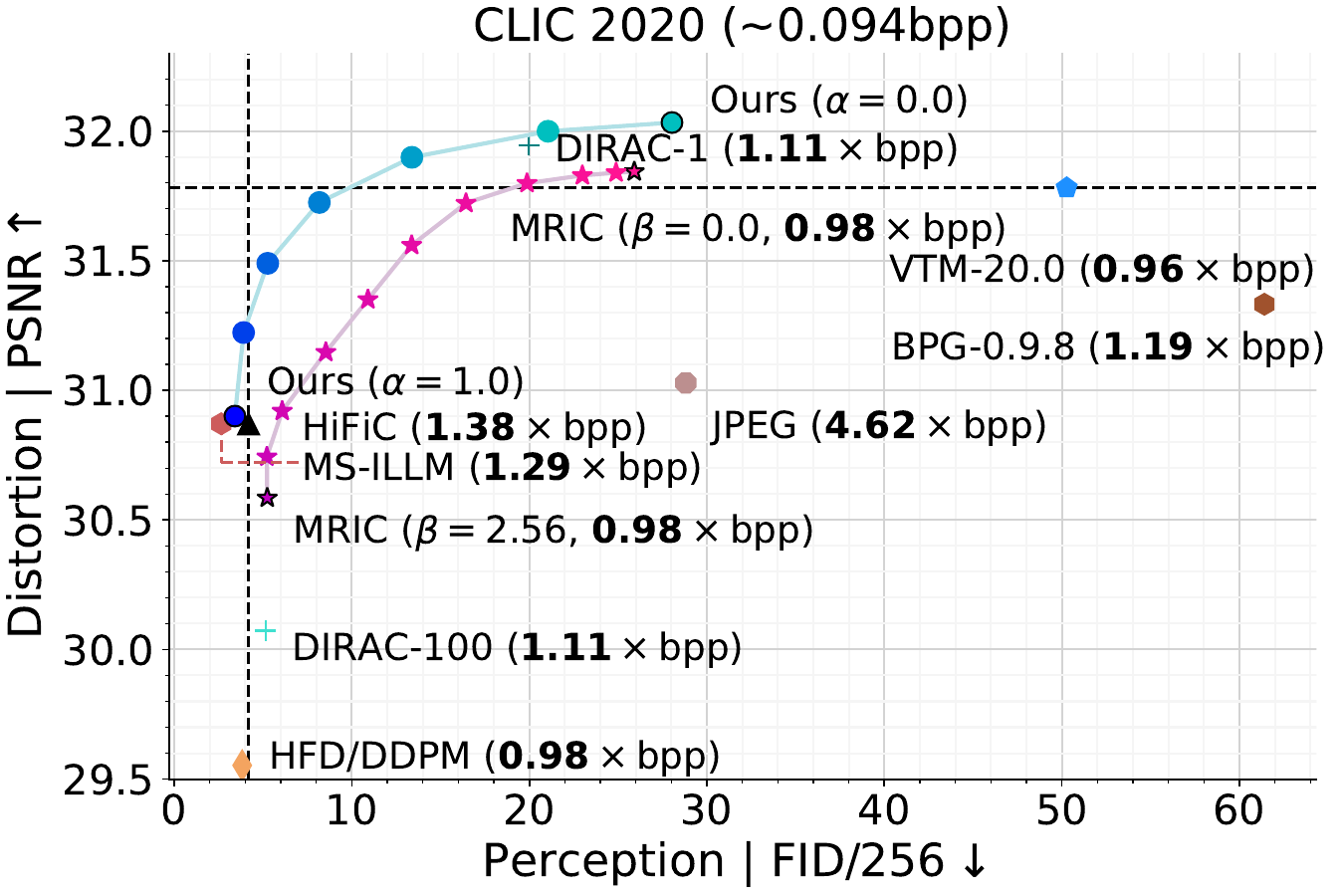}
    \caption{Distortion-perception comparison (top left is best)}
    \label{fig:teaser}
\end{figure}

\end{abstract}

\section{Introduction}\label{sec:intro}

Neural image compression methods incorporating generative models (\eg, Generative Adversarial Networks, short GANs~\cite{NIPS2014_5ca3e9b1}) have been able to achieve comparable perceptual quality at considerably lower bit-rates~\cite{agustsson2019generative, mentzer2020high}, hence being a promising direction for storage-efficient and bandwidth-constrained applications. Their underlying principle is that missing information can be realistically synthesized (\eg, textures), therefore allowing more control over highly-sensitive information. Formally, these methods fall into the category of lossy compression with high perception~\cite{ICML-2019-BlauM, 2021PerceptualReconstruction, 2022ControllablePerceptualCompression}, \ie, we are interested in the lowest possible distortion for a given bit-rate with the constraint that the reconstructions follow the underlying data distribution. Note that this definition implies that low distortion alone does not per se yield good perceptual quality. In fact, it has been shown that perception and distortion are at odds with each other~\cite{ICML-2019-BlauM}. 

With the rise of diffusion models~\cite{NEURIPS2021_49ad23d1}, there has been increasing efforts to rival GANs in the context of generative image compression~\cite{theis2023lossy, Yang2023, noor_2023, Hoogeboom_2023}. While these models have garnered attention for their better training dynamics and for their competitive or even higher image sample quality, they often come at considerably increased computational cost and inference latency. For example,  the denoising network DIRAC-$n$~\cite{noor_2023} requires additional $108.4$M parameters and up to $n=100$ sampling steps compared to the base codec. In HFD/DDPM~\cite{Hoogeboom_2023}, Hoogeboom~\etal even considered a larger setup, using more than $1$B additional model parameters and up to $250$ sampling steps. In this work, we challenge the prevailing belief in the superiority of diffusion models over GANs for generative image compression.


We propose EGIC, an \textbf{E}nhanced \textbf{G}enerative \textbf{I}mage \textbf{C}ompression method. EGIC is based on a novel conditional pre-trained semantic segmentation-guided discriminator (OASIS-C), which provides both spatially and semantically-aware gradient feedback to the generator, conditioned on the latent image distribution. Semantic segmentation-guided discriminators have been originally proposed for semi-supervised semantic segmentation~\cite{8237868} and later been adopted for the task of semantic image synthesis~\cite{schoenfeld2021you}. Both fields have played an important role for generative image compression early on. For example, the extreme learned image compression method GC ($D^{+}$)~\cite{agustsson2019generative} largely builds upon the success of pix2pixHD~\cite{pix2pix2017, wang2018pix2pixHD}, a powerful image-to-image translation method. In~\cite{agustsson2019generative}, ($D^{+}$) further refers to a specific configuration, in which the discriminator $D$ is presented with semantic labels as additional side information. Back then, however, only restrictive image domains were considered (the Cityscapes dataset~\cite{Cordts2016Cityscapes}). In this work, we go one step further and demonstrate that semantic segmentation-guided discriminators, through careful design, can also considerably boost the generative compression performance across the image domain. 

A common criticism of generative image compression methods is the lack of transparency in the underlying generation process. As pointed out by Agustsson~\etal~\cite{Agustsson_2023_CVPR}, users might worry that the reconstructions deviate too much from the original image. Multi-realism image compression algorithms try to address this by providing the user with a choice: from a single compressed representation, we can either obtain a reconstruction that resembles more the traditional compression setting (low distortion), a visually appealing reconstruction (high perception), or anything in between. Unfortunately, existing solutions~\cite{Wang_2019_CVPR, 9412185, 2022ControllablePerceptualCompression, noor_2023, Dosovitskiy2020You, Agustsson_2023_CVPR, Iwai_2024_WACV} typically come at the expense of considerably increased model size, decoding latency and/ or reduced overall performance.

We propose Output Residual Prediction (ORP), an efficient retrofit solution for multi-realism image compression. ORP is inspired by recent theoretical findings that simple image interpolation between an MSE-optimized decoder and a perfect perceptual decoder is sufficient to achieve any point on the distortion-perception (D-P) curve~\cite{2022ControllablePerceptualCompression}. The main idea is to (implicitly) predict the residual $R$ between an MSE-optimized and GAN-optimized decoder output, which allows control over the synthesis process by adjusting the impact of the residual ($\alpha \in [0, 1]$) on the GAN-based reconstruction, see~\cref{fig:teaser}. Compared to existing solutions, ORP only requires a fraction of additional model parameters (\eg, $0.15\times$ compared to MRIC~\cite{Agustsson_2023_CVPR}) and only requires a single inference-cycle (as opposed to DIRAC-$n$~\cite{noor_2023}), revealing that more sophisticated methods may in fact be unnecessary. In summary our contributions are:

\begin{enumerate}
  \item We introduce EGIC, a novel generative image compression method that allows traversing the D-P curve efficiently from a single model. EGIC is based on two core building blocks (\cref{sec:our_approach}): i) OASIS-C, a conditional pre-trained semantic segmentation-guided discriminator, and ii) ORP, a lightweight retrofit solution for multi-realism image compression.
  \item We conduct a thorough study to identify suitable discriminator architectures/ GAN formulations for the task of generative compression (\cref{chapter:preliminary_study}).
  \item We empirically evaluate the effectiveness of our method on both convolutional (HiFiC~\cite{mentzer2020high}) and transformer-based (SwinT-ChARM~\cite{zhu2022transformerbased}) backbones, on three challenging benchmark datasets (\cref{sec:comp_soa}). We find that EGIC is particularly well-suited for the low bit range. On the perception end, EGIC outperforms a wide-variety of diffusion and GAN-based methods (\eg, HiFiC, MS-ILLM, DIRAC-$100$), while being considerably more storage-efficient (\eg, $0.03\times$ model parameters compared to HFD/DDPM). On the distortion end, EGIC almost matches VTM-20.0, the state-of-the-art for non-learned image codecs, while providing excellent interpolation characteristics for all other operating modes in between.
\end{enumerate}

\section{Related Work}

\textbf{Generative image compression.} Agustsson \etal~\cite{agustsson2019generative} demonstrated that an extreme learned image compression method combined with a multi-scale PatchGAN discriminator~\cite{pix2pix2017} can achieve compression rates far beyond the prior state-of-the-art while maintaining similar perceptual quality. Their work was later refined and extended by a hyper-prior~\cite{ballé2018variational}, formally known as HiFiC~\cite{mentzer2020high}. In MRIC~\cite{Agustsson_2023_CVPR}, the authors further pushed the rate-distortion-perception frontier by incorporating more powerful building blocks into their system~\cite{He_2022_CVPR, 9190935}.

Yan~\etal~\cite{2021PerceptualReconstruction} proposed an allegedly optimal training framework that achieves the lowest possible distortion under the perfect perception constraint for a given bit-rate. Essentially, the authors state that a perceptual decoder can be trained using solely a GAN conditioned on an encoder optimized under the traditional rate-distortion objective. In~\cite{2021PerceptualReconstruction}, WGAN-GP~\cite{pmlr-v70-arjovsky17a, NIPS2017_892c3b1c} is employed using the vanilla concatenation-based conditioning scheme presented in~\cite{mirza_2014}. While theoretically appealing, the authors have only been able to demonstrate superior performance on the MNIST dataset. Their ideas were later refined in~\cite{2022ControllablePerceptualCompression}, but still did not reach the performance of HiFiC. From both works, it appears that their success is highly dependent on the underlying conditional GAN framework; it is interesting to note that most current works~\cite{mentzer2020high, Agustsson_2023_CVPR, He_2022_CVPRW, 2021PerceptualReconstruction, 2022ControllablePerceptualCompression, pmlr-v202-muckley23a} use a concatenation-based conditioning scheme, which is known to be inferior to projection~\cite{miyato2018cgans}.

Although there are numerous other works~\cite{NEURIPS2018_801fd8c2, He_2022_CVPRW, pmlr-v70-rippel17a, 9412185, 8456298, ICML-2019-BlauM, 2022ControllablePerceptualCompression, 2021PerceptualReconstruction}, we argue that the fundamental GAN principles have barely changed\footnote{An exception is MS-ILLM~\cite{pmlr-v202-muckley23a}, a concurrent work which we became aware of only during the completion of this work. We provide a short comparison in~\cref{sec:our_approach}.}. For example, in PO-ELIC, a recent work, He \etal~\cite{He_2022_CVPRW} use the same PatchGAN discriminator architecture as in HiFiC and MRIC, but with hinge-loss. Recent advances in the field of generative image compression can therefore mainly be attributed to improved building blocks, but not to more powerful generative models. 

An exception to this line of work are diffusion-based methods~\cite{theis2023lossy, Yang2023, noor_2023, Hoogeboom_2023}. Diffusion models have recently rivaled GANs~\cite{NEURIPS2021_49ad23d1}, often achieving competitive or higher image sample quality. While this direction is promising, their practical use is currently hindered by the high computational cost. 

Finally, a common criticism of generative image compression methods is the lack of transparency in the underlying generation process. As pointed out by Agustsson~\etal~\cite{Agustsson_2023_CVPR}, users might worry that the reconstructions deviate too much from the original image. However, this concern is not a limitation in general and can be addressed via universal rate-distortion-perception representations~\cite{zhang2021universal}. Existing solutions have considered image/ weight interpolation~\cite{Wang_2019_CVPR, 9412185, 2022ControllablePerceptualCompression, Xu2023}, denoising diffusion probabilistic models for residual prediction~\cite{noor_2023} and loss-conditional training~\cite{Dosovitskiy2020You, Agustsson_2023_CVPR, Iwai_2024_WACV}.

\textbf{Semantic image synthesis/ generative models.} Semantic image synthesis has played an important role in generative image compression from the very beginning~\cite{agustsson2019generative, wang2018pix2pixHD}. While its use has been primarily advertised for constrained application domains with semantic label maps available (\eg, the Cityscapes dataset~\cite{Cordts2016Cityscapes}), recent work~\cite{park2019SPADE, schoenfeld2021you, NEURIPS2021_4eff0720} as well as better semantic segmentation models~\cite{cheng2020panoptic, deeplabv3plus2018, xie2021segformer}, show its generation ability across the image domain. Of particular importance to us are semantic segmentation-guided discriminators, which have been originally proposed for semi-supervised semantic segmentation~\cite{8237868} and later been adopted for the task of semantic image synthesis~\cite{schoenfeld2021you}. The general idea is to convert the discriminator to a multi-class classifier, where the additional classes correspond to regular semantic labels. 

Other promising candidates we consider in this work are the SESAME~\cite{10.1007/978-3-030-58542-6_24}, the U-Net~\cite{schonfeld2020u}, and the projected discriminators~\cite{Sauer2021NEURIPS}. The SESAME discriminator has been introduced as a multi-scale and improved variant of the PatchGAN discriminator, while the use of U-Net and projected discriminators have led to significant advances over BigGAN~\cite{brock2018large} and StyleGAN~\cite{NEURIPS2020_8d30aa96}, respectively, arguably the two most popular GAN families.

Another interesting line of work are frequency-aware GANs~\cite{jiang2021focal, NEURIPS2021_96bf57c6, gal2021swagan}. These methods are based on the observation that the statistics of GAN-generated images often differ considerably from real images in the frequency domain. In this work, we use the Focal Frequency Loss (FFL)~\cite{jiang2021focal} as a tool to quantify the frequency awareness of each method.

\section{Background}

\textbf{Traditional rate-distortion trade-off.} We follow the same notation as in previous works~\cite{mentzer2020high}: a neural image compression method consists of three components, an encoder $E$, a decoder $G$ (hereafter referred to as generator) and an entropy model $P$. Specifically, $E$ encodes $x$ to a quantized latent representation $y=E(x)$, while $G$ creates a reconstruction of the original image $x'=G(y)$. The learning objective is to minimize the rate-distortion trade-off~\cite{cover2012elements}, with $\lambda > 0$:

\begin{equation}\label{eq:rd_objective}
\mathcal{L}_{RD}=\mathop{\mathbb{E}_{x\sim p_X}}[\lambda r(y) + d(x, x')].
\end{equation}

In \cref{eq:rd_objective}, the bit-rate is estimated using the cross entropy $r(y)=-\log{P(y)}$, where $P$ represents a probability model of $y$ and $d(x, x')$ is a full-reference metric. In practice, an entropy coding method based on $P$ is used to obtain the final bit representation, \eg, using adaptive arithmetic coding. For a more general overview of neural compression, we refer the interested reader to~\cite{Yang2022a}.

\textbf{Rate-distortion-perception trade-off.} In Mentzer \etal~\cite{mentzer2020high}, a discriminator $D$ is further added to navigate the triple trade-off~\cite{ICML-2019-BlauM} using the non-saturating loss~\cite{NIPS2014_5ca3e9b1}:

\begin{equation}\label{eq:generator_objective}
\mathcal{L}_{RDP}=\mathop{\mathbb{E}_{x\sim p_X}}[\lambda r(y) + d(x, x') - \beta \log(D(x', y))],
\end{equation}
\begin{equation}\label{eq:discriminator_objective}
\mathcal{L}_{disc}=\mathop{\mathbb{E}_{x\sim p_X}}[-\log(1-D(x', y))] + \mathop{\mathbb{E}_{x\sim p_X}}[-\log(D(x, y))].
\end{equation}

In \cref{eq:generator_objective}, $d(x, x')$ is decomposed into $d=k_{M}$MSE + $k_{P}$LPIPS~\cite{zhang2018perceptual}, where $k_M$ and $k_P$ are hyper-parameters. We keep this formulation to make use of the same hyper-parameters as in HiFiC. It is worth noting that the discriminator $D$ is conditioned on $y$, identical to the formulation under the optimal training framework~\cite{2021PerceptualReconstruction}. In both lines of work, a concatenation-based conditioning scheme~\cite{mirza_2014} is chosen to model $P_{X|Y}$. We will study this design decision later on.

\section{Our Approach}\label{sec:our_approach}

\textbf{Learning objective.} Inspired by recent advances in the field of semi-supervised semantic segmentation/ semantic image synthesis~\cite{8237868, schoenfeld2021you}, we redesign the discriminator to a ($N{+1}$)-class semantic segmentation task:

\begin{equation}\label{eq:generator_objective_ours}
\mathcal{L}_{ours}=\mathop{\mathbb{E}_{x\sim p_{X}}}[\lambda r(y) + d(x, x') + \beta \mathcal{L}_{wce}(x', y)],
\end{equation}
\begin{equation}\label{eq:discriminator_objective_ours}
\mathcal{L}_{seg}=\mathop{\mathbb{E}_{x\sim p_{X}}}[\mathcal{L}_{wce}(x, y)] + \mathop{\mathbb{E}_{x\sim p_{X}}}[-\sum_{i,j}^{H\times W}\log D(x', y)_{i,j,c_{ij}=N{+1}}].
\end{equation}

In our formulation, $D \in \mathbb{R}^{H\times W\times N{+1}}$ represents a probability distribution over all semantic classes $\{1, .., N{+1}\}$, with $N{+1}$ being the fake label. Note that $D$ is conditioned on $y$ following theoretical and empirical results in~\cite{mentzer2020high, 2021PerceptualReconstruction}. $\mathcal{L}_{wce}(x, y)= -\sum_{i,j}^{H\times W}w_{ij}\log D(x, y)_{i,j,c_{ij}}$ denotes the weighted ($N{+1}$)-class cross entropy loss over all pixel locations $(i,j) \in H\times W$, where $c_{ij}$ is the index of the prediction for the correct semantic class and $w_{ij}$ is a pixel weighting scheme. Different from the approach in~\cite{schoenfeld2021you}, however, we employ the more commonly used pixel loss weighting scheme presented in~\cite{Yang2019DeeperLabSI}, which puts more emphasis on small instances ($w_{ij}=3$ for area size smaller than $64 \times 64$ px, $w_{ij}=1$ everywhere else). This change is primarily due to practical considerations which will be motivated later on.

Similar to the non-saturating loss, $G$ tries to fool $D$ by generating realistic and semantically correct reconstructions, whereas $D$ tries to differentiate between $x$ and $x'$. This is essentially achieved by assigning the fake label as correct semantic class ($c_{ij}=N{+1}$). 

We additionally regularize the discriminator in~\cref{eq:discriminator_objective_ours} with the LabelMix (LM) consistency loss~\cite{schoenfeld2021you}, adapted to the compression setting:
\begin{equation}\label{eq:label_mix}
\mathcal{L}_{cons}=\|D_{\text{logits}}(\text{LM}(x, x', M), y) - \text{LM}(D_{\text{logits}}(x, y), D_{\text{logits}}(x', y), M)\|_{2}^{2},
\end{equation}

with LM($x, x', M)= M\odot x + (1-M)\odot x'$. In~\cref{eq:label_mix}, $M$ is a randomly generated binary mask that respects the underlying semantic boundaries of $x$ and LM($x, x', M$) corresponds to the resulting mixed real-fake image. The discriminator predictions are constrained to be equivariant under the LM operation, \ie, the discriminator prediction of the mixed image ($D_{\text{logits}}(\text{LM}(x, x', M), y)$) should be identical to the mixed discriminator predictions of the real and fake images, respectively ($\text{LM}(D_{\text{logits}}(x, y), D_{\text{logits}}(x', y), M)$), thus forcing the discriminator to focus more on content and structure. This is essentially achieved by applying the L2 norm on the unnormalized discriminator predictions $D_{\text{logits}}$. We use a fixed LM coefficient of $10$ for all experiments.

Our approach shares some similarities with the discriminator presented in MS-ILLM~\cite{pmlr-v202-muckley23a}, with the main difference being the labels. We directly employ human-annotated semantic labels, whereas Muckley~\etal propose to replace these with codebook indices from a pre-trained VQ-VAE~\cite{NIPS2017_7a98af17, NEURIPS2019_5f8e2fa1} model. Our work is motivated by recent findings that pixel-level supervision of the discriminator is crucial in obtaining artifact-free images~\cite{NEURIPS2021_96bf57c6}. Codebook entries of VQ-VAEs on the other hand refer to patch-based supervision ($32\times32$); the codebook size also typically exceeds the number of semantic labels by at least one order of magnitude (\eg, 134 vs 1024), which can be more challenging to train. Both approaches do not require labels during inference and can thus be considered as an enhanced version of the GC ($D^{+}$)-variant introduced in \cite{agustsson2019generative}.

\textbf{Training strategies.} We employ a two-stage training strategy. In the first stage, we use~\cref{eq:generator_objective_ours} without adversarial supervision (\ie,  $\mathcal{L}_{RD}=\mathop{\mathbb{E}_{x\sim p_{X}}}[\lambda r(y) + d(x, x')]$), using the same configuration as in the original work~\cite{mentzer2020high}. For the second stage, we consider two variants: strategy-I denotes the full learning objective as described in~\cref{eq:generator_objective_ours} and~\cref{eq:discriminator_objective_ours}, whereas strategy-II is based on pure adversarial supervision inspired by~\cite{2021PerceptualReconstruction}. In both cases, we only fine-tune $G$ and fix the pre-trained $E, P$ from stage one. The latter is motivated by the observation that an encoder trained under the traditional rate-distortion optimization is also well-suited for the perceptual compression task~\cite{2021PerceptualReconstruction}. The outputs of our training procedure are shared weights for $E$, $P$, $G_1$, and $G_2$, from stages one and two.

\begin{figure}[tb]
  \centering
  \includegraphics[height=3.5cm]{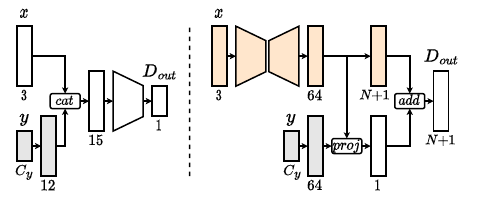}
  \caption{Schematic comparison between PatchGAN (l.) and OASIS-C (r.)
  }
  \label{fig:disc_comp}
\end{figure}

\textbf{OASIS-C.} In this section, we introduce OASIS-C, our novel conditional pre-trained semantic segmentation-guided discriminator. The name OASIS-C encapsulates both its origin (we base our work on the U-Net architecture~\cite{RFB15a} presented in OASIS~\cite{schoenfeld2021you}) and its target (\textbf{C}ompression). We provide a visual comparison to the widely adopted PatchGAN discriminator (HiFiC~\cite{mentzer2020high}) in~\cref{fig:disc_comp}. 

In HiFiC $y$ is pre-processed by a $3\times3$ Convolution-SpectralNorm-LeakyReLU layer with $12$ ﬁlters and stride $1$, upsampled and concatenated with $x$ and subsequently fed into a PatchGAN discriminator~\cite{pix2pix2017}. In OASIS-C, i) we use the same latent pre-processing blocks (highlighted gray) but instead employ a pixel-wise projection-based conditioning scheme. ii) We replace spectral norm with weight norm, which increases the overall model capacity. iii) We pre-train the (unconditional) discriminator using DeepLab2~\cite{deeplab2_2021} to accelerate training (highlighted orange); the projection layers are initialized randomly. Note that this is motivated by the fact that we also start from a pre-trained $E$, $G_1$, $P$ state (training stage two). Our formulation shares some characteristics with projected GANs~\cite{Sauer2021NEURIPS}, which projects samples to a pre-trained feature space, prior to classification. While in~\cite{Sauer2021NEURIPS} the pre-trained feature space is fixed,  we fine-tune the whole model to learn the conditional distribution $P_{X|Y}$. We justify all our design decisions in more detail in~\cref{chapter:preliminary_study}.

\begin{figure}[tb]
  \centering
  \includegraphics[height=3.5cm]{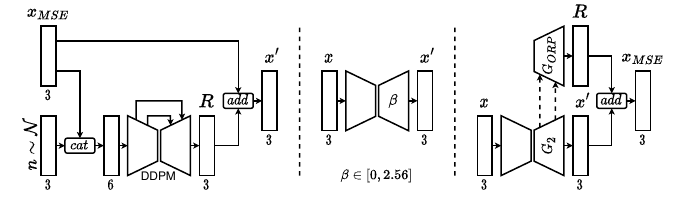}
  \caption{DIRAC-$n$~\cite{noor_2023} vs Beta Conditioning (MRIC)~\cite{Agustsson_2023_CVPR} vs ORP
  }
  \label{fig:mric_comp}
\end{figure}

\textbf{Output residual prediction (ORP).} We start by discussing existing solutions for multi-realism image compression, see \cref{fig:mric_comp} for an overview.

Dirac-$n$~\cite{noor_2023} uses a conditional denoising probabilistic model~\cite{NEURIPS2020_4c5bcfec} (DDPM) for residual prediction $R$, conditioned on an initial reconstruction $x_{MSE}$ (\eg, JPEG or learned codec). By adjusting the number of sampling steps, $n \in [1, 100]$, DIRAC-$n$ smoothly interpolates the D-P curve. Despite good performance, Dirac-$n$ has two main drawbacks: parameter overhead and speed (up to $100$ sampling steps), caused by the additional denoising network.

MRIC~\cite{Agustsson_2023_CVPR} proposes a variant of loss-conditional training~\cite{Dosovitskiy2020You} (in their paper referred to as Beta conditioning) to target different operating modes. Intuitively, the idea is to train on a distribution of losses, rather than on some specific configuration. At test time, the trained model can then be conditioned to generate outputs based on user-preferences ($\beta$). In order to work well, loss-conditional training requires a sophisticated conditioning mechanism. In practice, it also may be difficult to cover the whole spectrum of learning objectives.

ORP on the other hand is a lightweight retrofit solution that is inspired by recent theoretical findings, that simple image interpolation between an MSE-optimized decoder and a perfect perceptual decoder is sufficient to achieve any point on the D-P curve~\cite{2022ControllablePerceptualCompression}. The main idea is to (implicitly) predict the residual $R$ between a MSE-optimized and GAN-optimized decoder output:

\begin{equation}\label{eq:orp1}
x' = G_{2}(x) + (1-\alpha) R,
\end{equation}

\begin{equation}\label{eq:orp2}
\mathcal{L}_{ORP}=\mathop{\mathbb{E}_{x\sim p_{X}}}[MSE(x,x')]
\end{equation}

with $R = G_{ORP}(F)$, where $F$ are feature maps extracted from $G_2(x)$. During training $\alpha=0$, which constraints x' to the traditional MSE-optimized decoder output. During inference, $\alpha \in [0, 1]$ allows traversing any point on the D-P curve using an implicitly encoded variant of image interpolation~\cite{9412185, 2022ControllablePerceptualCompression}. Compared to existing solutions, ORP has several key advantages: i) The ORP is model agnostic and can be added to any pre-trained generative image compression method. ii) For $\alpha=1$, we always get the regular GAN-based output ($R$ is canceled out), which in practice is more difficult to obtain. iii) We only need to finetune $G_{ORP}$ (we keep $E$, $P$, $G_2$ frozen), which considerably speeds up training. 

\section{Exploring GANs for Compression}\label{chapter:preliminary_study}

In this section, we study and compare suitable discriminator architectures/ GAN formulations for the task of generative image compression. We consider PatchGAN~\cite{pix2pix2017}, SESAME~\cite{10.1007/978-3-030-58542-6_24}, U-Net~\cite{schonfeld2020u}, projected~\cite{Sauer2021NEURIPS}, and OASIS~\cite{schoenfeld2021you} discriminators. We employ training strategy-II for most parts, motivated by the observation that a perceptual decoder can be trained using solely a GAN conditioned on an encoder optimized under the traditional rate-distortion objective~\cite{2021PerceptualReconstruction}. Intuitively, a good candidate should be able to generate high-fidelity reconstructions based on pure adversarial supervision.

\textbf{Setup.} For a fair comparison, we base our experiments on the official code base of HiFiC~\cite{mentzer2020high} and DeepLab2~\cite{deeplab2_2021}, a TensorFlow library for deep labeling\footnote{We partially translate DeepLab2 to TensorFlow 1.15 to directly integrate the code base into HiFiC.}. We use the same encoder, decoder, and entropy architecture as in~\cite{mentzer2020high}, and only change the discriminator/ GAN learning objective. The latter is motivated by recent theoretical arguments that the critic is decisive in matching the distribution of the training data~\cite{Sauer2021NEURIPS}. All our experiments start from stage two, using the same pre-trained $E$, $G_1$, and $P$. As a baseline, we use the official training configuration from~\cite{mentzer2020high}, \ie, we use adversarial supervision\texttt{+}distortion terms, however with fixed $E$ and $P$ to enable identical bit-rates across all experiments.

\textbf{Datasets.} We use the Coco2017 panoptic dataset~\cite{Lin2014MicrosoftCC} with 118,287
training images and 133 semantic classes for stage one and our main experiments. To evaluate the generalization ability across the image domain, we use the following benchmarks: DIV2K~\cite{Agustsson_2017_CVPR_Workshops}, CLIC 2020~\cite{CLIC2020}, and Kodak~\cite{kodak}. DIV2K and CLIC 2020 are both high-resolution image datasets, which contain 100 and 428 images, respectively (see \cite[A.9]{mentzer2020high}); Kodak contains 24 images and is widely used as an image compression benchmark. For our preliminary study, we additionally consider a down-sized (factor $2$) version of Cityscapes~\cite{Cordts2016Cityscapes}, which contains 19 semantic classes, 2975 training, and 500 validation images, respectively.

\textbf{Training and evaluation.} Again, for a fair comparison, we use the same hyper-parameters as in HiFiC, except the learning rate, which we fix to $1\mathrm{e}{-5}$ for stage two. For strategy-II, we set $\beta=1$ to rebalance $G$ and $D$. We further reduce the number of optimization steps on Cityscapes from 1M to 150k, considering the reduced size and complexity. We use PSNR and the FID-score~\cite{NIPS2017_8a1d6947} as a measure for distortion and perception, following recent work~\cite{mentzer2020high, Agustsson_2023_CVPR}. For our preliminary experiments, we compute the patched FID-score (FID/256~\cite[A.7]{mentzer2020high}) based on the clean-FID implementation \cite{parmar2021cleanfid}; for our main experiments, we have switched to torch-fidelity~\cite{obukhov2020torchfidelity} to ease comparison to recent methods, where a recalculation based on clean-FID is prevented due to restrictive data access. As common in the literature, we pad all images and crop the resulting reconstructions.

\begin{table*}[tb]
\caption{Comparing GAN approaches on Cityscapes (0.092bpp)}
\begin{center}
\begin{tabular}{lllllclc}
\toprule
 Method     & Disc. ($D$)                                 & Cond.      & GAN objective             & \multicolumn{2}{c}{Distortion}  & \multicolumn{2}{c}{Perception}    \\ \cmidrule{5-8}
            &                                               &                   &                           & PSNR $\uparrow$ & rel-PSNR      & FID $\downarrow$  & rel-FID       \\ \midrule
 baseline   & PatchGAN~\cite{pix2pix2017}    & cat            & non-saturating            & \textbf{32.71}  & -             & \textbf{10.62}    & -             \\ \midrule
 conf-a   & PatchGAN                                      & cat            & non-saturating            & 24.32           & -25.6\%       & 112.29            & +957.3\%      \\
 conf-b   & SESAME~\cite{10.1007/978-3-030-58542-6_24}    & cat            & hinge                     & 29.43           & -10.0\%       & 75.65             & +612.3\%      \\
 conf-c   & U-Net~\cite{schonfeld2020u}                   & cat            & non-saturating            & 29.46           & -9.9\%        & 87.02             & +719.4\%      \\
 conf-d   & Projected~\cite{Sauer2021NEURIPS}             & cat            & hinge                     & 29.64           & -9.5\%        & 50.66             & +377.0\%      \\ 
 conf-e   & OASIS~\cite{schoenfeld2021you}                & cat            & CE ($N{+1}$)   & 30.03           & -8.2\%        & 16.50             & +55.4\%       \\ 
 conf-f   & DeepLabV3+~\cite{cheng2020panoptic}           & cat            & CE ($N{+1}$)   & 21.53           & -34.2\%       & 20.61             & +94.1\%       \\\midrule  
 conf-c   & U-Net                                         & proj.       & non-saturating            & 29.38           & -10.2\%       & 30.80             & +190.0\%      \\
 conf-e   & OASIS                                         & proj.       & CE ($N{+1}$)   & 29.79           & -8.9\%        & 13.54             & +27.5\%       \\\midrule 
 conf-e   & OASIS                                         & impl.         & CE ($N{+1}$)   & 29.90           & -8.6\%        & 15.30             & +44.1\%       \\\bottomrule
\end{tabular}
\end{center}
\label{table:comp_gan_formulations}
\end{table*}

\begin{figure*}[tb]
    \setlength{\tabcolsep}{1pt}
    \renewcommand{\arraystretch}{0.5}
    \centering
    \begin{tabular}{cccccc}
        \toprule
        input       & conf-a~\cite{pix2pix2017}/                     & conf-b~\cite{10.1007/978-3-030-58542-6_24}/                                     & conf-c~\cite{schonfeld2020u}/                     & conf-d~\cite{Sauer2021NEURIPS}/                         & conf-e~\cite{schoenfeld2021you}/ \\
                    & PatchGAN   & SESAME    & U-Net   & Projected & OASIS \\
        \midrule
        \begin{subfigure}{0.16\textwidth}
            \centering
            \includegraphics[width=\linewidth]{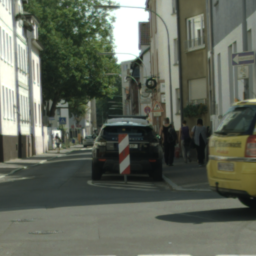}
        \end{subfigure}
        & 
        \begin{subfigure}{0.16\textwidth}
            \centering
            \includegraphics[width=\linewidth]{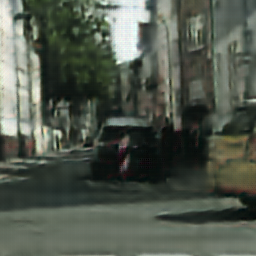}
        \end{subfigure} 
        & 
        \begin{subfigure}{0.16\textwidth}
            \centering
            \includegraphics[width=\linewidth]{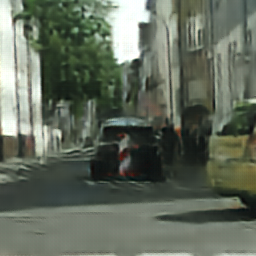}
        \end{subfigure} 
        & 
        \begin{subfigure}{0.16\textwidth}
            \centering
            \includegraphics[width=\linewidth]{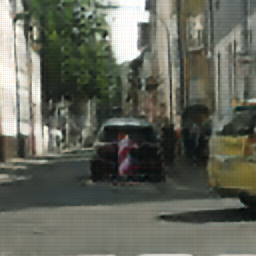}
        \end{subfigure} 
        & 
        \begin{subfigure}{0.16\textwidth}
            \centering
            \includegraphics[width=\linewidth]{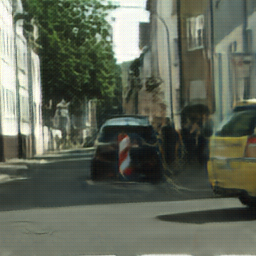}
        \end{subfigure}
        & 
        \begin{subfigure}{0.16\textwidth}
            \centering
            \includegraphics[width=\linewidth]{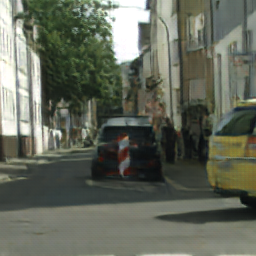}
        \end{subfigure}\\
        \begin{subfigure}{0.16\textwidth}
            \centering
            \includegraphics[width=\linewidth]{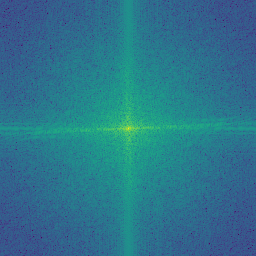}
        \end{subfigure}
        & 
        \begin{subfigure}{0.16\textwidth}
            \centering
            \includegraphics[width=\linewidth]{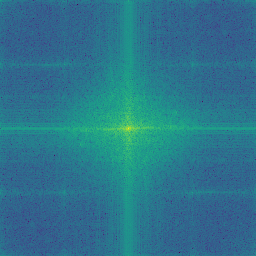}
        \end{subfigure} 
        & 
        \begin{subfigure}{0.16\textwidth}
            \centering
            \includegraphics[width=\linewidth]{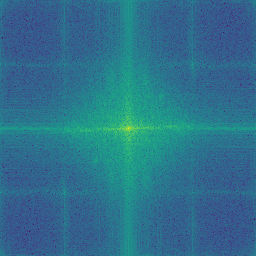}
        \end{subfigure} 
        & 
        \begin{subfigure}{0.16\textwidth}
            \centering
            \includegraphics[width=\linewidth]{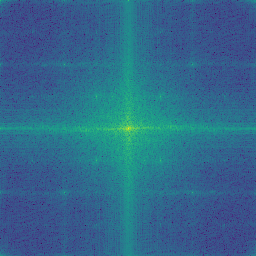}
        \end{subfigure} 
        & 
        \begin{subfigure}{0.16\textwidth}
            \centering
            \includegraphics[width=\linewidth]{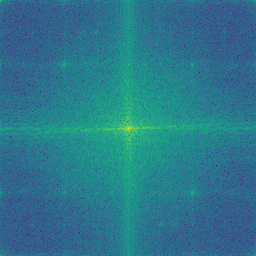}
        \end{subfigure}
        & 
        \begin{subfigure}{0.16\textwidth}
            \centering
            \includegraphics[width=\linewidth]{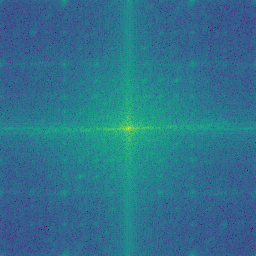}
        \end{subfigure}\\
        \begin{subfigure}{0.16\textwidth}
            \centering
            \includegraphics[width=\linewidth]{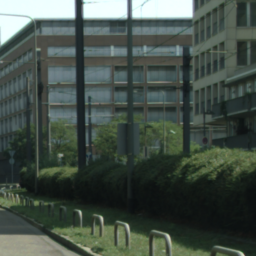}
        \end{subfigure}
        & 
        \begin{subfigure}{0.16\textwidth}
            \centering
            \includegraphics[width=\linewidth]{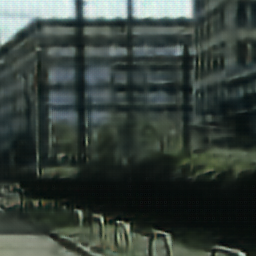}
        \end{subfigure} 
        & 
        \begin{subfigure}{0.16\textwidth}
            \centering
            \includegraphics[width=\linewidth]{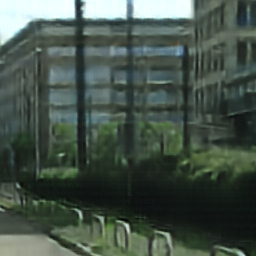}
        \end{subfigure} 
        & 
        \begin{subfigure}{0.16\textwidth}
            \centering
            \includegraphics[width=\linewidth]{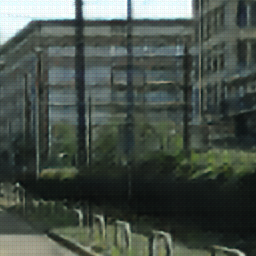}
        \end{subfigure} 
        & 
        \begin{subfigure}{0.16\textwidth}
            \centering
            \includegraphics[width=\linewidth]{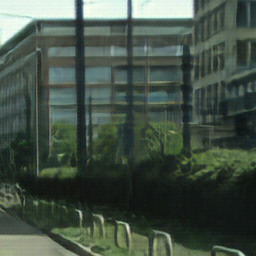}
        \end{subfigure}
        & 
        \begin{subfigure}{0.16\textwidth}
            \centering
            \includegraphics[width=\linewidth]{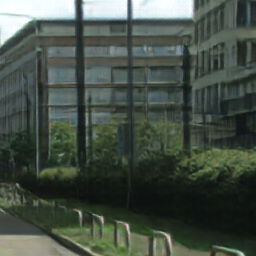}
        \end{subfigure}\\
        \begin{subfigure}{0.16\textwidth}
            \centering
            \includegraphics[width=\linewidth]{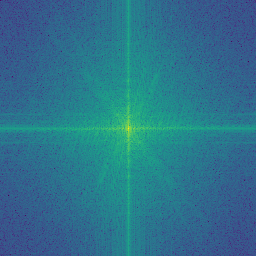}
        \end{subfigure}
        & 
        \begin{subfigure}{0.16\textwidth}
            \centering
            \includegraphics[width=\linewidth]{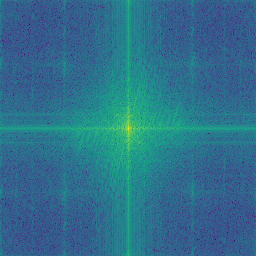}
        \end{subfigure} 
        & 
        \begin{subfigure}{0.16\textwidth}
            \centering
            \includegraphics[width=\linewidth]{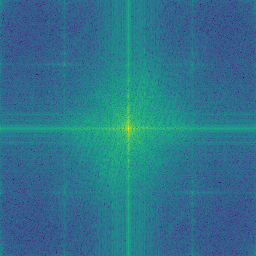}
        \end{subfigure} 
        & 
        \begin{subfigure}{0.16\textwidth}
            \centering
            \includegraphics[width=\linewidth]{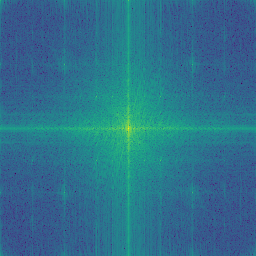}
        \end{subfigure} 
        & 
        \begin{subfigure}{0.16\textwidth}
            \centering
            \includegraphics[width=\linewidth]{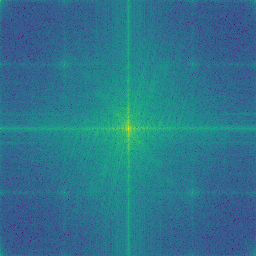}
        \end{subfigure}
        & 
        \begin{subfigure}{0.16\textwidth}
            \centering
            \includegraphics[width=\linewidth]{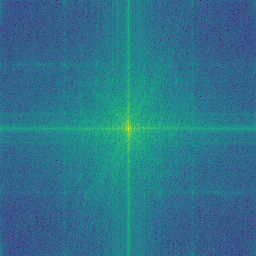}
        \end{subfigure}\\
    \end{tabular}
   \caption{Comparing various purely adversarially optimized generative image compression methods at low bit-rate (HiFiC-lo). Rows one and three show examples of reconstructed cropped images ($256\times256$), while rows two and four show the corresponding spectra of the images. \textbf{Best viewed electronically.}}
    \label{table:vis_gan_formulations}
\end{figure*}

\subsection{Comparing GAN Approaches} 
Fairly comparing discriminator architectures/ GAN formulations for generative image compression is difficult due to the large variety in model design, conditioning scheme and regularization terms. It is also important to note that these methods were primarily co-designed with their respective generator structures, while we consider them in isolation. We make no claim to the superiority of one method over another, but rather are interested in their general suitability for generative image compression. 

We consider the low bit range (HiFiC-lo), where the influence of generative models is arguably greatest. For each method, we use the best configuration following the original work, including regularization terms (see supplementary material). We start our experiments by applying the same HiFiC-based conditioning scheme in all model variants, if possible. We make the following observations: unsurprisingly, no pure GAN-based configuration exceeds the performance of the baseline method ($32.71$dB PSNR, $10.62$ FID-score). The performance gaps range from $-34.2\%$ to $-8.2\%$ and $+957.3\%$ to $+94.1\%$ decrease in PSNR and FID-score, respectively. The PatchGAN discriminator performs worst, which is to be expected since it was designed primarily to penalize high-frequency structure in addition to the commonly used L1/L2 loss functions~\cite{pix2pix2017}. Indeed, when paired with an additional distortion loss, this formulation works remarkably well as demonstrated by the baseline configuration as well as by previous work~\cite{agustsson2019generative, 9412185, mentzer2020high, Agustsson_2023_CVPR, He_2022_CVPRW}.

We find that both the SESAME, U-Net, and projected discriminators produce similar strong PSNR values ($29.43-29.64$dB), with varying degrees of artifacts (see~\cref{table:vis_gan_formulations}). The SESAME discriminator improves upon the PatchGAN variant due to its inherent multi-scale nature as well as access to additional semantic side information. For projected GANs, we find that the perceptual quality largely depends on the image resolution of the efficientnet-lite feature extractors. We suppose that the (well hidden) gridding artifacts are due to a resolution mismatch; \ie, the efficientnet-lite variants are based on low resolutions images, \eg, $224\times224$px for efficientnet-lite0, whereas HiFiC mostly targets high-resolution images up to $2000\times2000$px.

The best purely adversarial method is achieved by OASIS, which considerably
exceeds all its competitors in terms of perception (FID-score of $16.50$). We attribute its better performance to the spatially and semantically-aware pixel-level supervision, which implicitly provides a strong conditioning mechanism (see~\cref{table:comp_gan_formulations}, conf-e/ impl.\ conditioning). Note that a stronger semantic segmentation model (\eg, DeeplabV3+~\cite{cheng2020panoptic}) does not per se lead to better performance (conf-f). This finding is consistent with the observation that a stronger feature extractor does not necessarily lead to lower FID scores~\cite{Sauer2021NEURIPS}.

\subsection{Investigating the Conditioning Scheme}
We use the same setup as before, but now replace the vanilla concatenation-based conditioning scheme with projection~\cite{miyato2018cgans} for some selected methods. For OASIS and U-Net, we employ a pixel-wise projection-based conditioning scheme; for U-Net we use an additional projection for the global output.

It can be observed that all projection-based configurations improve their base configurations while largely reducing image artifacts (see supplementary material). For optimization strategies with distortion, these considerations probably play a minor role, since $d$ already provides for a strong implicit conditioning mechanism. However, for approaches based on pure adversarial optimization, such as in the case of the optimal training framework~\cite{2021PerceptualReconstruction, 2022ControllablePerceptualCompression}, our findings shed some light on the lack of generalizability beyond the MNIST dataset.

\begin{table}[tb]
\caption{Improving OASIS~\cite{schoenfeld2021you} step-by-step}
\begin{center}
\begin{tabular}{lll}
\toprule
 Method                                                             & PSNR $\uparrow$   & FID $\downarrow$  \\\midrule
 OASIS~\cite{schoenfeld2021you}                                     & 29.90             & 15.30             \\
 \phantom{+} + pre-trained                                         & 30.01             & 9.75              \\
 \phantom{++} + weight norm~\cite{NIPS2016_ed265bc9}                & 30.20             & 7.96              \\
 \phantom{+++} + projection~\cite{miyato2018cgans} (ours w/o $d$)    & 29.97             & 7.74              \\\midrule
 ours w/ $d$                                                        & \textbf{32.24}    & \textbf{6.36}     \\\bottomrule
 
\end{tabular}
\end{center}
\label{table:improving_oasis}
\end{table}

\subsection{Improving OASIS}
A major concern we had prior to the adoption was the tremendous amount of hardware resources required for training OASIS. Specifically, Sch{\"o}nfeld \etal trained their model on COCO-stuff for 4 weeks, in a multi-GPU environment ($4\times$ Tesla V100 GPUs). Instead, we target a single-GPU setup.

We attribute the slow training process to the spectral norm in its default configuration, which we have found to severely hinder learning progress. We note that similar observations have been reported recently, \eg, Lee~\etal~\cite{lee2022vitgan} proposed to multiply the normalized weight matrix with the spectral norm at initialization, which increases the often untuned Lipschitz constant and hence the overall model capacity. For our specific use case, we have found that weight normalization~\cite{NIPS2016_ed265bc9} combined with a pre-trained discriminator produces a good training speed/ stability/ compression performance trade-off (\cref{table:improving_oasis}). In the supplementary material we further show that simply replacing the PatchGAN discrimiantor with OASIS alone is not sufficient to improve over the state-of-the-art. This is especially true for highly complex learning tasks (Coco2017), where OASIS exhibits sever training instabilities.

\section{Comparison to the State-of-the-Art}\label{sec:comp_soa}

\begin{figure*}[bt]
    \centering
    \includegraphics[height=6cm]{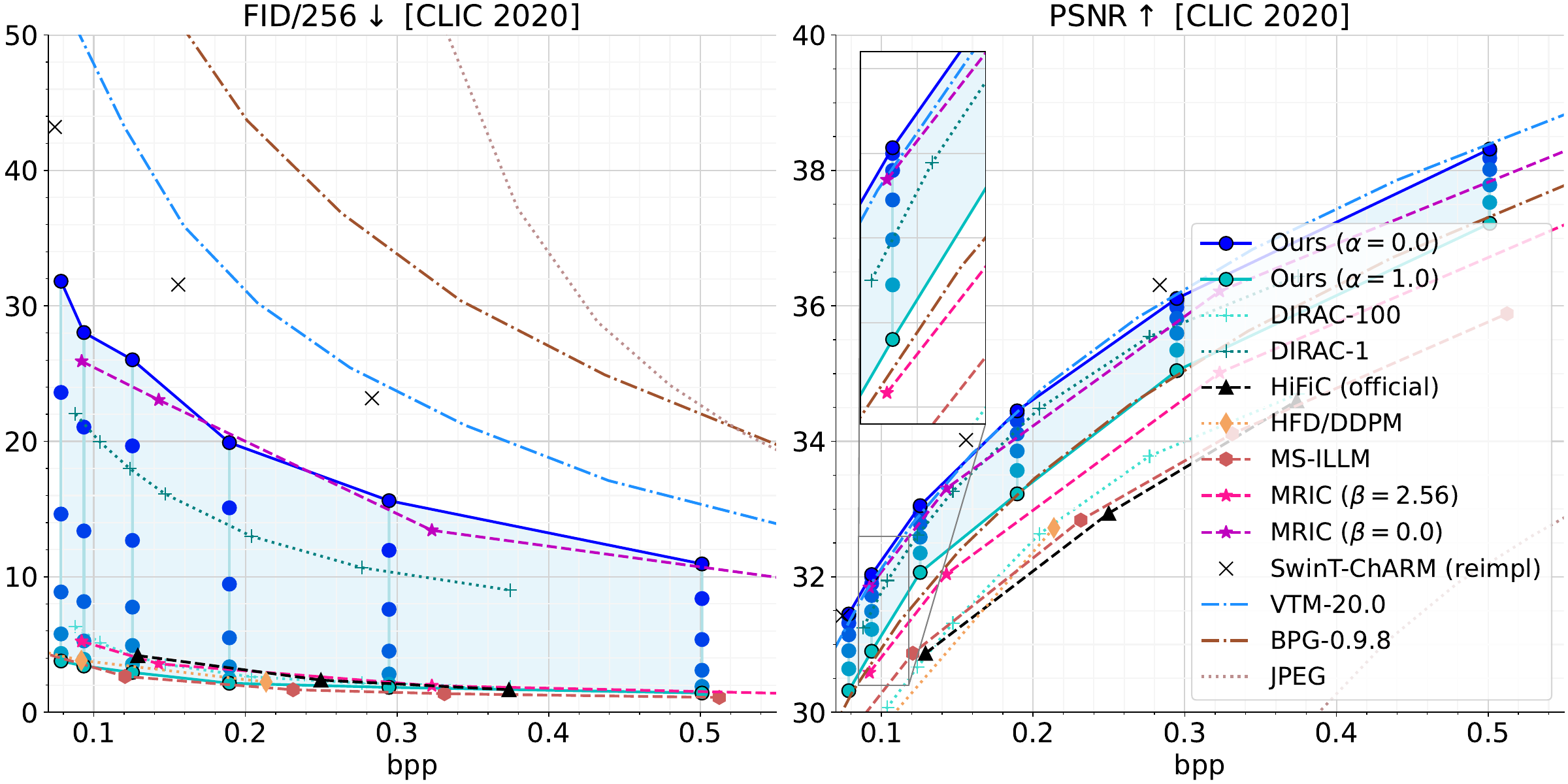}
   \caption{Comparison to the state-of-the-art on CLIC 2020}
    \label{fig:soa_comp}
\end{figure*}

For our main experiments, we use SwinT-ChARM~\cite{zhu2022transformerbased} as backbone architecture with an additional prediction head (we clone the last Swin-transformer block~\cite{liu2021Swin} of the pre-trained $G_2$, which corresponds to $G_{ORP}$ in~\cref{sec:our_approach}). Exact model configurations as well as extended experiments on HiFiC and additional datasets (Kodak and DIV2K) are summarized in the supplementary material.

\textbf{Baselines.} We consider both diffusion (SwinT-ChARM/~DIRAC-$100$, short DIRAC-$100$~\cite{noor_2023}, HFD/DDPM~\cite{Hoogeboom_2023}) and GAN-based (HiFiC~\cite{mentzer2020high}, MS-ILLM~\cite{pmlr-v202-muckley23a}, MRIC ($\beta=2.56$)) methods for perception, \ie, Ours ($\alpha=1.0$). Similarly, we consider VTM-20.0, BPG-0.9.8, JPEG, SwinT-ChARM/~DIRAC-$1$, short DIRAC-$1$~\cite{noor_2023}, and MRIC ($\beta=0.0$)~\cite{Agustsson_2023_CVPR} for distortion, \ie, Ours ($\alpha=0.0$). We further add SwinT-ChARM (reimpl), 
our TensorFlow-reimplementation of SwinT-ChARM, which can be considered an upper bound for distortion.

\begin{figure}[tb]
  \centering
  \includegraphics[height=6.9cm]{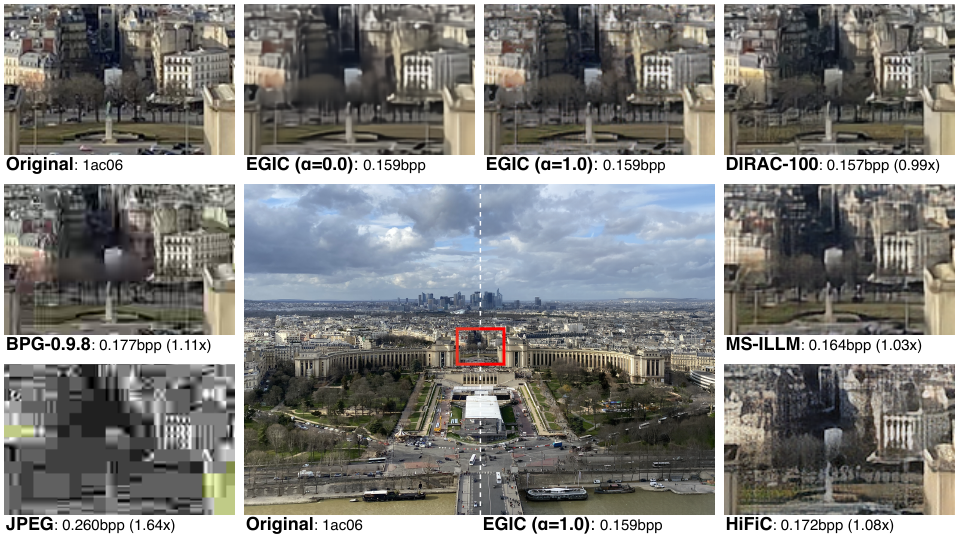}
  \caption{Visual comparison of EGIC ($\alpha \in \{0.0, 1.0\}$) with state-of-the-art distortion-oriented (l.) and perception-oriented (r.) codecs. Please visit the supplementary material for more impressions. \textbf{Best viewed electronically.}}
  \label{fig:vis_comp}
\end{figure}

We provide objective and subjective comparisons on CLIC 2020 in~\cref{fig:teaser,fig:soa_comp,fig:vis_comp}. We observe that Ours ($\alpha=1.0$) is most effective in the low to medium bit range, being competitive or outperforming strong baselines (HiFiC, MS-ILLM, MRIC ($\beta=2.56$), DIRAC-$100$, HFD/DDPM) in terms of FID, while having considerably better PSNR-scores in all cases. This becomes even more pronounced on DIV2K, where Ours ($\alpha=1.0$) dominates MS-ILLM in terms of perception. Noteworthy, EGIC outperforms HiFiC-lo, the long standing previous state-of-the-art, even when using $30$\% fewer bits. For Ours ($\alpha=0.0$), we find that ORP is surprisingly effective. In terms of distortion, our method outperforms MRIC and DIRAC, while (almost) matching VTM-20.0, the state-of-the-art for non-learned codecs. This is despite using only a fraction of additional model parameters (\eg, $0.15\times$ compared to Beta conditioning in MRIC~\cite{Agustsson_2023_CVPR}) and only requiring a single inference-cycle (as opposed to DIRAC-$n$~\cite{noor_2023}), revealing that more sophisticated methods may in fact be unnecessary. In the supplementary material we further discuss other formulations for ORP and provide a detailed comparison to more involved image and weight interpolation techniques.

Finally, it is worth mentioning that EGIC is considerably more storage-efficient compared to all other methods. EGIC only requires a fraction of the number of model parameters compared to HFD/DDPM ($0.03\times$), HiFiC/ MS-ILLM ($0.18\times$), DIRAC ($0.24\times$), and MRIC ($0.55\times$); in contrast to DIRAC and HFD/DDPM, EGIC also only requires a single inference cycle.

\section{Conclusion}

We have developed EGIC, a novel generative image compression method that allows traversing the D-P curve efficiently from a single model on the receiver side. We find that EGIC is highly competitive, outperforming a wide-variety of diffusion and GAN-based methods (\eg, HiFiC, MS-ILLM, DIRAC-$100$), while operating almost on par with VTM-20.0 on the distortion-oriented end. EGIC enjoys a simple and lightweight design with excellent interpolation characteristics, which makes it a promising candidate for practical applications targeting the low bit range. Our code will be made publicly available\footnote{\url{https://github.com/Nikolai10/EGIC}} upon publication to facilitate further research.


\textbf{Limitations.} Gradient feedback from OASIS-C is currently applied to the whole image, which in some cases might lead to sub-optimal preservation of small faces and text. This can be addressed via content-weighted learning mechanisms~\cite{Li_2018_CVPR, 10.1007/978-3-031-19800-7_37}. Furthermore, EGIC at this stage requires large labeled training data, which can be alleviated by switching to semi-supervised approaches or by incorporating a powerful prior for semantic segmentation tasks, \eg, SAM~\cite{Kirillov_2023_ICCV, chen2023semantic}. Finally, despite promising quantitative results, we find that HFD/DDPM and DIRAC-$100$ produce in some cases more pleasing (not necessarily more accurate) reconstructions. Whether this can be attributed to the superiority of diffusion models in general, or simply to the substantial difference in model sizes ($1$B in the case of HFD/DDPM), remains an open subject of debate.

\section*{Acknowledgements}

This work was supported by the German Federal Ministry of Education and Research under the funding program Forschung an Fachhochschulen - FKZ 13FH019KI2. The authors would further like to thank Auke Wiggers, Matthew Muckley, Eirikur Agustsson and Lucas Theis for providing evaluation data.

%
%
\bibliographystyle{splncs04}
\bibliography{egbib}

\clearpage
\appendix

\section{Supplementary Material}

\subsection{OASIS-C Architecture}

We summarize our discriminator architecture in~\cref{tbl:our_disc}. Our architecture is split into two parts. The first part is identical to OASIS (Sch{\"o}nfeld \etal, 2021), except that we replace spectral norm with weight norm. The output \texttt{out} is a $256\times256\times{N+1}$ prediction map. In the second part, we adopt a pixel-wise projection-based conditioning scheme. We use a similar latent pre-processing block as in HiFiC, but use $64$ instead of $12$ filters. The pre-processed latent feature map \texttt{y\textunderscore prep} has an identical shape as \texttt{out} and is subsequently incorporated into the discriminator, using projection (element-wise multiplication and sum across the channel dimension). The projected feature map \texttt{proj} is finally replicated and added back to \texttt{out}.

\begin{table*}
\caption{OASIS-C Architecture}
        \begin{center}
        \begin{tabular}{llllll}
            \toprule
            Operation       & Input                                                     & Size                      & Output                               & Size                      \\\midrule
            ResBlock-Down   & \texttt{image ($x$)}                                        & $256\times256\times3$     & \texttt{d1}        & $128\times128\times128$   \\
            ResBlock-Down   & \texttt{d1}                             & $128\times128\times128$   & \texttt{d2}        & $64\times64\times128$     \\
            ResBlock-Down   & \texttt{d2}                             & $64\times64\times128$     & \texttt{d3}        & $32\times32\times256$     \\
            ResBlock-Down   & \texttt{d3}                             & $32\times32\times256$     & \texttt{d4}        & $16\times16\times256$     \\
            ResBlock-Down   & \texttt{d4}                             & $16\times16\times256$     & \texttt{d5}        & $8\times8\times512$       \\
            ResBlock-Down   & \texttt{d5}                             & $8\times8\times512$       & \texttt{d6}        & $4\times4\times512$       \\
            ResBlock-Up   & \texttt{d6}                             & $4\times4\times512$       & \texttt{u1}          & $8\times8\times512$       \\
            ResBlock-Up     & \texttt{cat(u1, d5)}    & $8\times8\times1024$      & \texttt{u2}          & $16\times16\times256$     \\
            ResBlock-Up     & \texttt{cat(u2, d4)}    & $16\times16\times512$     & \texttt{u3}          & $32\times32\times256$     \\
            ResBlock-Up     & \texttt{cat(u3, d3)}    & $32\times32\times512$     & \texttt{u4}          & $64\times64\times128$     \\
            ResBlock-Up     & \texttt{cat(u4, d2)}    & $64\times64\times256$     & \texttt{u5}          & $128\times128\times128$   \\
            ResBlock-Up     & \texttt{cat(u5, d1)}    & $128\times128\times256$   & \texttt{u6}          & $256\times256\times64$    \\
            Conv2D          & \texttt{u6}                               & $256\times256\times64$    & \texttt{out}                         & $256\times256\times{N+1}$ \\\midrule
            Conv2D, Resize & \texttt{latent ($y$)}                                       & $16\times16\times{C_y}$     & \texttt{y\textunderscore prep}  & $256\times256\times64$    \\
            Projection      & \texttt{(u6, y\textunderscore prep)} & $256\times256\times64$    & \texttt{proj}                        & $256\times256\times1$     \\
            Add             & \texttt{(out, proj)}                                      & $256\times256\times{N+1}$ & \texttt{$D_{out}$}       & $256\times256\times{N+1}$ \\\bottomrule
        \end{tabular}
        \end{center}
        \label{tbl:our_disc}
    \end{table*}

\subsection{Pre-trained Semantic Segmentation Performance}

We pre-train the first discriminator block (\cref{fig:disc_comp}, highlighted orange) using DeepLab2\footnote{We base our experiments on the panoptic configurations presented in \url{https://github.com/google-research/deeplab2/blob/main/g3doc/projects/panoptic_deeplab.md} (ResNet-50).}. The resulting semantic segmentation performance measured by the mean intersection over union (mIoU) is summarized in~\cref{table:pretrained_seg_models}.

\begin{table}[bt]
\caption{Pre-trained Semantic Segmentation Performance}
\begin{center}
\begin{tabular}{ccccc}
\toprule
 Dataset & Crop Size & Batch Size & Steps & mIoU $\uparrow$ \\\midrule
 Cityscapes (Cordts~\etal, 2016) & $256\times256$ & $16$ &$320$k & $0.67$ \\
 Coco2017 (Lin~\etal, 2014) & $256\times256$ & $16$ & $1$M & $0.41$ \\\bottomrule
\end{tabular}
\end{center}
\label{table:pretrained_seg_models}
\end{table}

\subsection{Additional Experimental Details}

\textbf{Preliminary study.} We use the official implementation for PatchGAN and translate the SESAME, U-Net, projected and OASIS discriminators carefully to TensorFlow, based on the official PyTorch implementations. For projected GANs, we use the efficientnet-lite4-variant (Tan~\etal, 2019) as a pre-trained feature network, which we have found to produce the best results. 

\begin{figure*}[bt]
  \centering
  \includegraphics[width=\textwidth]{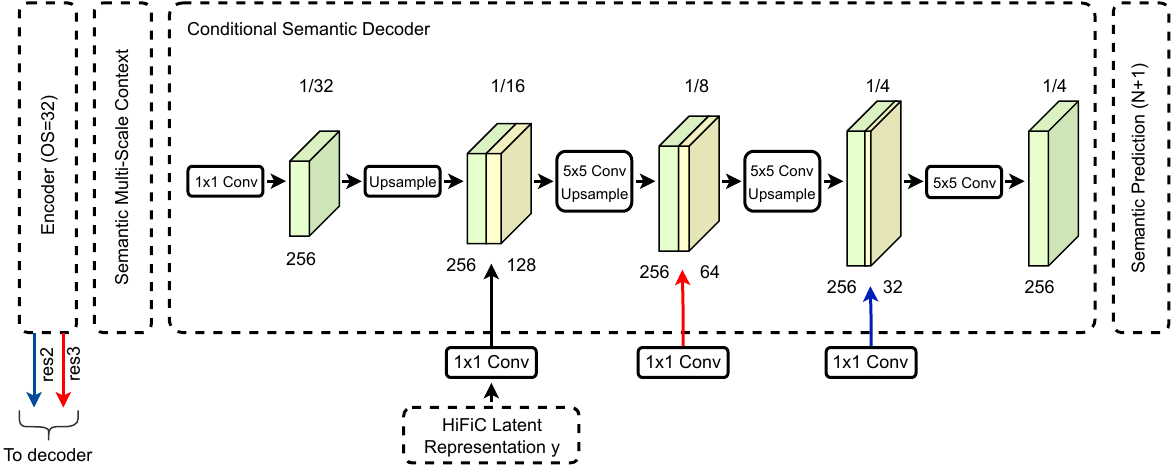}
  \caption{Conditional panoptic DeepLab-based semantic decoder (conf-f)}
  \label{fig:deeplabv3+}
\end{figure*}

To maintain the advantages of having pre-trained feature extractors, we have used a slightly different concatenation-based conditioning scheme for conf-d and conf-f; for conf-d, we pre-process and concatenate the latent features with the efficientnet-lite4-based feature maps at each scale separately. For conf-f, we integrate $y$ using a similar HiFiC-based concatenation scheme (see~\cref{fig:deeplabv3+}).

For conf-c (projection), we use two separate latent pre-processing blocks with channel dimensions $64$ and $4$, corresponding to the local and global outputs, respectively. We use no resize operation for the latter to match the feature dimension prior to classification ($16*16*4=1024$).

\textbf{Main study.} We train six models for 2+1M optimization steps, using $\lambda \in \{2, 1.5, 1, 0.5, 0.25, 0.1\}$, a crop size of $256$ and a batch size of $16$ and $8$ for stage one and two, respectively. We use the Adam optimizer with default settings ($\beta_{1}=0.9, \beta_{2}=0.999$). For stage one, we use a learning rate of $1\mathrm{e}{-4}$ for the first $1.8$M steps and subsequently decay the learning rate to $1\mathrm{e}{-5}$, similar to previous work. For stage two, we use the same settings as in Ours w/ $d$ (HiFiC), \ie, training strategy-I with a fixed learning rate of $1\mathrm{e}{-5}$, except for $\beta$ (\cref{eq:generator_objective_ours}), which we increase from $0.15$ to $0.30$. 

\textbf{ORP.} We finetune $G_{ORP}$ for additional 2M steps. In practice we have found it slightly more efficient to directly predict the MSE-optimized decoder output $MSE_{pred}$ and to calculate $R=MSE_{pred} - G_2(x)$.

Note that ORP is a general formulation for multi-realim image compression, which allows for different model parameterizations. By increasing the model capacity of $G_{ORP}$ up to $G_2$, we can approach the performance of traditional image/ weight interpolation techniques (see \cref{subsec:iw_interpol}).

\subsection{Comparing Normalization Strategies}

\begin{figure*}[bt]
    \setlength{\tabcolsep}{1pt}
    \renewcommand{\arraystretch}{0.5}
    \centering
    \begin{tabular}{ccc}
        \toprule
        Input & Semantic Prediction & Error Map \\
        \midrule
        & \multicolumn{2}{l}{OASIS w/ weight norm} \\  
        \begin{subfigure}{0.325\textwidth}
            \centering
            \includegraphics[width=\linewidth]{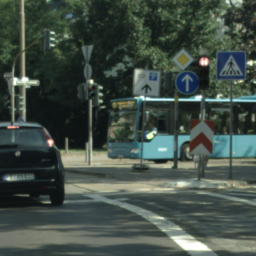}
        \end{subfigure}
        &
        \begin{subfigure}{0.325\textwidth}
            \centering
            \includegraphics[width=\linewidth]{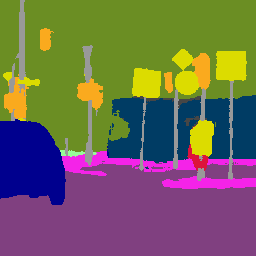}
        \end{subfigure}
        &
        \begin{subfigure}{0.325\textwidth}
            \centering
            \includegraphics[width=\linewidth]{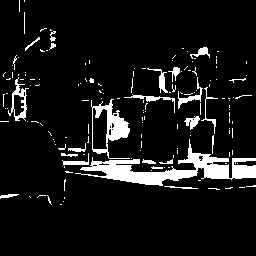}
        \end{subfigure}\\
        & \multicolumn{2}{l}{OASIS w/ spectral norm} \\
        \begin{subfigure}{0.325\textwidth}
            \centering
            \includegraphics[width=\linewidth]{figures/appendix/SNvsWN/000009_image_input_256x256.png}
        \end{subfigure}
        &
        \begin{subfigure}{0.325\textwidth}
            \centering
            \includegraphics[width=\linewidth]{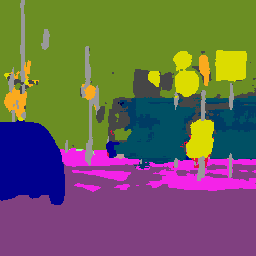}
        \end{subfigure}
        &
        \begin{subfigure}{0.325\textwidth}
            \centering
            \includegraphics[width=\linewidth]{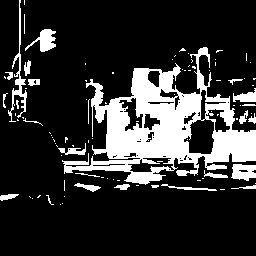}
        \end{subfigure}
        \\
    \end{tabular}
    \caption{Comparing the semantic segmentation performance of OASIS w/ weight normalization and OASIS w/ spectral normalization. Black pixels in the error map correspond to perfect prediction, white pixels highlight deviations from the ground truth.}
    \label{tbl:oasis_sn_vs_wn}
\end{figure*}

We summarize some of the normalization methods we tried for OASIS in~\cref{table:comp_norm_strategies} and~\cref{tbl:oasis_sn_vs_wn}. For spectral normalization, we have found that tuning the Lipschitz constant is indeed helpful. However, we did not find a configuration that exceeded the performance of weight normalization and hence omit it here. For layer normalization we had to reduce the batch size to $8$, due to out-of-memory issues.

\begin{table}[tb]
\caption{The effect of different normalization strategies on the semantic segmentation and its resulting compression performance.}
\begin{center}
\begin{tabular}{ccccc}
\toprule
 Method & mIoU $\uparrow$ & PSNR $\uparrow$ & FID $\downarrow$ & Batch Size \\\midrule
 Spectral norm (Miyato~\etal, 2018) & 0.49 & 30.01 & 9.75 & 16    \\
 Weight norm (Salimans~\etal, 2016)  & 0.67 & \textbf{30.20} & \textbf{7.96} & 16      \\
 Layer norm (Ba~\etal, 2016) & \textbf{0.68} & 29.49 & 9.00 & 8 \\\bottomrule
\end{tabular}
\end{center}
\label{table:comp_norm_strategies}
\end{table}

\subsection{LabelMix Regularization}\label{subsec:labelmix}

\begin{figure*}[bt]
    \setlength{\tabcolsep}{1pt}
    \renewcommand{\arraystretch}{0.5}
    \centering
    \begin{tabular}{ccccc}
        input $x$ & rec $x'$ & $\text{LM}_{(x, x', M)}$ & Mask $M$ & pixel weights $w$ \\
        \begin{subfigure}{0.19\textwidth}
            \centering
            \includegraphics[width=\linewidth]{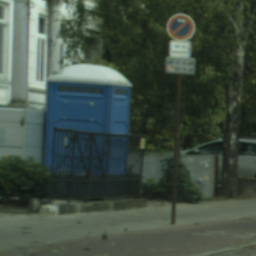}
        \end{subfigure}
        &
        \begin{subfigure}{0.19\textwidth}
            \centering
            \includegraphics[width=\linewidth]{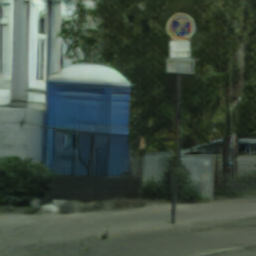}
        \end{subfigure}
        &
        \begin{subfigure}{0.19\textwidth}
            \centering
            \includegraphics[width=\linewidth]{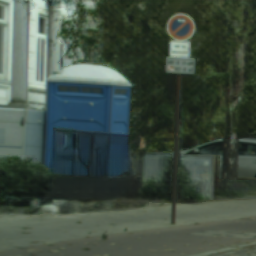}
        \end{subfigure}
        &
        \begin{subfigure}{0.19\textwidth}
            \centering
            \includegraphics[width=\linewidth]{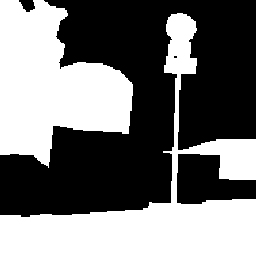}
        \end{subfigure}
        &
        \begin{subfigure}{0.19\textwidth}
            \centering
            \includegraphics[width=\linewidth]{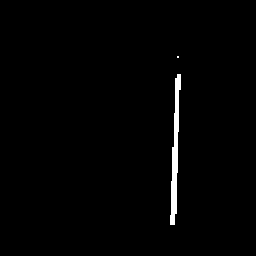}
        \end{subfigure}
        \\
        label map & ${D_{(x, y)}}$ & ${D_{(x', y)}}$ & $D_{\text{(LM}{(x, x', M), y)}}$ & $\text{LM}_{disc}$ \\
        \begin{subfigure}{0.19\textwidth}
            \centering
            \includegraphics[width=\linewidth]{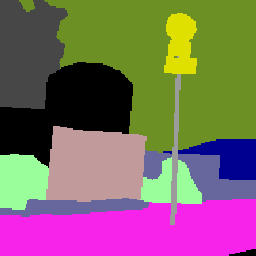}
        \end{subfigure}
        &
        \begin{subfigure}{0.19\textwidth}
            \centering
            \includegraphics[width=\linewidth]{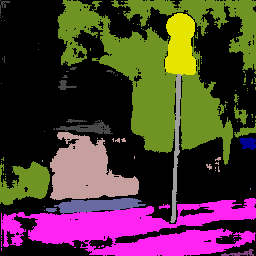}
        \end{subfigure}
        &
        \begin{subfigure}{0.19\textwidth}
            \centering
            \includegraphics[width=\linewidth]{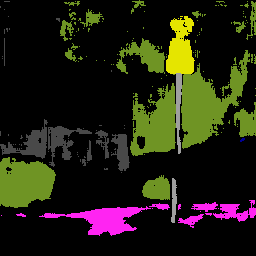}
        \end{subfigure}
        &
        \begin{subfigure}{0.19\textwidth}
            \centering
            \includegraphics[width=\linewidth]{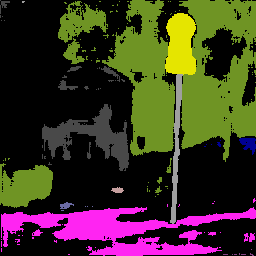}
        \end{subfigure}
        &
        \begin{subfigure}{0.19\textwidth}
            \centering
            \includegraphics[width=\linewidth]{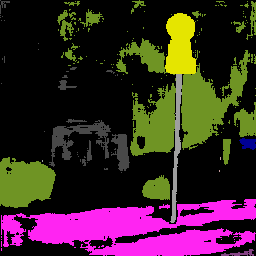}
        \end{subfigure}
        \\\midrule
        input $x$ & rec $x'$ & $\text{LM}_{(x, x', M)}$ & Mask $M$ & pixel weights $w$ \\
        \begin{subfigure}{0.19\textwidth}
            \centering
            \includegraphics[width=\linewidth]{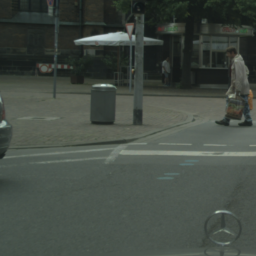}
        \end{subfigure}
        &
        \begin{subfigure}{0.19\textwidth}
            \centering
            \includegraphics[width=\linewidth]{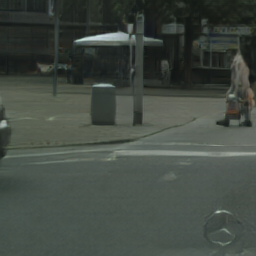}
        \end{subfigure}
        &
        \begin{subfigure}{0.19\textwidth}
            \centering
            \includegraphics[width=\linewidth]{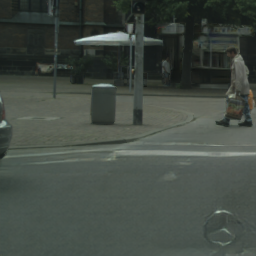}
        \end{subfigure}
        &
        \begin{subfigure}{0.19\textwidth}
            \centering
            \includegraphics[width=\linewidth]{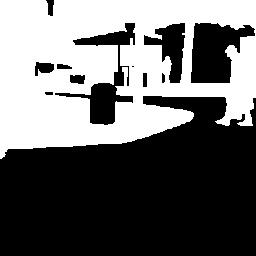}
        \end{subfigure}
        &
        \begin{subfigure}{0.19\textwidth}
            \centering
            \includegraphics[width=\linewidth]{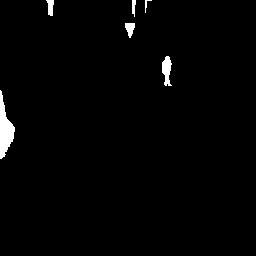}
        \end{subfigure}
        \\
        label map & ${D_{(x, y)}}$ & ${D_{(x', y)}}$ & $D_{\text{(LM}{(x, x', M), y)}}$ & $\text{LM}_{disc}$ \\
        \begin{subfigure}{0.19\textwidth}
            \centering
            \includegraphics[width=\linewidth]{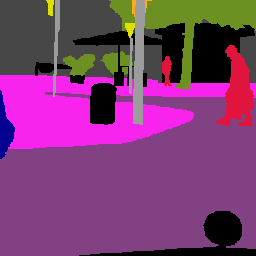}
        \end{subfigure}
        &
        \begin{subfigure}{0.19\textwidth}
            \centering
            \includegraphics[width=\linewidth]{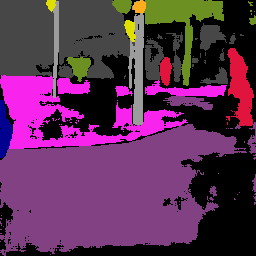}
        \end{subfigure}
        &
        \begin{subfigure}{0.19\textwidth}
            \centering
            \includegraphics[width=\linewidth]{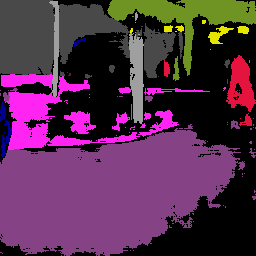}
        \end{subfigure}
        &
        \begin{subfigure}{0.19\textwidth}
            \centering
            \includegraphics[width=\linewidth]{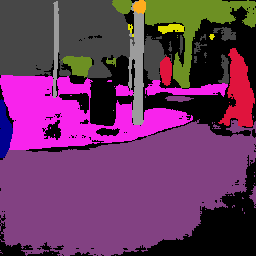}
        \end{subfigure}
        &
        \begin{subfigure}{0.19\textwidth}
            \centering
            \includegraphics[width=\linewidth]{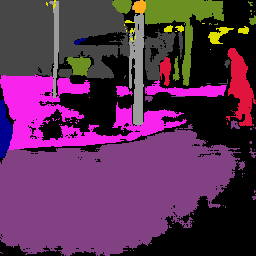}
        \end{subfigure}
        \\
    \end{tabular}
    \caption{Visualizing the discriminator loss components at step 33.5k (top row) and 35.2k (bottom row) on the Cityscapes dataset. We use an abbreviated notation in some cases due to space constraints; $D_{\text{(LM}{(x, x', M), y)}}$ and $\text{LM}_{disc}$ correspond to $D_{\text{logits}}(\text{LM}(x, x', M), y)$ and $\text{LM}(D_{\text{logits}}(x, y), D_{\text{logits}}(x', y), M)$ in~\cref{eq:label_mix}. We additionally visualize the pixel weight masks $w$ introduced in~\cref{sec:our_approach} that highlight small object instances, as well as the corresponding label maps and the discriminator predictions for ($x, y$) and ($x', y$), respectively. The colorized discriminator predictions are obtained by $\argmax(D)$. The black color corresponds to the fake class.}
    \label{tbl:add_disc_obj}
\end{figure*}

\begin{figure*}[bt]
    \setlength{\tabcolsep}{1pt}
    \renewcommand{\arraystretch}{0.5}
    \centering
    \begin{tabular}{ccccc}
        input $x$ & rec $x'$ & $\text{LM}_{(x, x', M)}$ & Mask $M$ & pixel weights $w$ \\
        \begin{subfigure}{0.19\textwidth}
            \centering
            \includegraphics[width=\linewidth]{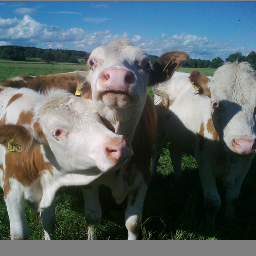}
        \end{subfigure}
        &
        \begin{subfigure}{0.19\textwidth}
            \centering
            \includegraphics[width=\linewidth]{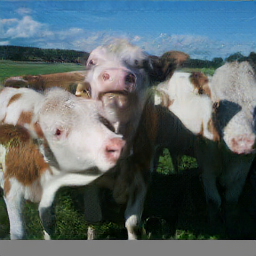}
        \end{subfigure}
        &
        \begin{subfigure}{0.19\textwidth}
            \centering
            \includegraphics[width=\linewidth]{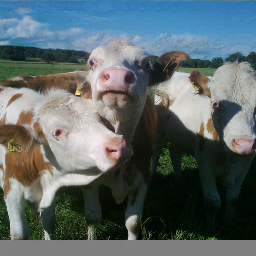}
        \end{subfigure}
        &
        \begin{subfigure}{0.19\textwidth}
            \centering
            \includegraphics[width=\linewidth]{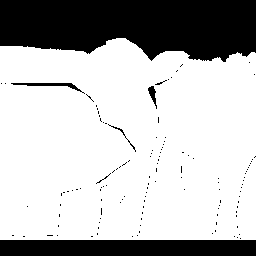}
        \end{subfigure}
        &
        \begin{subfigure}{0.19\textwidth}
            \centering
            \includegraphics[width=\linewidth]{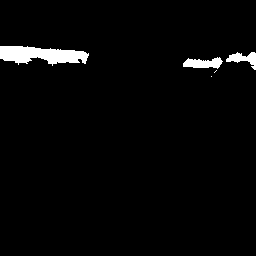}
        \end{subfigure}
        \\
        label map & ${D_{(x, y)}}$ & ${D_{(x', y)}}$ & $D_{\text{(LM}{(x, x', M), y)}}$ & $\text{LM}_{disc}$ \\
        \begin{subfigure}{0.19\textwidth}
            \centering
            \includegraphics[width=\linewidth]{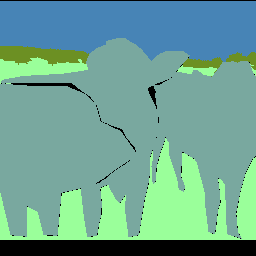}
        \end{subfigure}
        &
        \begin{subfigure}{0.19\textwidth}
            \centering
            \includegraphics[width=\linewidth]{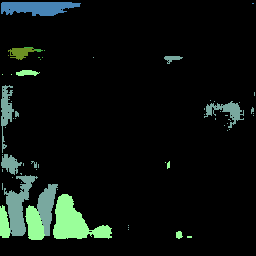}
        \end{subfigure}
        &
        \begin{subfigure}{0.19\textwidth}
            \centering
            \includegraphics[width=\linewidth]{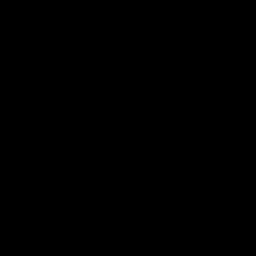}
        \end{subfigure}
        &
        \begin{subfigure}{0.19\textwidth}
            \centering
            \includegraphics[width=\linewidth]{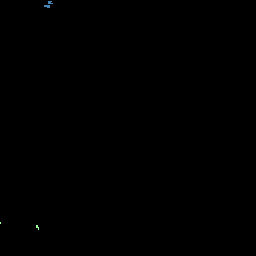}
        \end{subfigure}
        &
        \begin{subfigure}{0.19\textwidth}
            \centering
            \includegraphics[width=\linewidth]{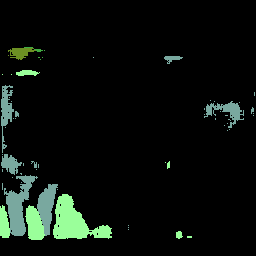}
        \end{subfigure}
        \\\midrule
        input $x$ & rec $x'$ & $\text{LM}_{(x, x', M)}$ & Mask $M$ & pixel weights $w$ \\
        \begin{subfigure}{0.19\textwidth}
            \centering
            \includegraphics[width=\linewidth]{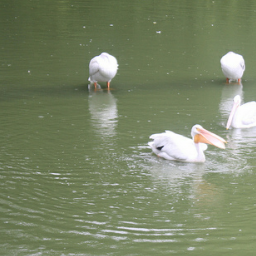}
        \end{subfigure}
        &
        \begin{subfigure}{0.19\textwidth}
            \centering
            \includegraphics[width=\linewidth]{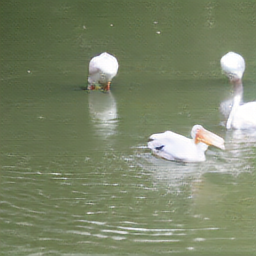}
        \end{subfigure}
        &
        \begin{subfigure}{0.19\textwidth}
            \centering
            \includegraphics[width=\linewidth]{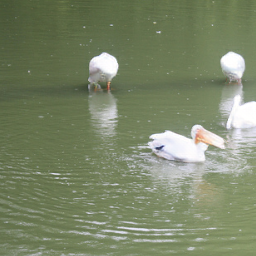}
        \end{subfigure}
        &
        \begin{subfigure}{0.19\textwidth}
            \centering
            \includegraphics[width=\linewidth]{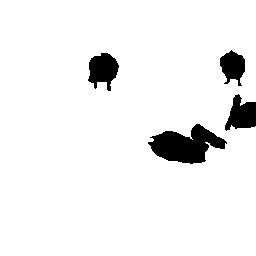}
        \end{subfigure}
        &
        \begin{subfigure}{0.19\textwidth}
            \centering
            \includegraphics[width=\linewidth]{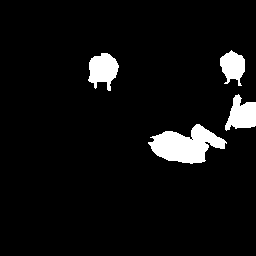}
        \end{subfigure}
        \\
        label map & ${D_{(x, y)}}$ & ${D_{(x', y)}}$ & $D_{\text{(LM}{(x, x', M), y)}}$ & $\text{LM}_{disc}$ \\
        \begin{subfigure}{0.19\textwidth}
            \centering
            \includegraphics[width=\linewidth]{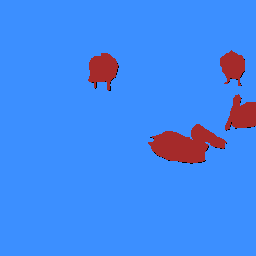}
        \end{subfigure}
        &
        \begin{subfigure}{0.19\textwidth}
            \centering
            \includegraphics[width=\linewidth]{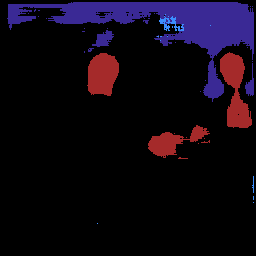}
        \end{subfigure}
        &
        \begin{subfigure}{0.19\textwidth}
            \centering
            \includegraphics[width=\linewidth]{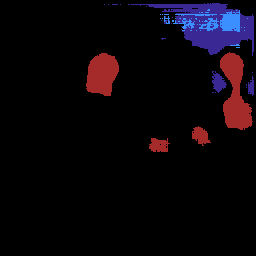}
        \end{subfigure}
        &
        \begin{subfigure}{0.19\textwidth}
            \centering
            \includegraphics[width=\linewidth]{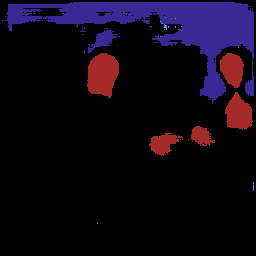}
        \end{subfigure}
        &
        \begin{subfigure}{0.19\textwidth}
            \centering
            \includegraphics[width=\linewidth]{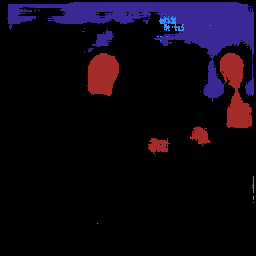}
        \end{subfigure}
        \\
    \end{tabular}
    \caption{Visualizing the discriminator loss components at step 1M (top row) and at an early training stage (160k, bottom row) on the Coco2017 dataset. See~\cref{tbl:add_disc_obj} for a detailed description.}
    \label{tbl:add_disc_obj_part_two}
\end{figure*}

As mentioned in the main paper (\cref{eq:label_mix}), we regularize the discriminator with the LabelMix consistency loss (Sch{\"o}nfeld~\etal, 2021), tailored to the compression setting. We provide additional intuition in~\cref{tbl:add_disc_obj,tbl:add_disc_obj_part_two}.

\subsection{Performance on DIV2K}
\begin{figure*}[h]
\begin{center}
\includegraphics[height=6cm]{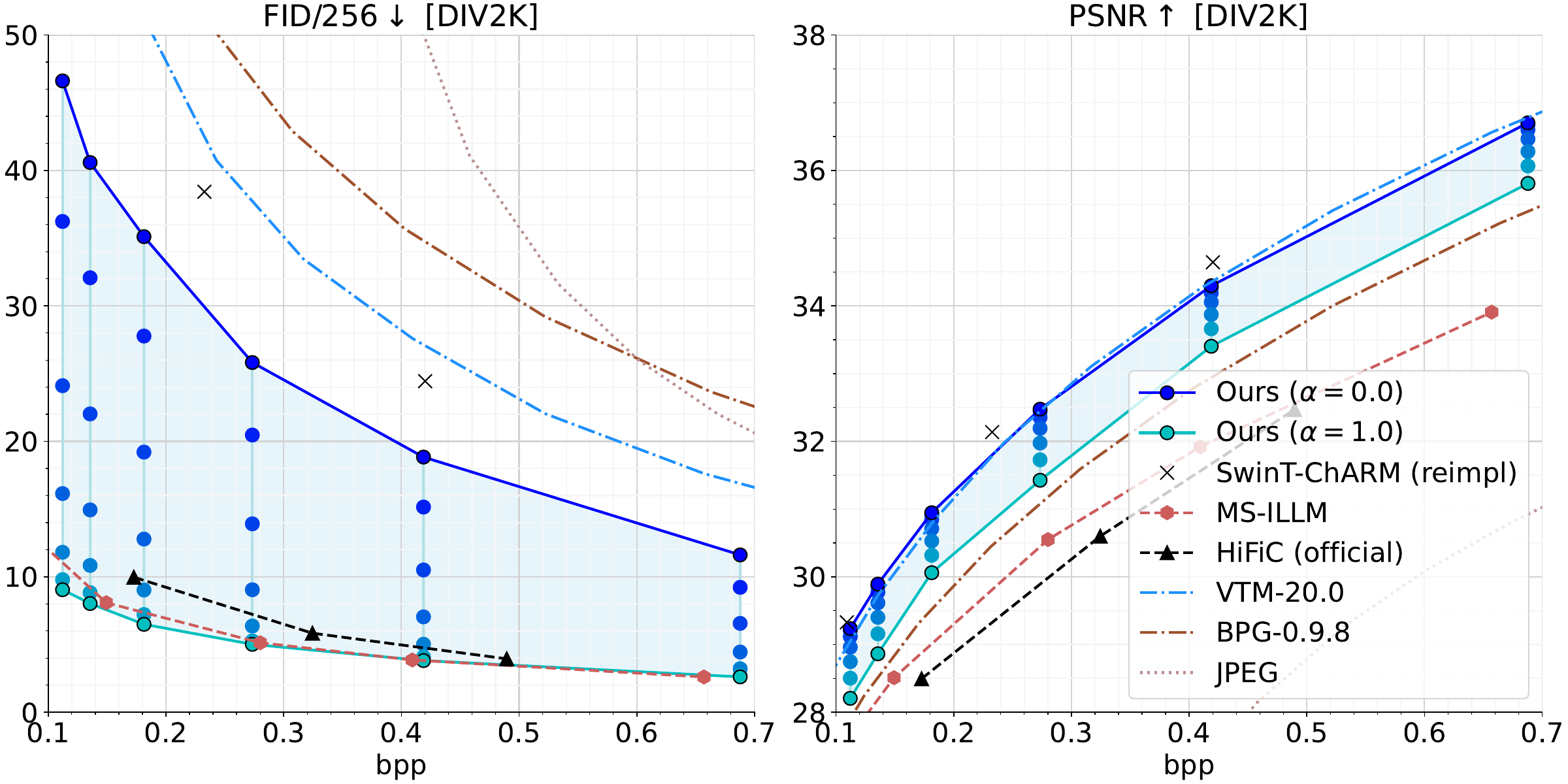}
\end{center}
   \caption{Comparison to the state-of-the-art on DIV2K}
\label{fig:soa_comp_div2k}
\end{figure*}

In \cref{fig:soa_comp_div2k} we provide an extended comparison to the state-of-the-art on DIV2K. We observe similar trends as discussed for CLIC 2020, except that our method outperforms MS-ILLM in terms of FID in the low bit range.

\subsection{Performance on Kodak}

\begin{figure}[bt]
\begin{center}
\includegraphics[height=6cm]{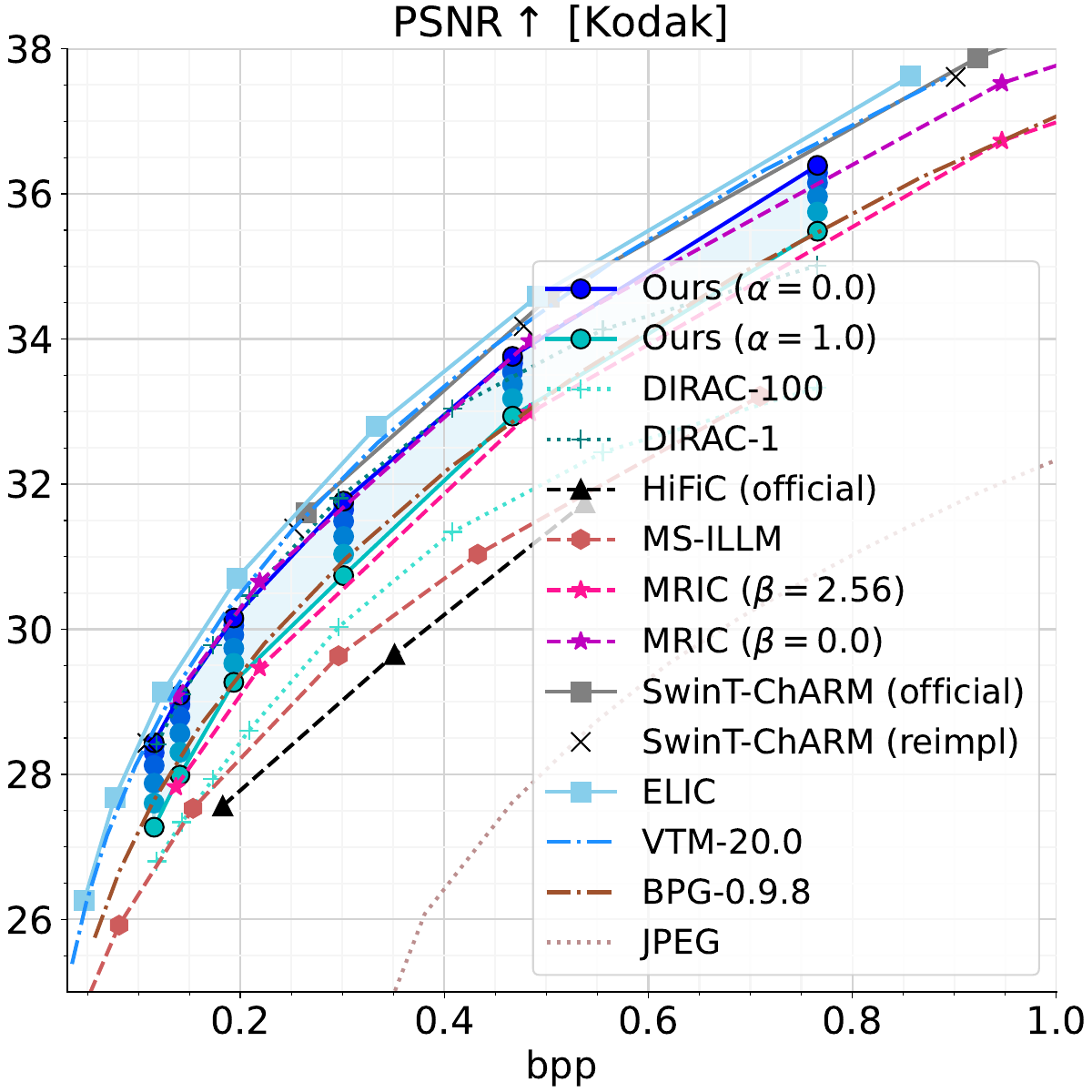}
\end{center}
\caption{Kodak rate-distortion plot}
\label{fig:soa_comp_kodak}
\end{figure}

In~\cref{fig:soa_comp_kodak} we provide the rate-distortion performance for the Kodak dataset. We add the official values of SwinT-ChARM (Zhu \etal, 2022) and ELIC (He \etal, 2022) for reference. Exact configurations for JPEG, BPG-0.9.8, and VTM-20.0 can be found in~\crefrange{subsec:vvc}{subsec:jpeg}.

Note that SwinT-ChARM is almost on par with the current state-of-the-art method ELIC in terms of PSNR and thus represents a good base model for our work. The marginal gap is due to ELIC's more powerful entropy model. We emphasize that both EGIC, MRIC, and DIRAC rely on some variant of the ChARM-entropy model (Minnen \etal, 2020). 

Similar to MRIC, we observe that introducing higher perception results in a $1-1.5$dB PSNR decrease.

\subsection{SwinT-ChARM Reimplementation}

In~\cref{fig:soa_comp_kodak}, we compare the compression performance of our SwinT-ChARM reimplementation (reimpl) to the official values, measured on the Kodak dataset. We optimized the reimplemented version for $2$M optimization steps on the CLIC 2020 training set, using $\lambda \in \{0.01, 0.003, 0.001, 0.0003\}$ and a batch size of $8$. We used a learning rate of $1\mathrm{e}{-4}$ for the first $1.8$M steps and subsequently decayed the learning rate to $1\mathrm{e}{-5}$.  We find that our reimplementation closely matches the official values (up to $0.1$dB tolerance), despite being trained from scratch and using less than two-thirds of the optimization steps. 

\subsection{Experiments on HiFiC}

\begin{figure}[bt]
\begin{center}
\includegraphics[height=6cm]{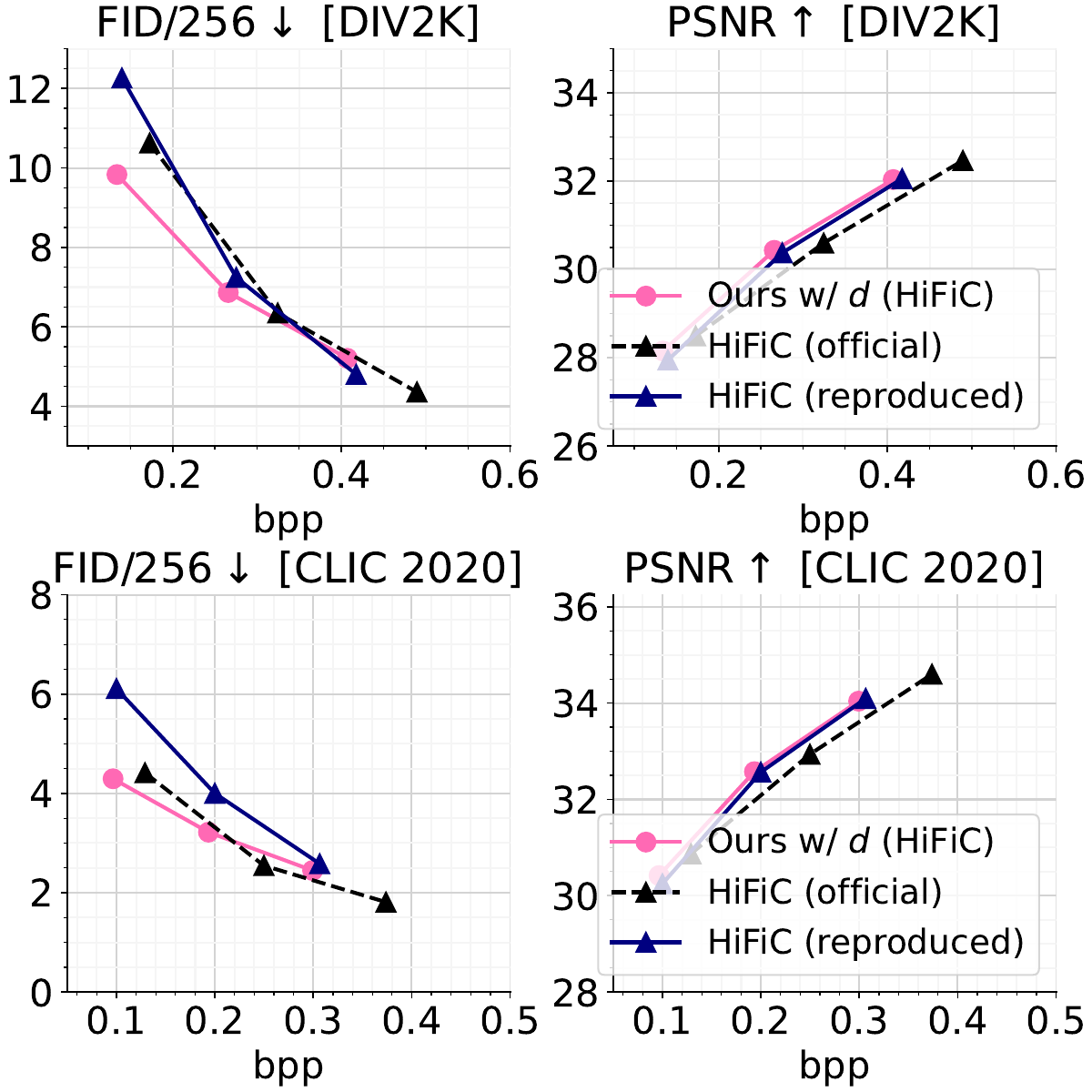}
\end{center}
\caption{Relative comparison to HiFiC}
\label{fig:rel_comp_hific}
\end{figure}

In \cref{fig:rel_comp_hific}, we compare Ours w/ $d$ (HiFiC) to HiFiC (reproduced). Note that HiFiC (official) was trained on an internal dataset is therefore only visualized for transparency reasons. We observe that Ours w/ $d$ (HiFiC) is most effective in the low to medium bit range, which is the key focus of our work. For HiFiC-lo, we achieve an improvement of up to 2 FID points with slightly better PSNR (+0.2dB), suggesting that our method is particularly well suited for the extremely low bit range $< 0.1$bpp. 

For HiFiC-hi, we experience a slight decrease in performance. We suspect that this is due to misaligned hyper-parameters; indeed recent work suggests that a separate set of hyper-parameters is required for various bit-rates (Muckley \etal, 2023).

\subsection{Comparing Training Dynamics}

\begin{figure*}[bt]
  \centering
  \includegraphics[width=\textwidth]{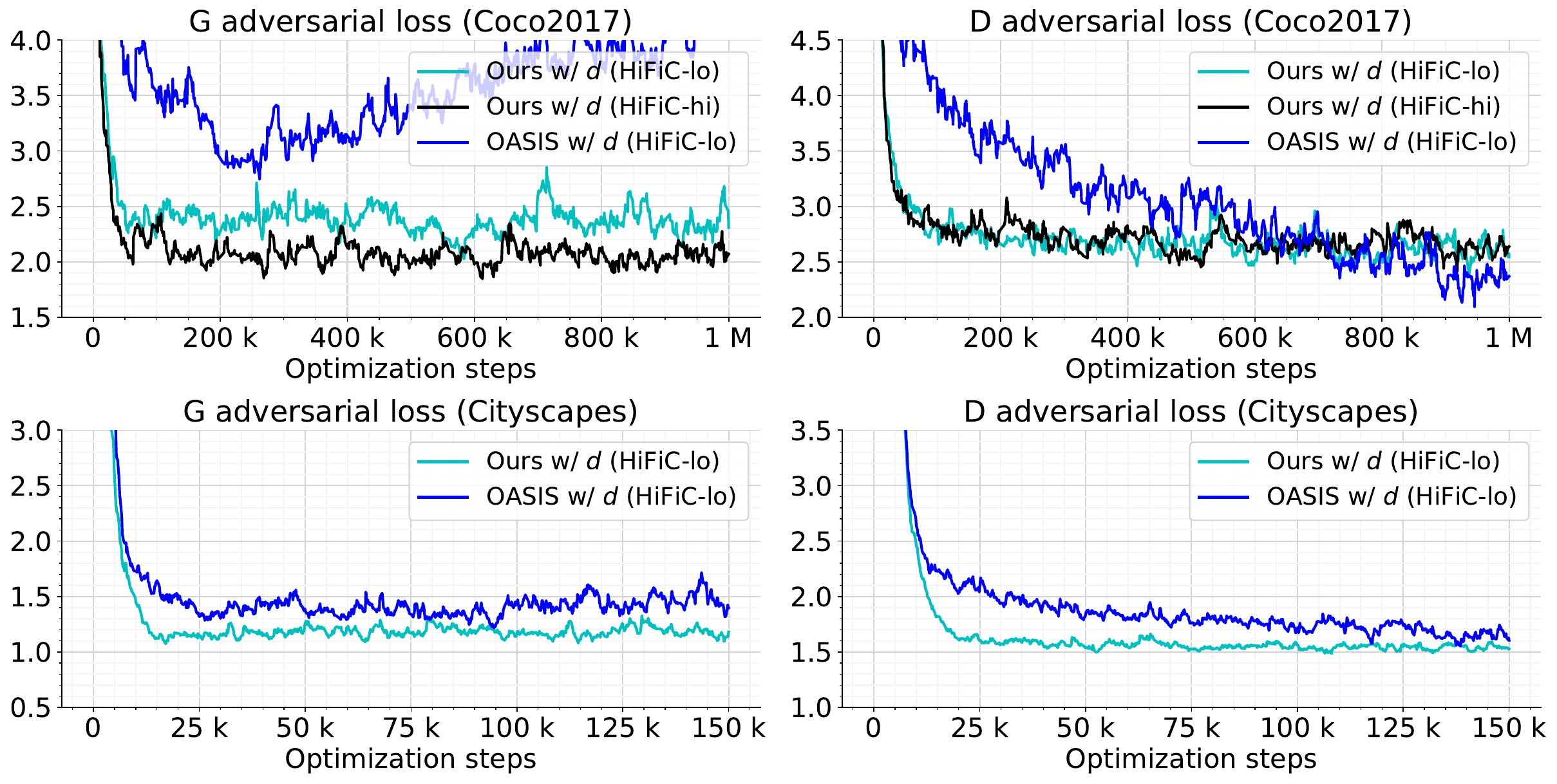}
  \caption{Comparing the training dynamics of OASIS w/ $d$ and Ours w/ $d$. "G adversarial" corresponds to the ($N{+1}$)-cross entropy loss and hence gives an idea of how realistic and semantically correct the resulting reconstructions are (lower is better). "D adversarial" includes regularization terms (lower is better).}
  \label{fig:comp_training_dyn}
\end{figure*}

In~\cref{fig:comp_training_dyn}, we compare the training dynamics of OASIS w/ $d$ and Ours w/ $d$. We find that OASIS with weight norm greatly increases model capacity, while pre-training accelerates training, resulting in superior compression performance. Note that our method provides robust and stable training across different compression rates, while OASIS w/ $d$ exhibits training instabilities that are particularly evident on complex datasets (Coco2017).

\begin{figure*}[bt]
  \centering
  \includegraphics[width=\textwidth]{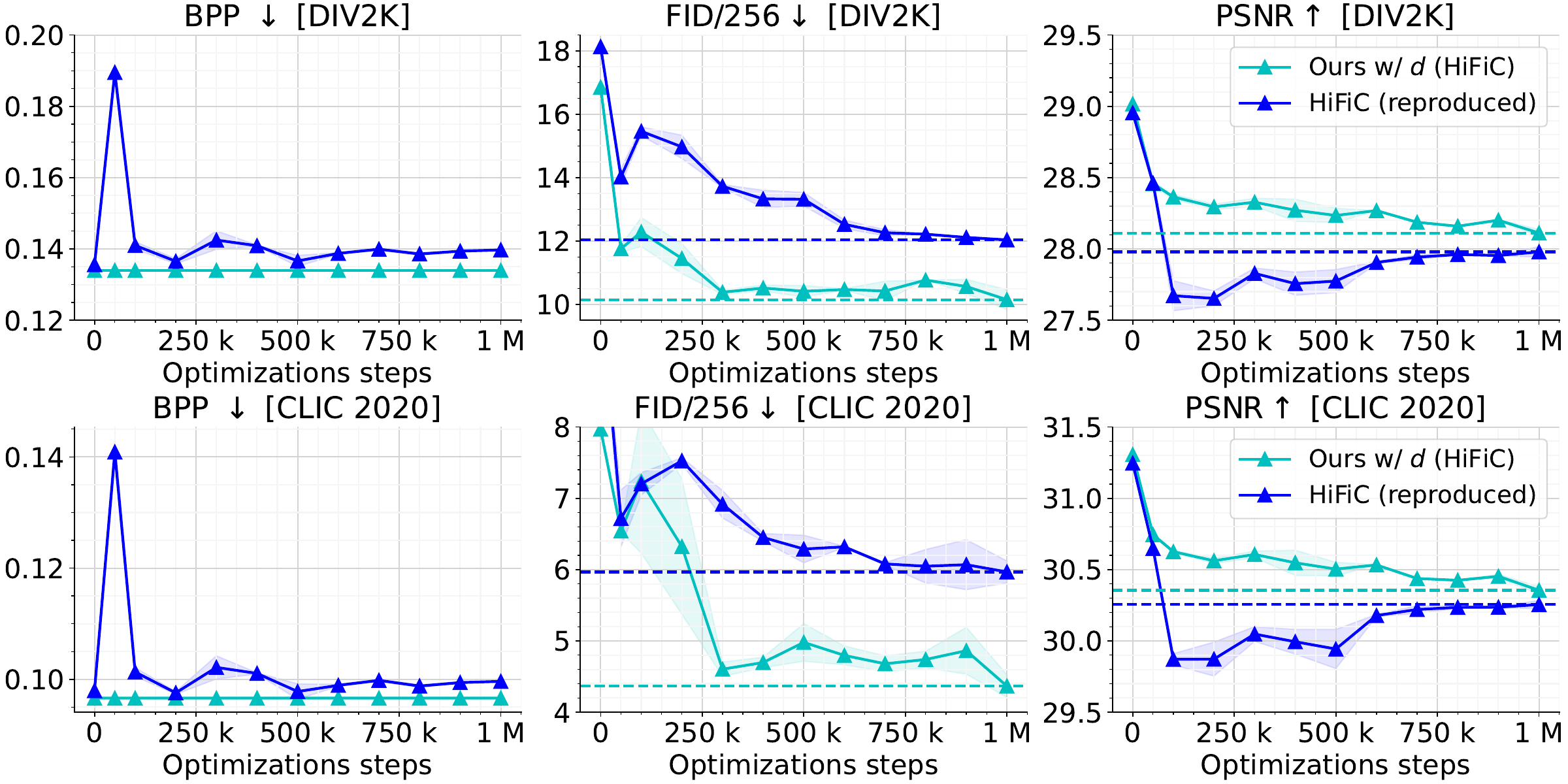}
  \caption{BPP, FID, and PSNR vs optimization steps for the second stage of Ours w/ $d$ (HiFiC) and HiFiC (reproduced). We show the mean and standard deviation across 2 runs per setting. Values at step $0$ correspond to the output values from stage one. We additionally show the values at step $50$k (warm-up phase in HiFiC).}
  \label{fig:comp_training_dyn_details}
\end{figure*}

In~\cref{fig:comp_training_dyn_details}, we provide further performance insights into the training dynamics of Ours w/ $d$ (HiFiC) and HiFiC (reproduced) for stage two. We report the means and standard deviations of BPP, PSNR, and FID as a function of the number of optimization steps across two test runs. Note that HiFiC's training procedure is divided into 3 phases: warm-up ($0-50$k), training with a learning rate of $1\mathrm{e}{-4}$ ($50-500$k) and $1\mathrm{e}{-5}$ ($500$k-$1$M), respectively, whereas, Ours w/ $d$ (HiFiC) uses the same learning rate and $\lambda$-schedule across all training steps.

We find that our method considerably
accelerates training progress, similar to projected GANs (Sauer~\etal, 2021). As can be seen, our method exceeds the performance of HiFiC after only $300$k optimization steps. The large deviations at the beginning of the training phase can be attributed to a sort of calibration phase in which the variables for the projection-based conditioning mechanism are learned from scratch.

\subsection{Impact of the Focal Frequency Loss}

In~\cref{table:gan_freq_bias}, we summarize the effect of the focal frequency loss (FFL, Jiang~\etal, 2021) on the concat base configurations. For that, we finetune all base models for additional 50k steps. We find that the FFL has the greatest impact on conf-a and conf-d, whereas it has little impact on the discriminators based on pixel-level supervision (conf-c and conf-e). We also find that the FFL cannot further improve ours w/o $d$, which reinforces the design decisions made in our work.

\begin{table}[bt]
\caption{Which method benefits the most from the FFL? The relative change (rel-PSNR, rel-FID) is here denoted over their respective concat-base configurations.}
\begin{center}
\begin{tabular}{llclc}
\toprule
 Method                     & \multicolumn{2}{c}{Distortion}  & \multicolumn{2}{c}{Perception}  \\ \cmidrule{2-5}
                            & PSNR $\uparrow$ & rel-PSNR    & FID $\downarrow$  & rel-FID       \\ \midrule
 conf-a w/ FFL                   & \textbf{32.53}  & +33.8\%     & 40.64             & -63.8\%       \\
 conf-b w/ FFL                   & 29.17           & -0.9\%      & 79.55             & +5.2\%        \\
 conf-c w/ FFL                   & 29.23           & -0.8\%      & 89.73             & +3.1\%        \\
 conf-d w/ FFL                   & 29.89           & +1.7\%      & 21.85             & -29.1\%       \\
 conf-e w/ FFL                   & 30.25           & +0.7\%      & 15.93             & -3.5\%        \\ \midrule
 ours w/o $d$               & 29.56           & -1.4\%      & \textbf{8.73}     & +12.8\%       \\ \bottomrule
\end{tabular}
\end{center}
\label{table:gan_freq_bias}
\end{table}

\subsection{Computational/ Model Complexity}

In~\cref{table:model_comp}, we compare the storage-efficiency of each model in terms of model parameters (in millions). For the generator, we further differentiate between the base model size and additional parameters required for traversing the D-P curve (denoted by $+$). The calculation for $P$ includes the hyper-analysis and hyper-synthesis transforms, as well as additional slice transforms in the case of ChARM.

It is worth noting that EGIC during inference is identical to SwinT-ChARM (neglecting ORP); latency numbers can be found in Zhu~\etal, 2022 (Tab.~3 and Sec.~D.3). GPU-memory overhead only incurs during training.

\begin{table*}[bt]
\caption{Model size comparison in millions of parameters (M)}
\begin{center}
\begin{tabular}{lllll}
\toprule
 Method                                     & $E$       & $G$             & $P$   & Total (M)    \\ \midrule
 HiFiC (Mentzer~\etal, 2020)                & 7.4       & 156.8           & 17.3  & 181.5   \\
 MS-ILLM (Muckley~\etal, 2023)              & 7.4       & 156.8           & 17.3  & 181.5   \\
 DIRAC (Ghouse~\etal, 2023)                 & 7.0       & 7.0 + 108.4     & 14.3 & 136.8    \\ 
 HFD/DDPM (Hoogeboom~\etal, 2023)           & 10.7       & 10.7 + 1033.9     & 36.4 & 1091.7    \\ 
 MRIC (Agustsson~\etal, 2023)               & 10.7      & 10.7 + 2.65     & 36.4  & 60.45   \\
 EGIC (Ours)                                & 9.1       & 9.1 + 0.4       & 14.4  & 33      \\ 
 \bottomrule 
\end{tabular}
\end{center}
\label{table:model_comp}
\end{table*}

\subsection{Image/ Weight Interpolation}\label{subsec:iw_interpol}

Image and weight interpolation (Wang~\etal 2019, Iwai~\etal 2021, Yan~\etal 2022) can be achieved using

\begin{equation}\label{eq:image_interpol}
x' = (1-\alpha) G_{1}(y)+\alpha G_{2}(y),
\end{equation}

\begin{equation}\label{eq:weight_interpol}
x' = G_{\theta}(y); \theta = (1-\alpha) \theta_{G_1}+\alpha \theta_{G_2}, 
\end{equation}

where $\theta$ and $\alpha$ correspond to the model parameters and interpolation weight, respectively. For our ablation study, we fine-tune the generator weights from stage one $G_1$ ($=G_{ORP}$) for additional 500k optimization steps. We use $\alpha \in \{0.0, 0.17, 0.33, 0.5, 0.67, 0.83, 1.0\}$, resulting in seven points per bit-rate.

\begin{figure*}[bt]
\begin{center}
\includegraphics[width=1\textwidth]{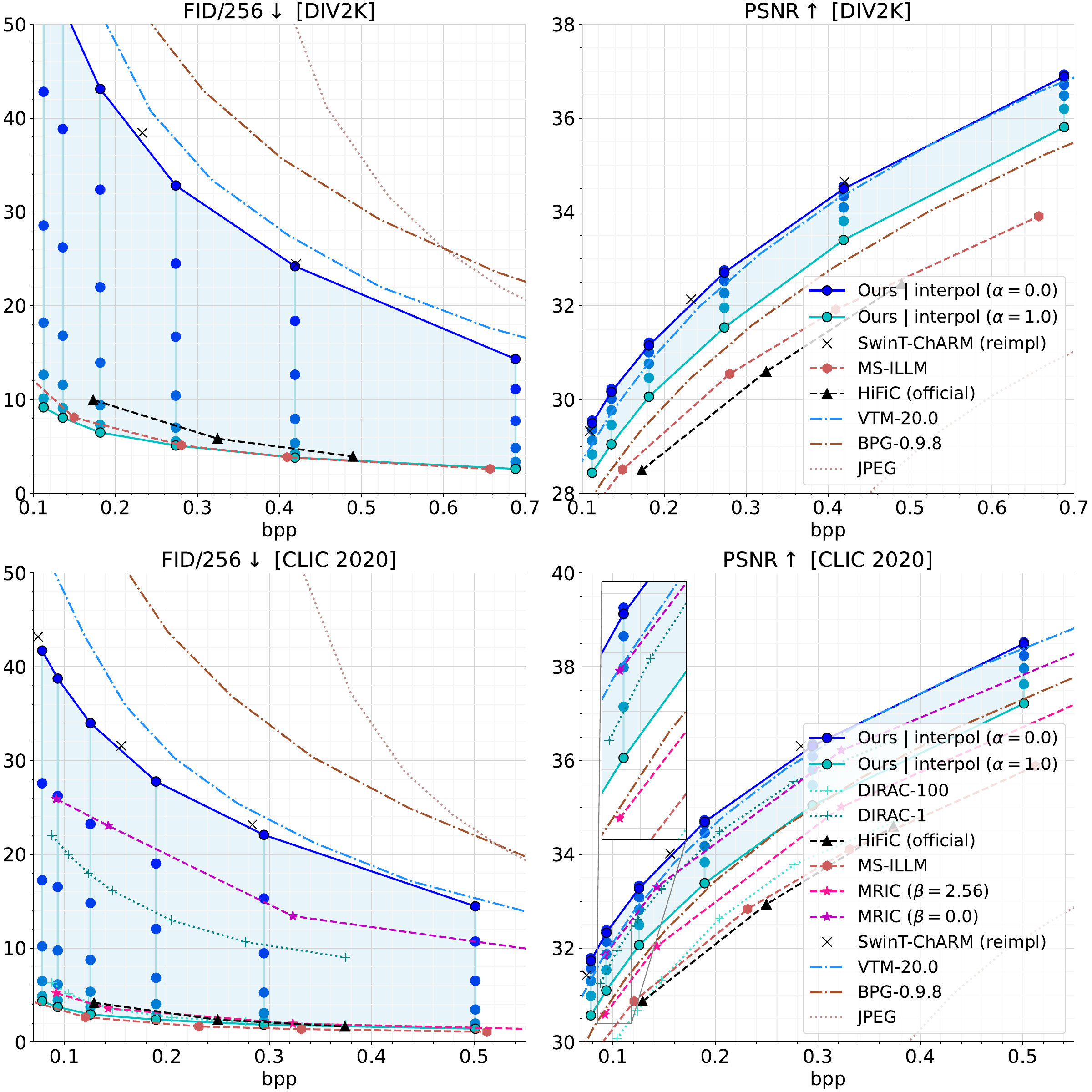}
\end{center}
   \caption{Traversing the rate-distortion-perception plane using image interpolation}
\label{fig:ext_mric_ii}
\end{figure*}

\begin{figure*}[bt]
\begin{center}
\includegraphics[width=1\textwidth]{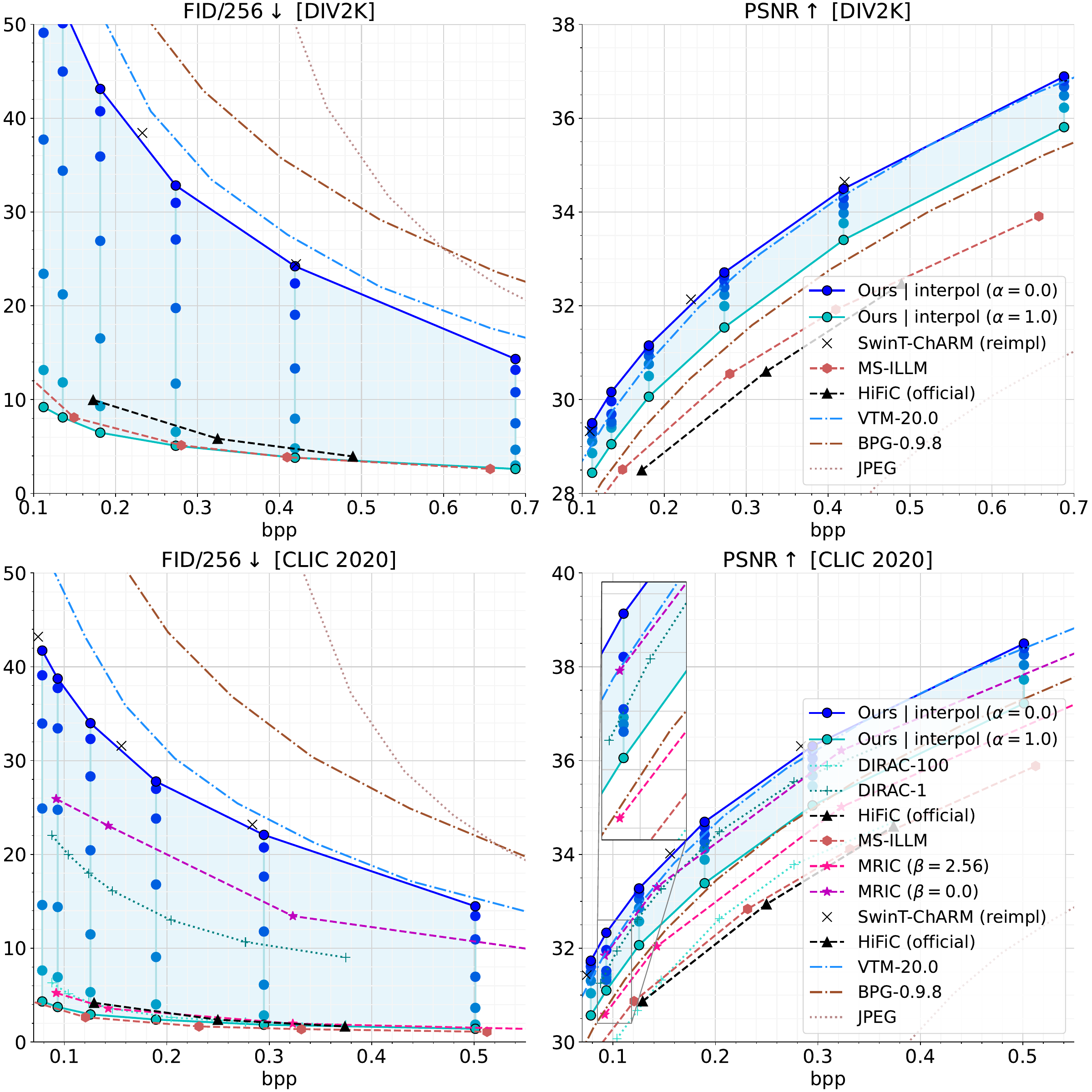}
\end{center}
   \caption{Traversing the rate-distortion-perception plane using weight interpolation}
\label{fig:ext_mric_wi}
\end{figure*}

Our results\footnote{The results are based on an early stage of EGIC, which produces a slightly different D-P trade-off. The overall logic remains however the same.} are summarized in~\cref{fig:ext_mric_ii} and~\cref{fig:ext_mric_wi}. As can be seen, both methods work reasonably well; for weight interpolation, we observe skewed interpolation characteristics in some cases (\eg, CLIC 2020 at low bit-rate). Noteworthy, Ours $\vert$ interpol ($\alpha=0.0$) almost matches the performance of SwinT-ChARM (reimpl), which can be considered an upper bound.

\subsection{Visual Comparison: Concat vs Projection}

\begin{figure*}[bt]
   \setlength{\tabcolsep}{1pt}
   \renewcommand{\arraystretch}{0.5}
    \centering
    \begin{tabular}{ccc}
        \toprule
        input   &conf-c/ & conf-c/ \\
                & concat & projection \\
        \midrule
        \begin{subfigure}{0.325\textwidth}
            \centering
            \includegraphics[width=\linewidth]{figures/comparing_gans/input_frankfurt_000000_000294_leftImg8bit_crop.png}
        \end{subfigure}
        & 
        \begin{subfigure}{0.325\textwidth}
            \centering
            \includegraphics[width=\linewidth]{figures/comparing_gans/config-c_frankfurt_000000_000294_leftImg8bit_crop.png}
        \end{subfigure}
        & 
        \begin{subfigure}{0.325\textwidth}
            \centering
            \includegraphics[width=\linewidth]{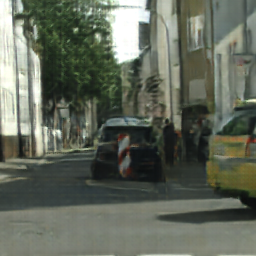}
        \end{subfigure} \\
        \begin{subfigure}{0.325\textwidth}
            \centering
            \includegraphics[width=\linewidth]{figures/comparing_gans/input_frankfurt_000000_000294_leftImg8bit_crop_spectra.png}
        \end{subfigure}
        &
        \begin{subfigure}{0.325\textwidth}
            \centering
            \includegraphics[width=\linewidth]{figures/comparing_gans/config-c_frankfurt_000000_000294_leftImg8bit_crop_spectra.png}
        \end{subfigure}
        & 
        \begin{subfigure}{0.325\textwidth}
            \centering
            \includegraphics[width=\linewidth]{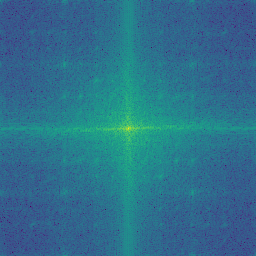}
        \end{subfigure} \\
    \end{tabular}
    \caption{Comparing U-Net (conf-c w/ concat) vs U-Net (conf-c w/ projection). Note that projection considerably reduces image artifacts.}
    \label{tbl:concat_vs_projection}
\end{figure*}

In~\cref{tbl:concat_vs_projection}, we provide additional visual impressions of the effect of various conditioning strategies. We find that projection greatly helps to reduce image artifacts.

\subsection{Pixel Weighting Schemes}

\begin{figure*}[bt]
   \setlength{\tabcolsep}{1pt}
   \renewcommand{\arraystretch}{0.5}
   \centering
   
   \begin{tabular}{cccc}
       input $x$ & label map & pixel weights $w$      & pixel weights $w$ \\
        &  & (Yang~\etal)     & (Sch{\"o}nfeld~\etal) \\
       \begin{subfigure}{0.24\textwidth}
        \centering
        \includegraphics[width=\linewidth]{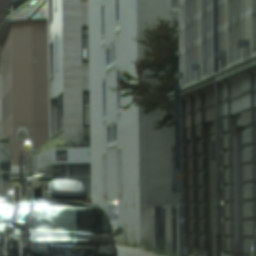}
       \end{subfigure}
       & 
       \begin{subfigure}{0.24\textwidth}
        \centering
        \includegraphics[width=\linewidth]{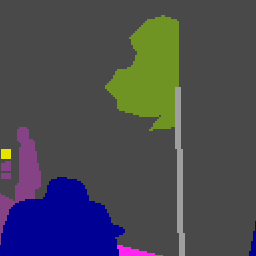}
       \end{subfigure}
       & 
       \begin{subfigure}{0.24\textwidth}
        \centering
        \includegraphics[width=\linewidth]{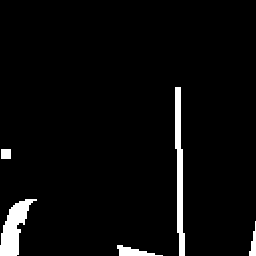}
        \end{subfigure}
        & 
       \begin{subfigure}{0.24\textwidth}
        \centering
        \includegraphics[width=\linewidth]{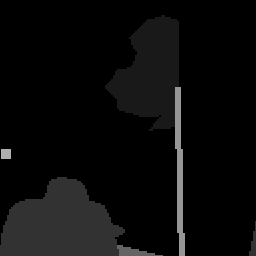}
        \end{subfigure}\\
       \begin{subfigure}{0.24\textwidth}
        \centering
        \includegraphics[width=\linewidth]{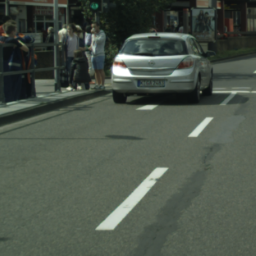}
       \end{subfigure}
       & 
       \begin{subfigure}{0.24\textwidth}
        \centering
        \includegraphics[width=\linewidth]{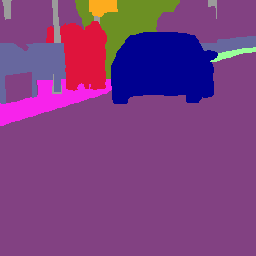}
       \end{subfigure}
       & 
       \begin{subfigure}{0.24\textwidth}
        \centering
        \includegraphics[width=\linewidth]{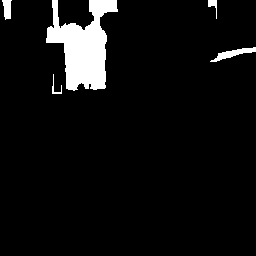}
        \end{subfigure}
        & 
       \begin{subfigure}{0.24\textwidth}
        \centering
        \includegraphics[width=\linewidth]{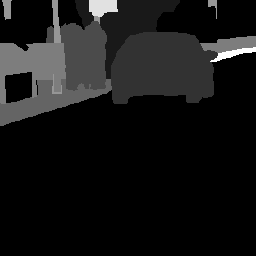}
        \end{subfigure}\\
       \begin{subfigure}{0.24\textwidth}
        \centering
        \includegraphics[width=\linewidth]{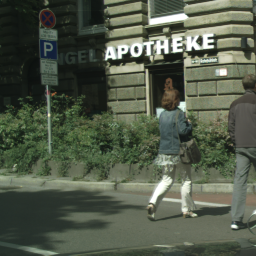}
       \end{subfigure}
       & 
       \begin{subfigure}{0.24\textwidth}
        \centering
        \includegraphics[width=\linewidth]{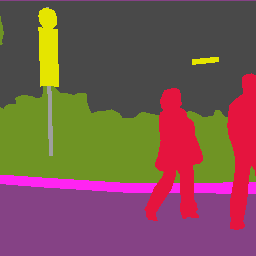}
       \end{subfigure}
       & 
       \begin{subfigure}{0.24\textwidth}
        \centering
        \includegraphics[width=\linewidth]{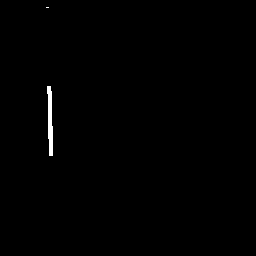}
        \end{subfigure}
        & 
       \begin{subfigure}{0.24\textwidth}
        \centering
        \includegraphics[width=\linewidth]{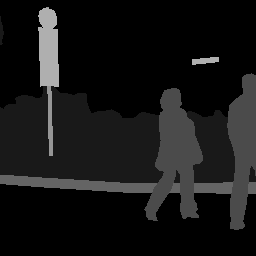}
        \end{subfigure}\\
       \begin{subfigure}{0.24\textwidth}
        \centering
        \includegraphics[width=\linewidth]{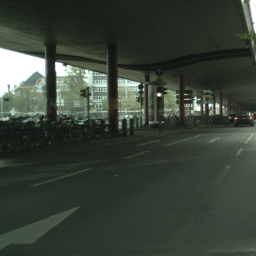}
       \end{subfigure}
       & 
       \begin{subfigure}{0.24\textwidth}
        \centering
        \includegraphics[width=\linewidth]{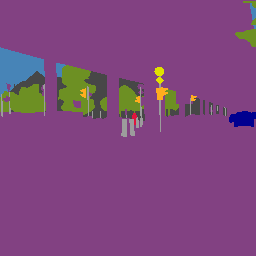}
       \end{subfigure}
       & 
       \begin{subfigure}{0.24\textwidth}
        \centering
        \includegraphics[width=\linewidth]{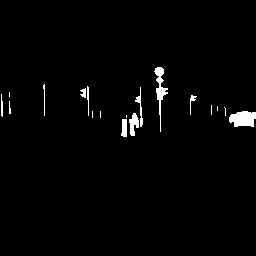}
        \end{subfigure}
        & 
       \begin{subfigure}{0.24\textwidth}
        \centering
        \includegraphics[width=\linewidth]{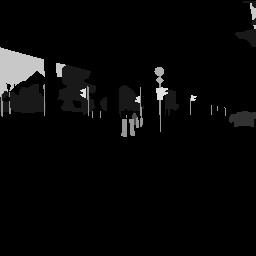}
        \end{subfigure}\\
   \end{tabular}
   \caption{Comparing pixel weighting schemes based on instance size (third column, Yang~\etal, 2019) and class imbalance (fourth column, Sch{\"o}nfeld~\etal, 2021). We map the pixel weights $w$ to a pre-defined color map for better visualization; the brighter the color, the larger the weight. Note that in Sch{\"o}nfeld~\etal (2021) identical class segments share the same pixel weights (same color).}
   \label{tbl:cls_pxl_weights}
\end{figure*}

Pixel weighting schemes have played a minor role in our work. As mentioned earlier, we use the simple instance size-based weighting scheme introduced in Yang~\etal (2019), whereas, in Sch{\"o}nfeld~\etal (2021), each semantic class is weighted by its inverse per-pixel frequency, computed over a batch of images. In~\cref{table:pixel_weight_schemes}, we show that this method is indeed effective and performs comparably to the more sophisticated approach of Sch{\"o}nfeld~\etal (2021). In~\cref{tbl:cls_pxl_weights}, we provide additional visual comparisons.

\begin{table}[tb]
\caption{Comparing different pixel weighting schemes}
\begin{center}
\begin{tabular}{lcc}
\toprule
 Method & PSNR $\uparrow$ & FID $\downarrow$ \\\midrule
 OASIS (instance size-oriented) & 29.90 & \textbf{15.30} \\
 OASIS (class-oriented) & 29.90 & 15.56 \\\bottomrule
\end{tabular}
\end{center}
\label{table:pixel_weight_schemes}
\end{table}

\subsection{Comparison to VVC-intra}\label{subsec:vvc}

The evaluation of the VVC standard (current state-of-the-art for non-learned image compression codecs) is based on VTM-20.0, a reference software provided by \url{https://vcgit.hhi.fraunhofer.de/jvet/VVCSoftware_VTM/-/releases/VTM-20.0}. Similar to previous work, we first convert the PNG images to YCbCr-format using ffmpeg \url{https://www.ffmpeg.org/}:

\begin{verbatim}
ffmpeg 
-i $PNGPATH -pix_fmt yuv444p $YUVPATH
\end{verbatim}

To compress/ decompress the images, we use:

\begin{verbatim}
# Encode
EncoderAppStatic 
-c encoder_intra_vtm.cfg -i $YUVPATH -q $Q, 
-o /dev/null -b $OUTPUT
--SourceWidth=$WIDTH
--SourceHeight=$HEIGHT
--FrameRate=1 --FramesToBeEncoded=1
--InputBitDepth=8
--InputChromaFormat=444
--ConformanceWindowMode=1

# Decode
DecoderAppStatic
-b $OUTPUT -o $RECON -d 8
\end{verbatim}

To convert the outputs back to PNG-format, we use:
\begin{verbatim}
ffmpeg
-f rawvideo -s $WIDTHx$HEIGHT 
-pix_fmt yuv444p -i $RECON $RECON_PNG
\end{verbatim}

PSNR is measured on the 8bit-decoded images and not on the floating point reconstructions, which is consistent with all our comparisons.

\subsection{Comparison to BPG}\label{subsec:bpg}

The evaluation of BPG-0.9.8 is based on the HEVC open video compression standard, provided by \url{https://bellard.org/bpg/}:

\begin{verbatim}
# Encode
bpgenc -o $OUTPUT -q $Q 
-f 444 -e x265 -b 8 $INPUT

# Decode
bpgdec -o $RECON $OUTPUT
\end{verbatim}

\subsection{Comparison to JPEG}\label{subsec:jpeg}

We use the Python Imaging Library (PIL) to obtain the JPEG encoded/ decoded images; (chroma) subsampling is set to $0$, which corresponds to $4:4:4$, the highest quality setting.

\begin{verbatim}
tmp = io.BytesIO()
img.save(tmp, format='jpeg', 
         subsampling=0, 
         quality=Q)
tmp.seek(0)
filesize = tmp.getbuffer().nbytes
bpp = filesize * float(8)/ 
      img.size[0] * img.size[1]
rec = Image.open(tmp)
\end{verbatim}

\subsection{Visual Comparison}
\begin{figure*}[bt]
   \setlength{\tabcolsep}{1pt}
   \renewcommand{\arraystretch}{0.5}
   \centering
   
   \begin{tabular}{ccccc}
         \multicolumn{5}{c}{\begin{subfigure}{0.98\textwidth}
        \centering
        \includegraphics[width=\linewidth]{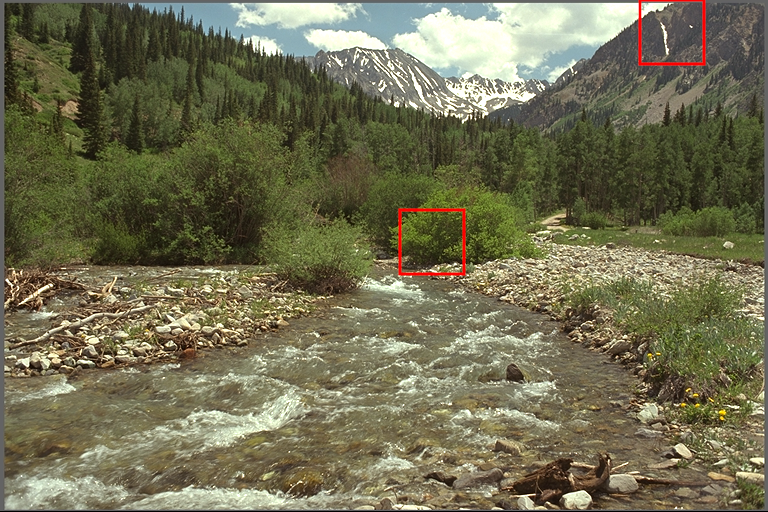}
       \end{subfigure}}\\
   \toprule
       JPEG & BPG-0.9.8 & VTM-20.0 & EGIC & EGIC \\
        &  &  & Ours ($\alpha=0.0$) & Ours ($\alpha=1.0$) \\
       \midrule
       $0.296$bpp & $0.333$bpp & $0.292$bpp & $0.282$bpp & $0.282$bpp \\
       ($1.05\times$) & ($1.18\times$) & ($1.04\times$) &  &  \\
       \begin{subfigure}{0.195\textwidth}
        \centering
        \includegraphics[width=\linewidth]{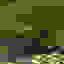}
       \end{subfigure}
       & 
       \begin{subfigure}{0.195\textwidth}
        \centering
        \includegraphics[width=\linewidth]{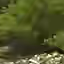}
       \end{subfigure}
       & 
       \begin{subfigure}{0.195\textwidth}
        \centering
        \includegraphics[width=\linewidth]{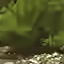}
        \end{subfigure}
        & 
       \begin{subfigure}{0.195\textwidth}
        \centering
        \includegraphics[width=\linewidth]{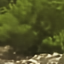}
        \end{subfigure}
        & 
       \begin{subfigure}{0.195\textwidth}
        \centering
        \includegraphics[width=\linewidth]{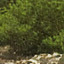}
        \end{subfigure}\\
       \begin{subfigure}{0.195\textwidth}
        \centering
        \includegraphics[width=\linewidth]{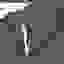}
       \end{subfigure}
       & 
       \begin{subfigure}{0.195\textwidth}
        \centering
        \includegraphics[width=\linewidth]{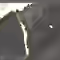}
       \end{subfigure}
       & 
       \begin{subfigure}{0.195\textwidth}
        \centering
        \includegraphics[width=\linewidth]{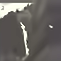}
        \end{subfigure}
        & 
       \begin{subfigure}{0.195\textwidth}
        \centering
        \includegraphics[width=\linewidth]{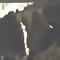}
        \end{subfigure}
        & 
       \begin{subfigure}{0.195\textwidth}
        \centering
        \includegraphics[width=\linewidth]{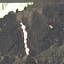}
        \end{subfigure}\\
    
   \end{tabular}
   \caption{Visual comparison of EGIC ($\alpha=\{0.0, 1.0\}$) with JPEG, BPG-0.9.8 and VTM-20.0 on the Kodak dataset (kodim13).}
   \label{tbl:vis_kodak_13}
\end{figure*}

\begin{figure*}[bt]
   \setlength{\tabcolsep}{1pt}
   \renewcommand{\arraystretch}{0.5}
   \centering
   
   \begin{tabular}{ccccc}
         \multicolumn{5}{c}{\begin{subfigure}{0.98\textwidth}
        \centering
        \includegraphics[width=\linewidth]{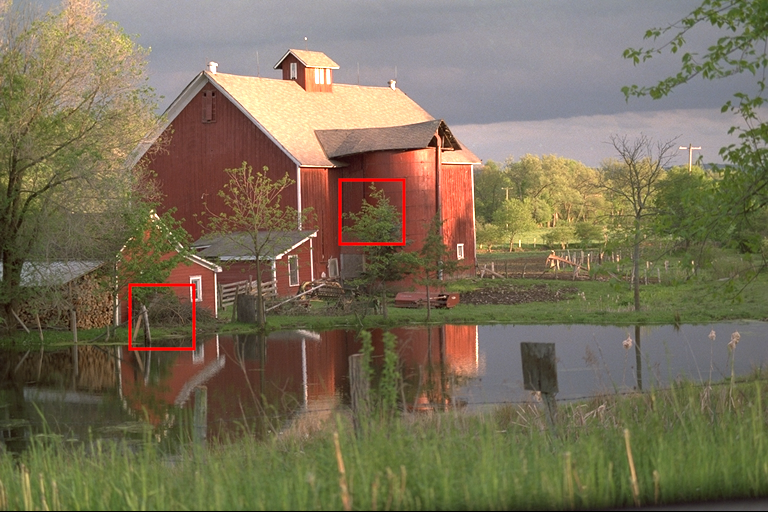}
       \end{subfigure}}\\
   \toprule
       JPEG & BPG-0.9.8 & VTM-20.0 & EGIC & EGIC \\
        &  &  & Ours ($\alpha=0.0$) & Ours ($\alpha=1.0$) \\
       \midrule
       $0.254$bpp & $0.126$bpp & $0.143$bpp & $0.127$bpp & $0.127$bpp \\
       ($2.0\times$) & ($0.99\times$) & ($1.13\times$) &  &  \\
       \begin{subfigure}{0.195\textwidth}
        \centering
        \includegraphics[width=\linewidth]{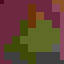}
       \end{subfigure}
       & 
       \begin{subfigure}{0.195\textwidth}
        \centering
        \includegraphics[width=\linewidth]{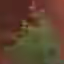}
       \end{subfigure}
       & 
       \begin{subfigure}{0.195\textwidth}
        \centering
        \includegraphics[width=\linewidth]{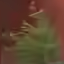}
        \end{subfigure}
        & 
       \begin{subfigure}{0.195\textwidth}
        \centering
        \includegraphics[width=\linewidth]{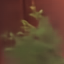}
        \end{subfigure}
        & 
       \begin{subfigure}{0.195\textwidth}
        \centering
        \includegraphics[width=\linewidth]{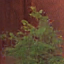}
        \end{subfigure}\\
       \begin{subfigure}{0.195\textwidth}
        \centering
        \includegraphics[width=\linewidth]{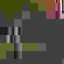}
       \end{subfigure}
       & 
       \begin{subfigure}{0.195\textwidth}
        \centering
        \includegraphics[width=\linewidth]{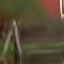}
       \end{subfigure}
       & 
       \begin{subfigure}{0.195\textwidth}
        \centering
        \includegraphics[width=\linewidth]{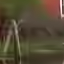}
        \end{subfigure}
        & 
       \begin{subfigure}{0.195\textwidth}
        \centering
        \includegraphics[width=\linewidth]{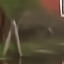}
        \end{subfigure}
        & 
       \begin{subfigure}{0.195\textwidth}
        \centering
        \includegraphics[width=\linewidth]{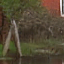}
        \end{subfigure}\\
   \end{tabular}
   \caption{Visual comparison of EGIC ($\alpha=\{0.0, 1.0\}$) with JPEG, BPG-0.9.8 and VTM-20.0 on the Kodak dataset (kodim22).}
   \label{tbl:vis_kodak_22}
\end{figure*}

\begin{figure*}[bt]
   \setlength{\tabcolsep}{1pt}
   \renewcommand{\arraystretch}{0.5}
   \centering
   
   \begin{tabular}{cccc}
   \toprule
       \multicolumn{2}{c}{input $x$} & \multicolumn{2}{c}{EGIC} \\
       \multicolumn{2}{c}{}  & \multicolumn{2}{c}{Ours ($\alpha=1.0$)} \\
    \midrule
    \multicolumn{2}{c}{kodim20} & \multicolumn{2}{c}{$0.077$bpp}  \\
       \begin{subfigure}{0.242\textwidth}
        \centering
        \includegraphics[width=\linewidth]{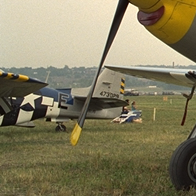}
       \end{subfigure}
       & 
       \begin{subfigure}{0.242\textwidth}
        \centering
        \includegraphics[width=\linewidth]{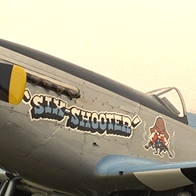}
       \end{subfigure}
       & 
       \begin{subfigure}{0.242\textwidth}
        \centering
        \includegraphics[width=\linewidth]{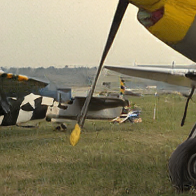}
        \end{subfigure}
        & 
       \begin{subfigure}{0.242\textwidth}
        \centering
        \includegraphics[width=\linewidth]{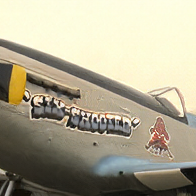}
        \end{subfigure}\\
     \multicolumn{2}{c}{\begin{subfigure}{0.49\textwidth}
        \centering
        \includegraphics[width=\linewidth]{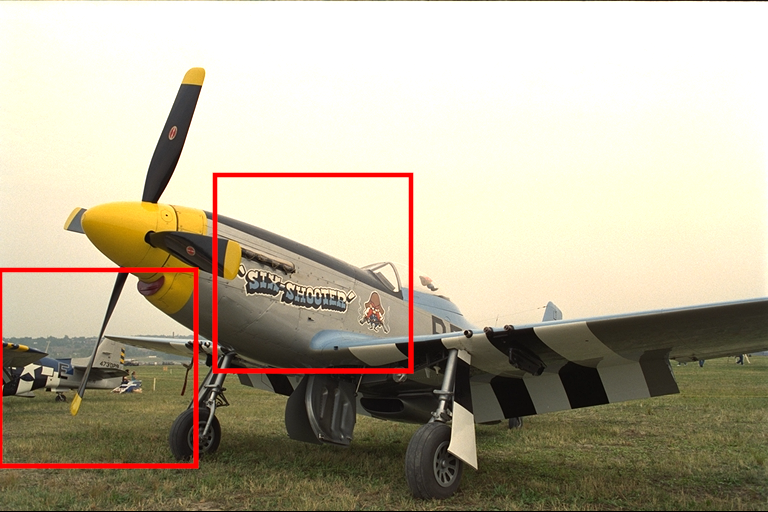}
       \end{subfigure}}
       & 
       \multicolumn{2}{c}{\begin{subfigure}{0.49\textwidth}
        \centering
        \includegraphics[width=\linewidth]{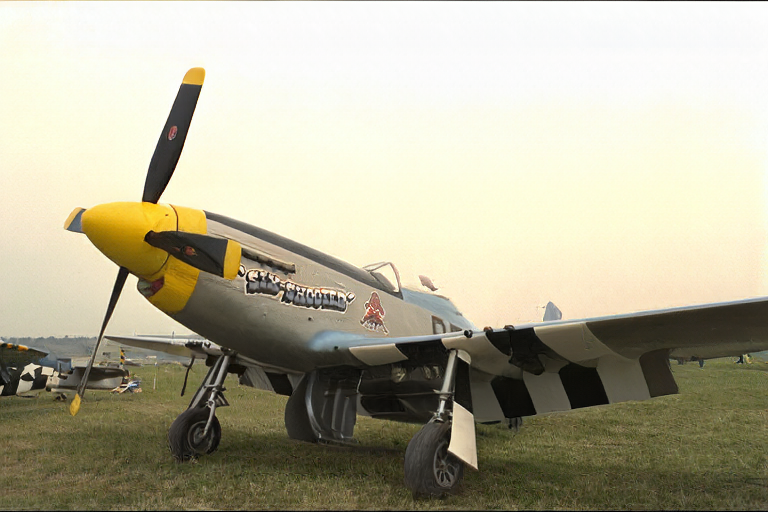}
       \end{subfigure}} \\
       \midrule
        \multicolumn{2}{c}{HiFiC} &  \multicolumn{2}{c}{MS-ILLM} \\
        \multicolumn{2}{c}{(Mentzer~\etal)} &  \multicolumn{2}{c}{(Muckley~\etal)} \\
       \midrule
    \multicolumn{2}{c}{$0.121$bpp ($1.57\times$)} & \multicolumn{2}{c}{$0.094$bpp ($1.22\times$)}  \\
       \begin{subfigure}{0.242\textwidth}
        \centering
        \includegraphics[width=\linewidth]{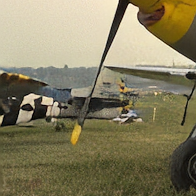}
       \end{subfigure}
       & 
       \begin{subfigure}{0.242\textwidth}
        \centering
        \includegraphics[width=\linewidth]{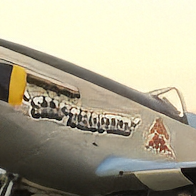}
       \end{subfigure}
       & 
       \begin{subfigure}{0.242\textwidth}
        \centering
        \includegraphics[width=\linewidth]{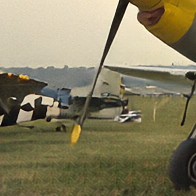}
        \end{subfigure}
        & 
       \begin{subfigure}{0.242\textwidth}
        \centering
        \includegraphics[width=\linewidth]{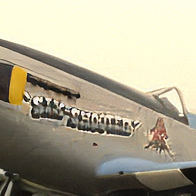}
        \end{subfigure}\\
     \multicolumn{2}{c}{\begin{subfigure}{0.49\textwidth}
        \centering
        \includegraphics[width=\linewidth]{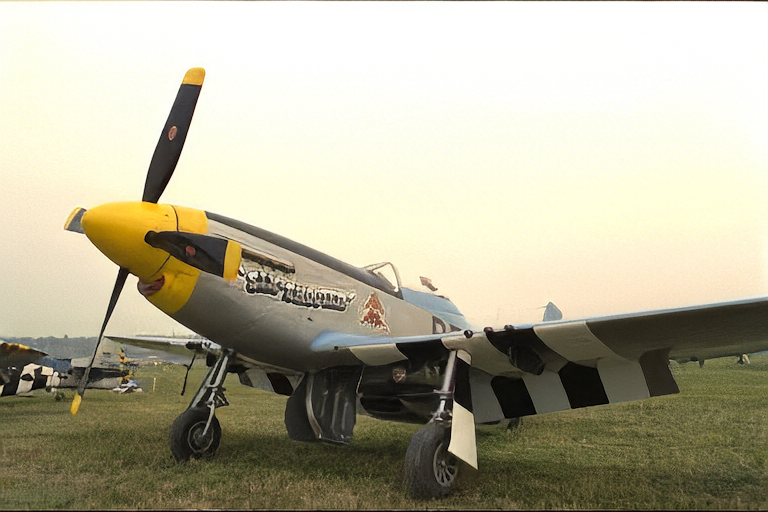}
       \end{subfigure}}
       & 
       \multicolumn{2}{c}{\begin{subfigure}{0.49\textwidth}
        \centering
        \includegraphics[width=\linewidth]{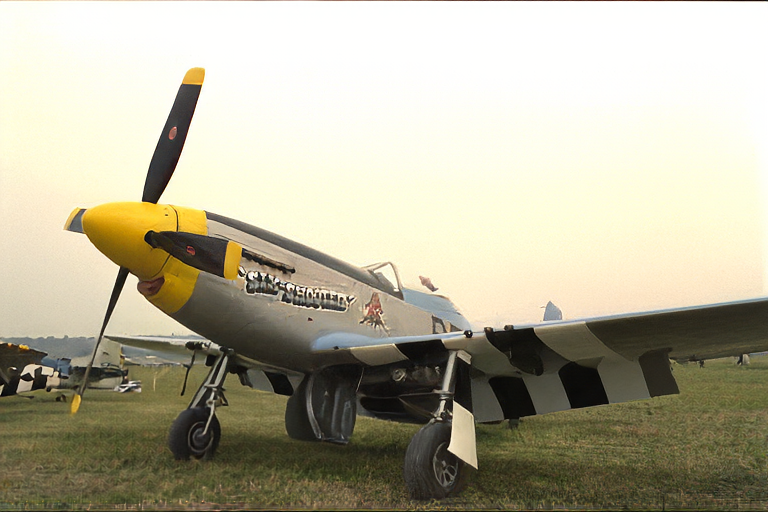}
       \end{subfigure}} \\
   \end{tabular}
   \caption{Visual comparison of EGIC ($\alpha=1.0$) with HiFiC and MS-ILLM on the Kodak dataset (kodim20). Note that our method better preserves textual information and texture (grass), despite using less bpp.}
   \label{tbl:vis_kodak_20}
\end{figure*}

\begin{figure*}[bt]
   \setlength{\tabcolsep}{1pt}
   \renewcommand{\arraystretch}{0.5}
   \centering
   
   \begin{tabular}{cccc}
   \toprule
       \multicolumn{2}{c}{input $x$} & \multicolumn{2}{c}{EGIC} \\
       \multicolumn{2}{c}{}  & \multicolumn{2}{c}{Ours ($\alpha=1.0$)} \\
    \midrule
    \multicolumn{2}{c}{kodim14} & \multicolumn{2}{c}{$0.184$bpp}  \\
       \begin{subfigure}{0.242\textwidth}
        \centering
        \includegraphics[width=\linewidth]{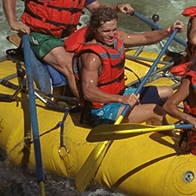}
       \end{subfigure}
       & 
       \begin{subfigure}{0.242\textwidth}
        \centering
        \includegraphics[width=\linewidth]{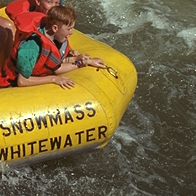}
       \end{subfigure}
       & 
       \begin{subfigure}{0.242\textwidth}
        \centering
        \includegraphics[width=\linewidth]{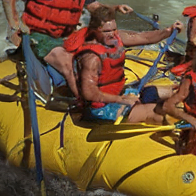}
        \end{subfigure}
        & 
       \begin{subfigure}{0.242\textwidth}
        \centering
        \includegraphics[width=\linewidth]{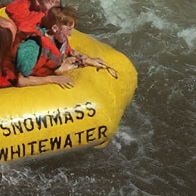}
        \end{subfigure}\\
     \multicolumn{2}{c}{\begin{subfigure}{0.49\textwidth}
        \centering
        \includegraphics[width=\linewidth]{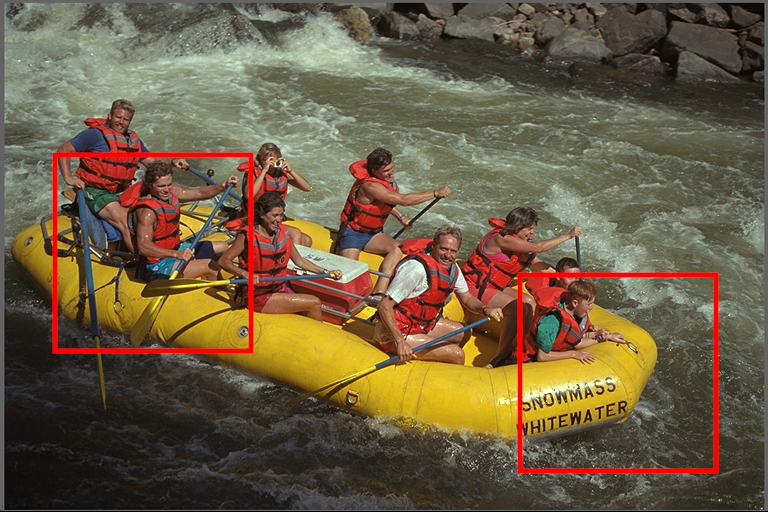}
       \end{subfigure}}
       & 
       \multicolumn{2}{c}{\begin{subfigure}{0.49\textwidth}
        \centering
        \includegraphics[width=\linewidth]{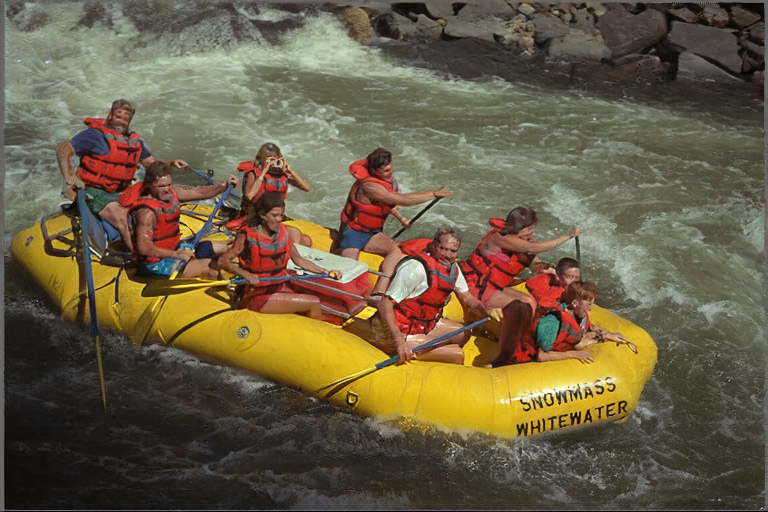}
       \end{subfigure}} \\
       \midrule
        \multicolumn{2}{c}{HiFiC} &  \multicolumn{2}{c}{MS-ILLM} \\
        \multicolumn{2}{c}{(Mentzer~\etal)} &  \multicolumn{2}{c}{(Muckley~\etal)} \\
       \midrule
    \multicolumn{2}{c}{$0.235$bpp ($1.28\times$)} & \multicolumn{2}{c}{$0.185$bpp}  \\
       \begin{subfigure}{0.242\textwidth}
        \centering
        \includegraphics[width=\linewidth]{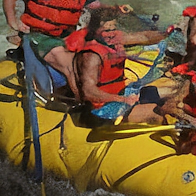}
       \end{subfigure}
       & 
       \begin{subfigure}{0.242\textwidth}
        \centering
        \includegraphics[width=\linewidth]{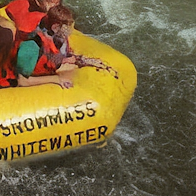}
       \end{subfigure}
       & 
       \begin{subfigure}{0.242\textwidth}
        \centering
        \includegraphics[width=\linewidth]{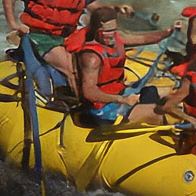}
        \end{subfigure}
        & 
       \begin{subfigure}{0.242\textwidth}
        \centering
        \includegraphics[width=\linewidth]{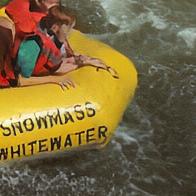}
        \end{subfigure}\\
     \multicolumn{2}{c}{\begin{subfigure}{0.49\textwidth}
        \centering
        \includegraphics[width=\linewidth]{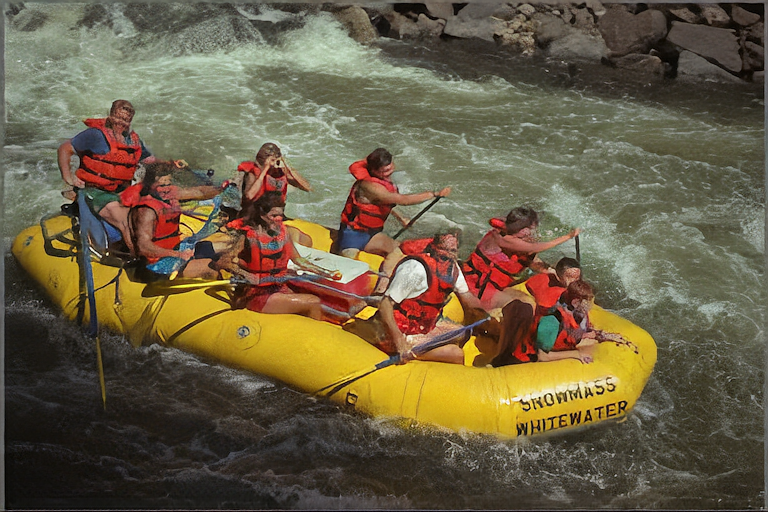}
       \end{subfigure}}
       & 
       \multicolumn{2}{c}{\begin{subfigure}{0.49\textwidth}
        \centering
        \includegraphics[width=\linewidth]{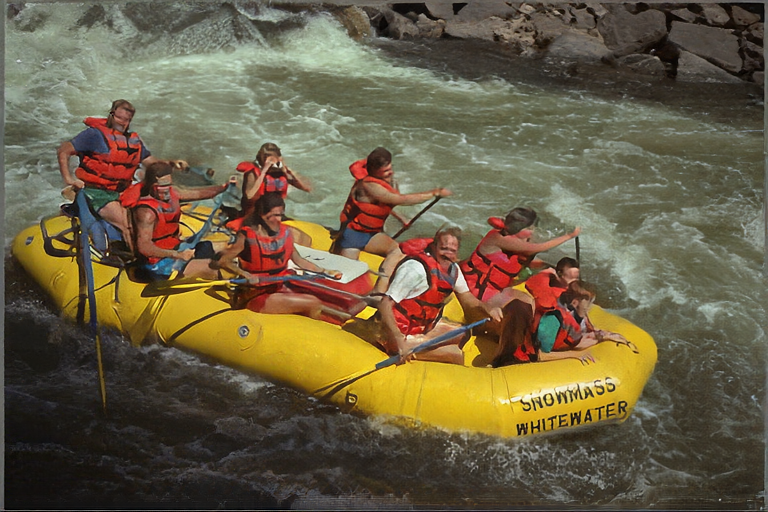}
       \end{subfigure}} \\
   \end{tabular}
   \caption{Visual comparison of EGIC ($\alpha=1.0$) with HiFiC and MS-ILLM on the Kodak dataset (kodim14). Note that our method better preserves small faces.}
   \label{tbl:vis_kodak_14}
\end{figure*}

\begin{figure*}[bt]
   \setlength{\tabcolsep}{1pt}
   \renewcommand{\arraystretch}{0.5}
   \centering
   
   \begin{tabular}{cccc}
   \toprule
       \multicolumn{2}{c}{input $x$} & \multicolumn{2}{c}{EGIC} \\
       \multicolumn{2}{c}{}  & \multicolumn{2}{c}{Ours ($\alpha=1.0$)} \\
    \midrule
    \multicolumn{2}{c}{kodim11} & \multicolumn{2}{c}{$0.134$bpp}  \\
       \begin{subfigure}{0.242\textwidth}
        \centering
        \includegraphics[width=\linewidth]{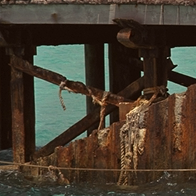}
       \end{subfigure}
       & 
       \begin{subfigure}{0.242\textwidth}
        \centering
        \includegraphics[width=\linewidth]{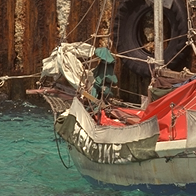}
       \end{subfigure}
       & 
       \begin{subfigure}{0.242\textwidth}
        \centering
        \includegraphics[width=\linewidth]{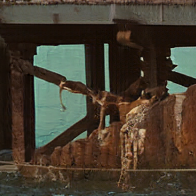}
        \end{subfigure}
        & 
       \begin{subfigure}{0.242\textwidth}
        \centering
        \includegraphics[width=\linewidth]{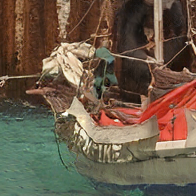}
        \end{subfigure}\\
     \multicolumn{2}{c}{\begin{subfigure}{0.49\textwidth}
        \centering
        \includegraphics[width=\linewidth]{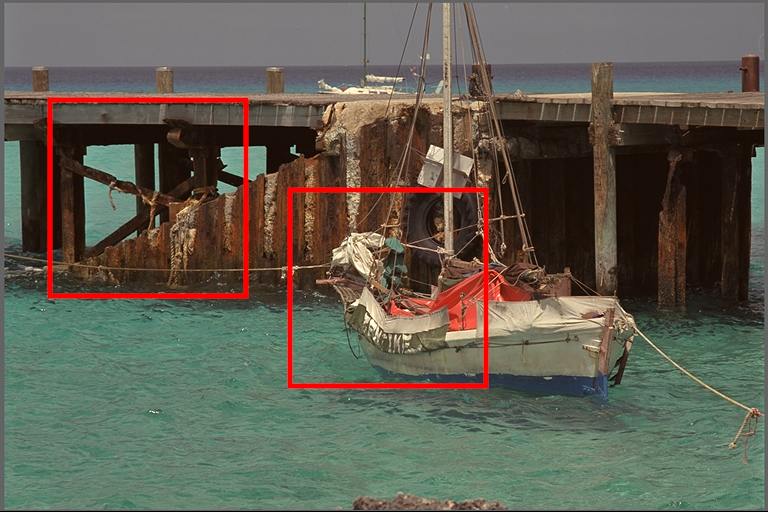}
       \end{subfigure}}
       & 
       \multicolumn{2}{c}{\begin{subfigure}{0.49\textwidth}
        \centering
        \includegraphics[width=\linewidth]{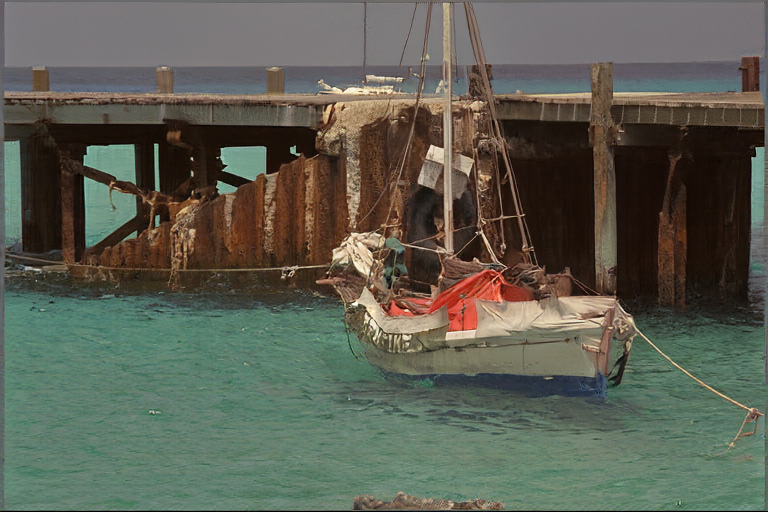}
       \end{subfigure}} \\
       \midrule
        \multicolumn{2}{c}{HiFiC} &  \multicolumn{2}{c}{MS-ILLM} \\
        \multicolumn{2}{c}{(Mentzer~\etal)} &  \multicolumn{2}{c}{(Muckley~\etal)} \\
       \midrule
    \multicolumn{2}{c}{$0.186$bpp ($1.39\times$)} & \multicolumn{2}{c}{$0.151$bpp ($1.13\times$)}  \\
       \begin{subfigure}{0.242\textwidth}
        \centering
        \includegraphics[width=\linewidth]{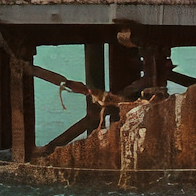}
       \end{subfigure}
       & 
       \begin{subfigure}{0.242\textwidth}
        \centering
        \includegraphics[width=\linewidth]{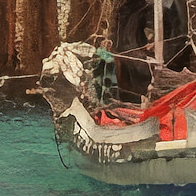}
       \end{subfigure}
       & 
       \begin{subfigure}{0.242\textwidth}
        \centering
        \includegraphics[width=\linewidth]{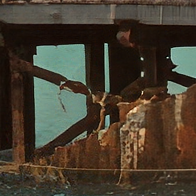}
        \end{subfigure}
        & 
       \begin{subfigure}{0.242\textwidth}
        \centering
        \includegraphics[width=\linewidth]{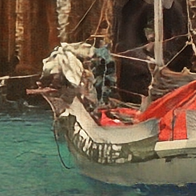}
        \end{subfigure}\\
     \multicolumn{2}{c}{\begin{subfigure}{0.49\textwidth}
        \centering
        \includegraphics[width=\linewidth]{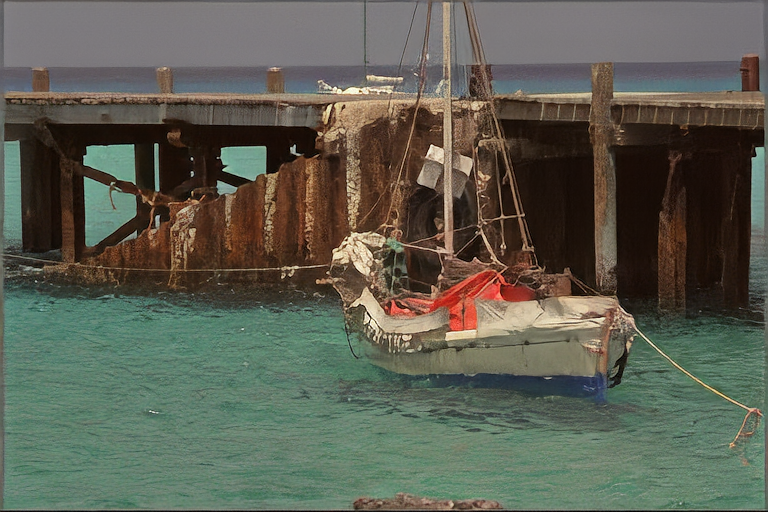}
       \end{subfigure}}
       & 
       \multicolumn{2}{c}{\begin{subfigure}{0.49\textwidth}
        \centering
        \includegraphics[width=\linewidth]{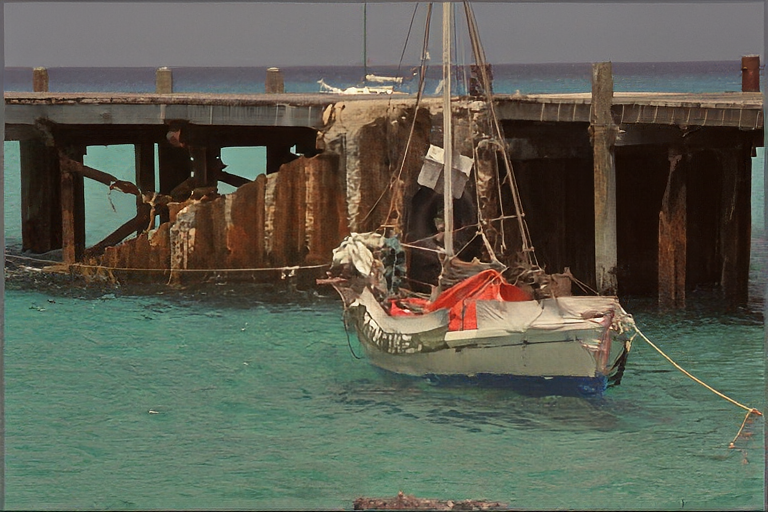}
       \end{subfigure}} \\
   \end{tabular}
   \caption{Visual comparison of EGIC ($\alpha=1.0$) with HiFiC and MS-ILLM on the Kodak dataset (kodim11). Note that our method better preserves small details (\eg, the rope in the left image), despite using less bpp.}
   \label{tbl:vis_kodak_11}
\end{figure*}

\begin{figure*}[bt]
   \setlength{\tabcolsep}{1pt}
   \renewcommand{\arraystretch}{0.5}
   \centering
   
   \begin{tabular}{cccc}
   \toprule
       \multicolumn{2}{c}{input $x$} & \multicolumn{2}{c}{EGIC} \\
       \multicolumn{2}{c}{}  & \multicolumn{2}{c}{Ours ($\alpha=1.0$)} \\
    \midrule
    \multicolumn{2}{c}{kodim21} & \multicolumn{2}{c}{$0.157$bpp}  \\
       \begin{subfigure}{0.242\textwidth}
        \centering
        \includegraphics[width=\linewidth]{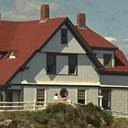}
       \end{subfigure}
       & 
       \begin{subfigure}{0.242\textwidth}
        \centering
        \includegraphics[width=\linewidth]{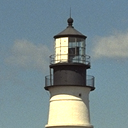}
       \end{subfigure}
       & 
       \begin{subfigure}{0.242\textwidth}
        \centering
        \includegraphics[width=\linewidth]{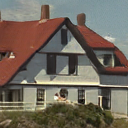}
        \end{subfigure}
        & 
       \begin{subfigure}{0.242\textwidth}
        \centering
        \includegraphics[width=\linewidth]{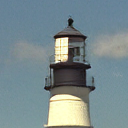}
        \end{subfigure}\\
     \multicolumn{2}{c}{\begin{subfigure}{0.49\textwidth}
        \centering
        \includegraphics[width=\linewidth]{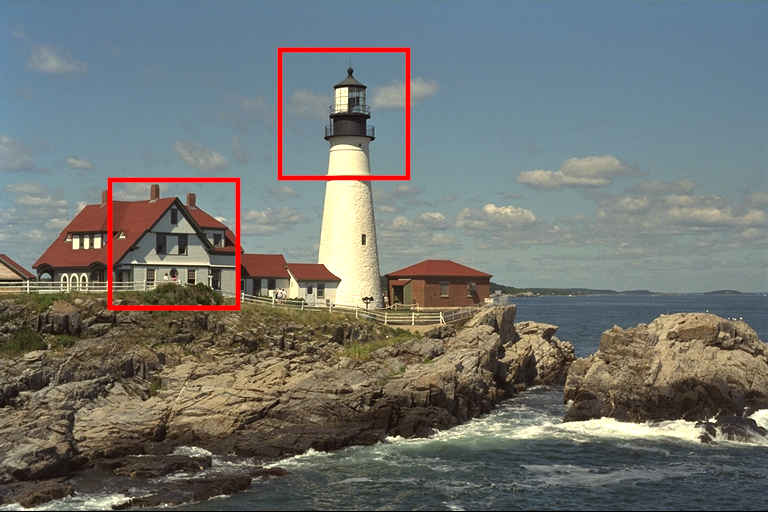}
       \end{subfigure}}
       & 
       \multicolumn{2}{c}{\begin{subfigure}{0.49\textwidth}
        \centering
        \includegraphics[width=\linewidth]{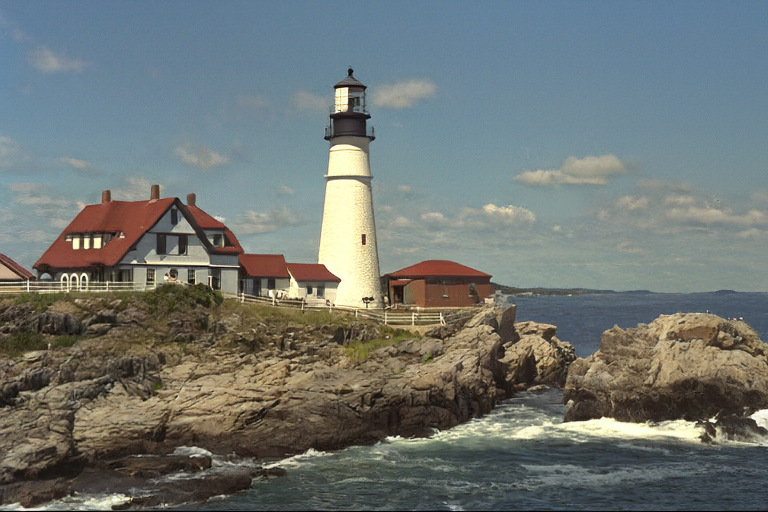}
       \end{subfigure}} \\
       \midrule
        \multicolumn{2}{c}{HiFiC} &  \multicolumn{2}{c}{MS-ILLM} \\
        \multicolumn{2}{c}{(Mentzer~\etal)} &  \multicolumn{2}{c}{(Muckley~\etal)} \\
       \midrule
    \multicolumn{2}{c}{$0.173$bpp ($1.1\times$)} & \multicolumn{2}{c}{$0.157$bpp}  \\
       \begin{subfigure}{0.242\textwidth}
        \centering
        \includegraphics[width=\linewidth]{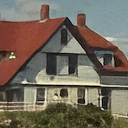}
       \end{subfigure}
       & 
       \begin{subfigure}{0.242\textwidth}
        \centering
        \includegraphics[width=\linewidth]{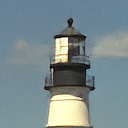}
       \end{subfigure}
       & 
       \begin{subfigure}{0.242\textwidth}
        \centering
        \includegraphics[width=\linewidth]{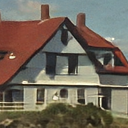}
        \end{subfigure}
        & 
       \begin{subfigure}{0.242\textwidth}
        \centering
        \includegraphics[width=\linewidth]{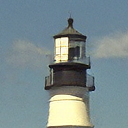}
        \end{subfigure}\\
     \multicolumn{2}{c}{\begin{subfigure}{0.49\textwidth}
        \centering
        \includegraphics[width=\linewidth]{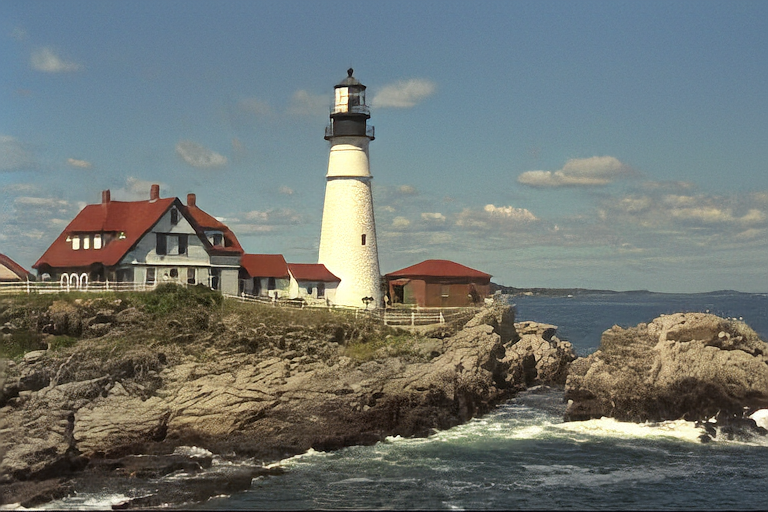}
       \end{subfigure}}
       & 
       \multicolumn{2}{c}{\begin{subfigure}{0.49\textwidth}
        \centering
        \includegraphics[width=\linewidth]{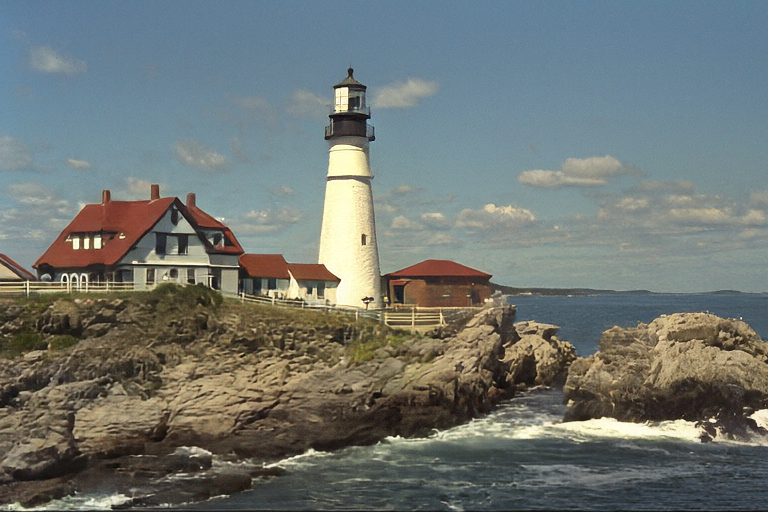}
       \end{subfigure}} \\
   \end{tabular}
   \caption{Visual comparison of EGIC ($\alpha=1.0$) with HiFiC and MS-ILLM on the Kodak dataset (kodim21). Note that our method better preserves small details (\eg, the people in the left image).}
   \label{tbl:vis_kodak_21}
\end{figure*}

\begin{figure*}[bt]
   \setlength{\tabcolsep}{1pt}
   \renewcommand{\arraystretch}{0.5}
   \centering
   
   \begin{tabular}{cccc}
         \multicolumn{4}{c}{\begin{subfigure}{0.98\textwidth}
        \centering
        \includegraphics[width=\linewidth]{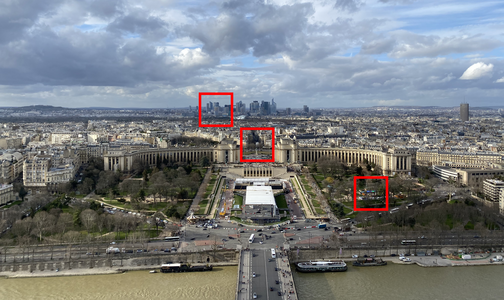}
   \end{subfigure}}\\
   \toprule
       input $x$ &  MRIC ($\beta=2.56$) &DIRAC-$100$      & EGIC \\
        &  (Agustsson~\etal)     & (Ghouse~\etal) & Ours ($\alpha=1.0$) \\
    \midrule
    1ac06 & $0.166$bpp ($1.04\times$) & $0.157$bpp ($0.99\times$)& $0.159$bpp \\
       \begin{subfigure}{0.24\textwidth}
        \centering
        \includegraphics[width=\linewidth]{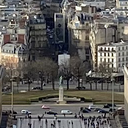}
       \end{subfigure}
       & 
       \begin{subfigure}{0.24\textwidth}
        \centering
        \includegraphics[width=\linewidth]{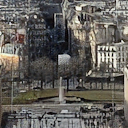}
       \end{subfigure}
       & 
       \begin{subfigure}{0.24\textwidth}
        \centering
        \includegraphics[width=\linewidth]{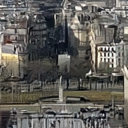}
        \end{subfigure}
        & 
       \begin{subfigure}{0.24\textwidth}
        \centering
        \includegraphics[width=\linewidth]{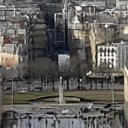}
        \end{subfigure}\\
       \begin{subfigure}{0.24\textwidth}
        \centering
        \includegraphics[width=\linewidth]{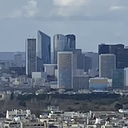}
       \end{subfigure}
       & 
       \begin{subfigure}{0.24\textwidth}
        \centering
        \includegraphics[width=\linewidth]{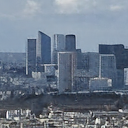}
       \end{subfigure}
       & 
       \begin{subfigure}{0.24\textwidth}
        \centering
        \includegraphics[width=\linewidth]{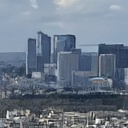}
        \end{subfigure}
        & 
       \begin{subfigure}{0.24\textwidth}
        \centering
        \includegraphics[width=\linewidth]{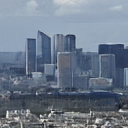}
        \end{subfigure}\\
       \begin{subfigure}{0.24\textwidth}
        \centering
        \includegraphics[width=\linewidth]{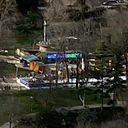}
       \end{subfigure}
       & 
       \begin{subfigure}{0.24\textwidth}
        \centering
        \includegraphics[width=\linewidth]{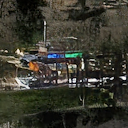}
       \end{subfigure}
       & 
       \begin{subfigure}{0.24\textwidth}
        \centering
        \includegraphics[width=\linewidth]{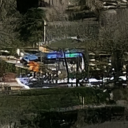}
        \end{subfigure}
        & 
       \begin{subfigure}{0.24\textwidth}
        \centering
        \includegraphics[width=\linewidth]{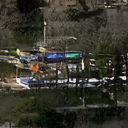}
        \end{subfigure}\\

   \end{tabular}
   \caption{Visual comparison of EGIC ($\alpha=1.0$) with MRIC and DIRAC-$100$ on the CLIC dataset (1ac06). Note that EGIC has less artifacts (compared to MRIC) and better retrains color (compared to DIRAC).}
   \label{tbl:vis_1ac06}
\end{figure*}

\begin{figure*}[bt]
   \setlength{\tabcolsep}{1pt}
   \renewcommand{\arraystretch}{0.5}
   \centering
   
   \begin{tabular}{cccc}
        \multicolumn{4}{c}{\begin{subfigure}{0.97\textwidth}
        \centering
        \includegraphics[width=\linewidth]{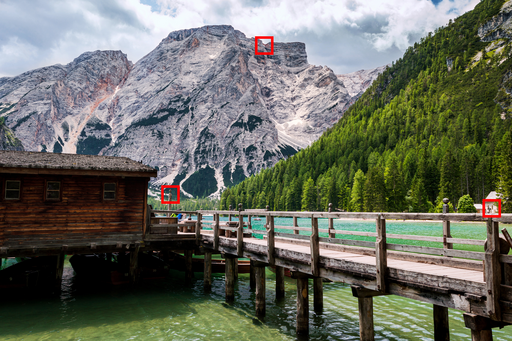}
       \end{subfigure}}\\
   \toprule
       input $x$ & MRIC ($\beta=2.56$) & DIRAC-$100$      & EGIC \\
        &  (Agustsson~\etal)     & (Ghouse~\etal) & Ours ($\alpha=1.0$) \\
    \midrule
    46c18 & $0.219$bpp ($1.08\times$) & $0.217$bpp ($1.07\times$)& $0.202$bpp \\
       \begin{subfigure}{0.24\textwidth}
        \centering
        \includegraphics[width=\linewidth]{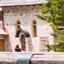}
       \end{subfigure}
       & 
       \begin{subfigure}{0.24\textwidth}
        \centering
        \includegraphics[width=\linewidth]{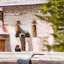}
       \end{subfigure}
       & 
       \begin{subfigure}{0.24\textwidth}
        \centering
        \includegraphics[width=\linewidth]{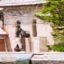}
        \end{subfigure}
        & 
       \begin{subfigure}{0.24\textwidth}
        \centering
        \includegraphics[width=\linewidth]{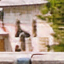}
        \end{subfigure}\\
       \begin{subfigure}{0.24\textwidth}
        \centering
        \includegraphics[width=\linewidth]{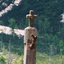}
       \end{subfigure}
       & 
       \begin{subfigure}{0.24\textwidth}
        \centering
        \includegraphics[width=\linewidth]{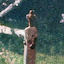}
       \end{subfigure}
       & 
       \begin{subfigure}{0.24\textwidth}
        \centering
        \includegraphics[width=\linewidth]{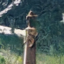}
        \end{subfigure}
        & 
       \begin{subfigure}{0.24\textwidth}
        \centering
        \includegraphics[width=\linewidth]{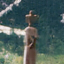}
        \end{subfigure}\\
       \begin{subfigure}{0.24\textwidth}
        \centering
        \includegraphics[width=\linewidth]{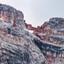}
       \end{subfigure}
       & 
       \begin{subfigure}{0.24\textwidth}
        \centering
        \includegraphics[width=\linewidth]{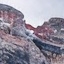}
       \end{subfigure}
       & 
       \begin{subfigure}{0.24\textwidth}
        \centering
        \includegraphics[width=\linewidth]{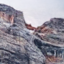}
        \end{subfigure}
        & 
       \begin{subfigure}{0.24\textwidth}
        \centering
        \includegraphics[width=\linewidth]{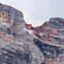}
        \end{subfigure}\\

   \end{tabular}
   \caption{Visual comparison of EGIC ($\alpha=1.0$) with MRIC and DIRAC-$100$ on the CLIC dataset (46c18). Note that we use less bpp.}
   \label{tbl:vis_46c18}
\end{figure*}

\begin{figure*}[bt]
   \setlength{\tabcolsep}{1pt}
   \renewcommand{\arraystretch}{0.5}
   \centering
   
   \begin{tabular}{ccc} 
   \toprule
       input $x$ & PO-ELIC & EGIC \\
        &  (He~\etal)  & Ours ($\alpha=1.0$) \\
    \midrule
    732bf & $0.068$bpp ($0.95\times$) & $0.0716$bpp \\
       \begin{subfigure}{0.33\textwidth}
        \centering
        \includegraphics[width=\linewidth]{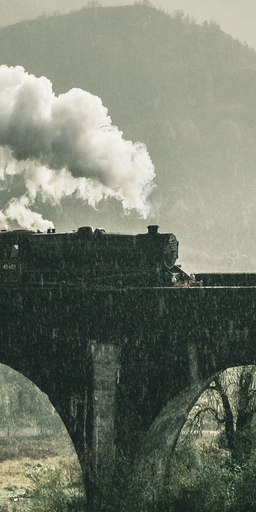}
       \end{subfigure}
       & 
       \begin{subfigure}{0.33\textwidth}
        \centering
        \includegraphics[width=\linewidth]{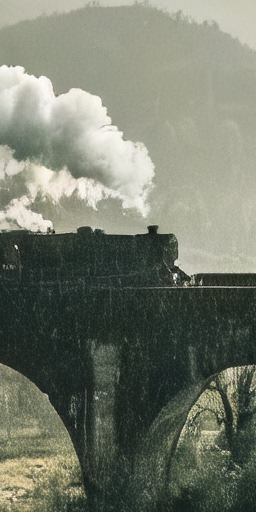}
       \end{subfigure}
       & 
       \begin{subfigure}{0.33\textwidth}
        \centering
        \includegraphics[width=\linewidth]{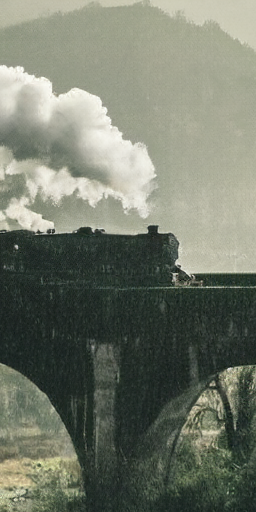}
       \end{subfigure}

   \end{tabular}
   \caption{Visual comparison of EGIC ($\alpha=1.0$) with PO-ELIC, the winning solution of the CLIC 2022 competition, using our lowest bit-rate setting.}
   \label{tbl:vis_732bf}
\end{figure*}

\begin{figure*}[bt]
   \setlength{\tabcolsep}{1pt}
   \renewcommand{\arraystretch}{0.5}
   \centering
   
   \begin{tabular}{cccc}
   \toprule
       \multicolumn{2}{c}{input $x$} & \multicolumn{2}{c}{EGIC} \\
       \multicolumn{2}{c}{}  & \multicolumn{2}{c}{Ours ($\alpha=1.0$)} \\
    \midrule
    \multicolumn{2}{c}{kodim24} & \multicolumn{2}{c}{$0.197$bpp}  \\
       \begin{subfigure}{0.242\textwidth}
        \centering
        \includegraphics[width=\linewidth]{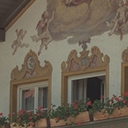}
       \end{subfigure}
       & 
       \begin{subfigure}{0.242\textwidth}
        \centering
        \includegraphics[width=\linewidth]{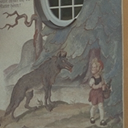}
       \end{subfigure}
       & 
       \begin{subfigure}{0.242\textwidth}
        \centering
        \includegraphics[width=\linewidth]{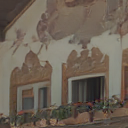}
        \end{subfigure}
        & 
       \begin{subfigure}{0.242\textwidth}
        \centering
        \includegraphics[width=\linewidth]{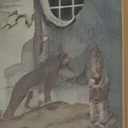}
        \end{subfigure}\\
     \multicolumn{2}{c}{\begin{subfigure}{0.49\textwidth}
        \centering
        \includegraphics[width=\linewidth]{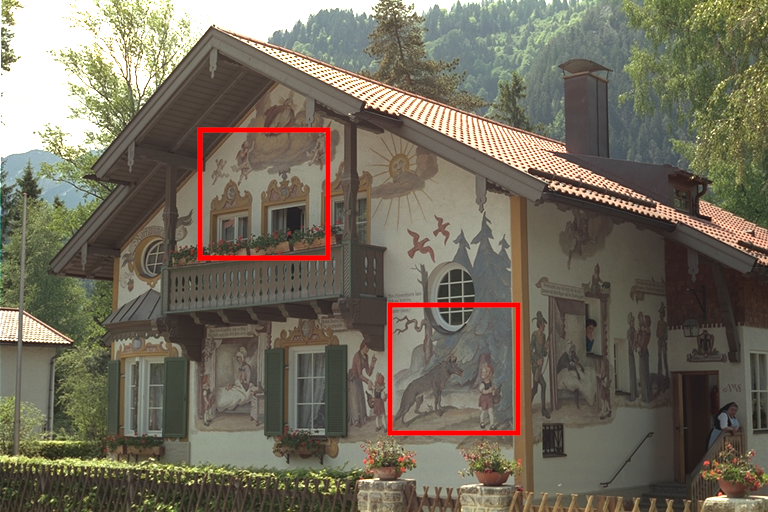}
       \end{subfigure}}
       & 
       \multicolumn{2}{c}{\begin{subfigure}{0.49\textwidth}
        \centering
        \includegraphics[width=\linewidth]{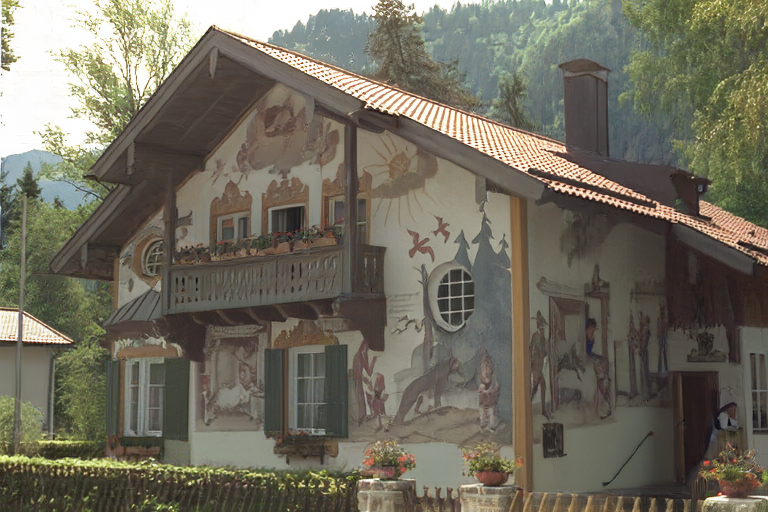}
       \end{subfigure}} \\
       \midrule
        \multicolumn{2}{c}{HFD/DDPM (ELIC)} &  \multicolumn{2}{c}{HFD/DDPM (250 steps)} \\
        \multicolumn{2}{c}{(Hoogeboom~\etal)} &  \multicolumn{2}{c}{(Hoogeboom~\etal)} \\
       \midrule
    \multicolumn{2}{c}{$0.196$bpp} & \multicolumn{2}{c}{$0.196$bpp}  \\
       \begin{subfigure}{0.242\textwidth}
        \centering
        \includegraphics[width=\linewidth]{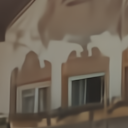}
       \end{subfigure}
       & 
       \begin{subfigure}{0.242\textwidth}
        \centering
        \includegraphics[width=\linewidth]{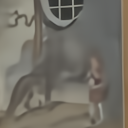}
       \end{subfigure}
       & 
       \begin{subfigure}{0.242\textwidth}
        \centering
        \includegraphics[width=\linewidth]{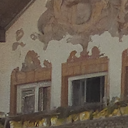}
        \end{subfigure}
        & 
       \begin{subfigure}{0.242\textwidth}
        \centering
        \includegraphics[width=\linewidth]{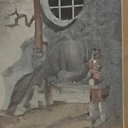}
        \end{subfigure}\\
     \multicolumn{2}{c}{\begin{subfigure}{0.49\textwidth}
        \centering
        \includegraphics[width=\linewidth]{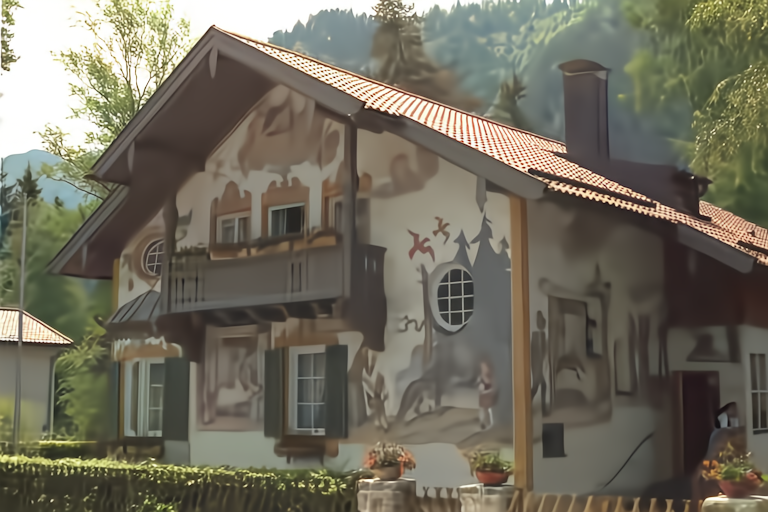}
       \end{subfigure}}
       & 
       \multicolumn{2}{c}{\begin{subfigure}{0.49\textwidth}
        \centering
        \includegraphics[width=\linewidth]{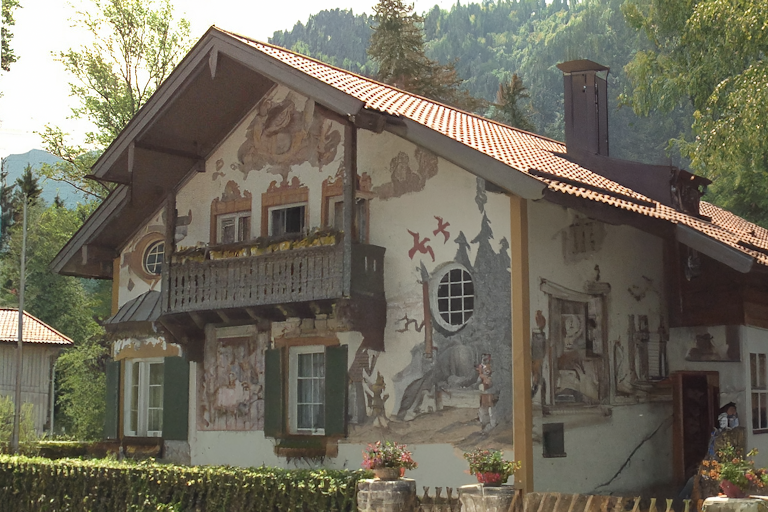}
       \end{subfigure}} \\
   \end{tabular}
   \caption{Visual comparison of EGIC ($\alpha=1.0$) with HFD/DDPM on the Kodak dataset (kodim24). Note that the quality of HFD/DDPM (250 steps) largely depends on the base reconstruction HFD/DDPM (ELIC).}
   \label{tbl:vis_hfd_ddpm_kodim24}
\end{figure*}

\begin{figure*}[bt]
   \setlength{\tabcolsep}{1pt}
   \renewcommand{\arraystretch}{0.5}
   \centering
   
   \begin{tabular}{cccc}
         \multicolumn{4}{c}{\begin{subfigure}{0.98\textwidth}
        \centering
        \includegraphics[width=\linewidth]{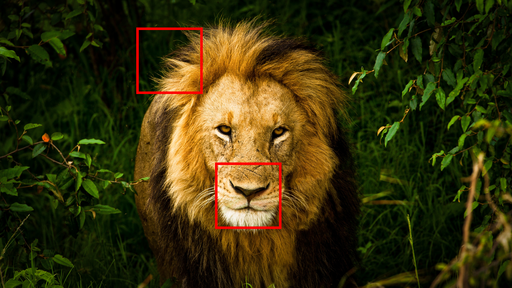}
       \end{subfigure}}\\
   \toprule
       \multicolumn{2}{c}{input $x$} & \multicolumn{2}{c}{EGIC} \\
       \multicolumn{2}{c}{}  & \multicolumn{2}{c}{Ours ($\alpha=1.0$)} \\
    \midrule
    \multicolumn{2}{c}{2ff70} & \multicolumn{2}{c}{$0.092$bpp}  \\
       \begin{subfigure}{0.242\textwidth}
        \centering
        \includegraphics[width=\linewidth]{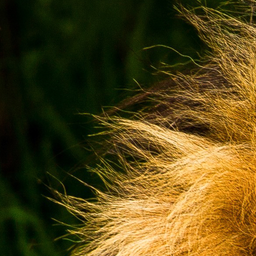}
       \end{subfigure}
       & 
       \begin{subfigure}{0.242\textwidth}
        \centering
        \includegraphics[width=\linewidth]{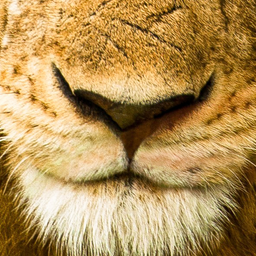}
       \end{subfigure}
       & 
       \begin{subfigure}{0.242\textwidth}
        \centering
        \includegraphics[width=\linewidth]{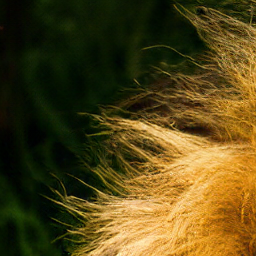}
        \end{subfigure}
        & 
       \begin{subfigure}{0.242\textwidth}
        \centering
        \includegraphics[width=\linewidth]{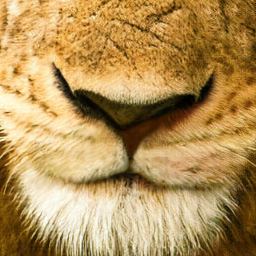}
        \end{subfigure}\\
       \midrule
        \multicolumn{2}{c}{HFD/DDPM (ELIC)} &  \multicolumn{2}{c}{HFD/DDPM (250 steps)} \\
        \multicolumn{2}{c}{(Hoogeboom~\etal)} &  \multicolumn{2}{c}{(Hoogeboom~\etal)} \\
       \midrule
    \multicolumn{2}{c}{$0.106$bpp ($1.15\times$)} & \multicolumn{2}{c}{$0.106$bpp ($1.15\times$)}  \\
       \begin{subfigure}{0.242\textwidth}
        \centering
        \includegraphics[width=\linewidth]{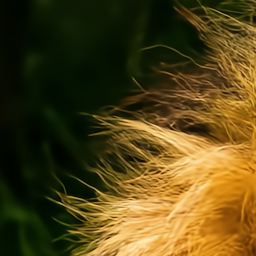}
       \end{subfigure}
       & 
       \begin{subfigure}{0.242\textwidth}
        \centering
        \includegraphics[width=\linewidth]{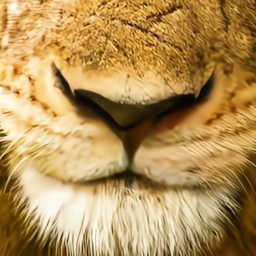}
       \end{subfigure}
       & 
       \begin{subfigure}{0.242\textwidth}
        \centering
        \includegraphics[width=\linewidth]{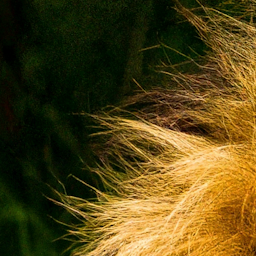}
        \end{subfigure}
        & 
       \begin{subfigure}{0.242\textwidth}
        \centering
        \includegraphics[width=\linewidth]{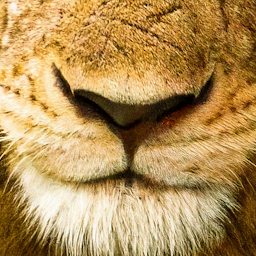}
        \end{subfigure}\\
   \end{tabular}
   \caption{Visual comparison of EGIC ($\alpha=1.0$) with HFD/DDPM on the CLIC 2022 dataset (2ff70). We leave the assessment to the reader.}
   \label{tbl:vis_hfd_ddpm_clic_2ff70}
\end{figure*}

We provide extensive visual comparison to JPEG, BPG-0.9.8 and VTM.20.0 in~\cref{tbl:vis_kodak_13,tbl:vis_kodak_22}, to HiFiC and MS-ILLM in~\cref{tbl:vis_kodak_20,tbl:vis_kodak_14,tbl:vis_kodak_11,tbl:vis_kodak_21}, to MRIC ($\beta=2.56$) and DIRAC-$100$ in~\cref{tbl:vis_1ac06,tbl:vis_46c18}, to PO-ELIC in~\cref{tbl:vis_732bf} and to HFD/DDPM in~\cref{tbl:vis_hfd_ddpm_kodim24,tbl:vis_hfd_ddpm_clic_2ff70}.

\clearpage

\clearpage

\clearpage

\clearpage

\clearpage

\clearpage

\clearpage

\clearpage

\clearpage

\clearpage

\clearpage

\clearpage

\end{document}